\documentclass[pra, amsmath, reprint, amssymb, aps, superscriptaddress, onecolumn]{revtex4-1}
\usepackage{graphicx}
\usepackage{amsmath, braket, amsfonts}

\usepackage{multirow}
\usepackage{mathrsfs,amsmath} 
\usepackage{fixmath}
\usepackage{graphicx}
\usepackage[colorlinks,linkcolor=blue,anchorcolor=red,citecolor=green]{hyperref}
\usepackage{placeins}
\usepackage{cancel}

\renewcommand\vec{\boldsymbol}

\begin{document}


\title{The universal shear conductivity of Fermi liquids and spinon Fermi surface states and its detection via spin qubit noise magnetometry}

\author{Jun Yong Khoo}
\affiliation{Max-Planck Institute for the Physics of Complex Systems, D-01187 Dresden, Germany}
\affiliation{Institute of High Performance Computing, Agency for Science, Technology, and Research, Singapore 138632}

\author{Falko Pientka}
\affiliation{
Institute of Theoretical Physics, Goethe University, 60438 Frankfurt a.M., Germany
}
\affiliation{Max-Planck Institute for the Physics of Complex Systems, D-01187 Dresden, Germany}

\author{Inti Sodemann}
\affiliation{Max-Planck Institute for the Physics of Complex Systems, D-01187 Dresden, Germany}
\affiliation{Department of Physics and Astronomy, University of California, Irvine, California 92697, USA}


\date{\today}

\begin{abstract}
We demonstrate a remarkable property of metallic Fermi liquids: the transverse conductivity assumes a universal value in the quasi-static ($\omega \rightarrow 0$) limit for wavevectors $q$ in the regime $l_{\rm mfp}^{-1} \ll q \ll p_{\rm F}$, where $l_{\rm mfp}$ is the mean free path and $p_{\rm F}$ is the Fermi momentum. This value is $(e^2/h) \mathcal{R}_{\rm FS}/q$ in two dimensions (2D), where $\mathcal{R}_{\rm FS}$ measures the local radius of curvature of the Fermi surface in momentum space. Even more surprisingly, we find that U(1) spin liquids with a spinon Fermi surface have the same universal transverse conductivity. This means such spin liquids behave effectively as metals in this regime, even though they appear insulating in standard transport experiments. Moreover, we show that transverse current fluctuations result in a universal low-frequency magnetic noise that can be directly probed by a spin qubit, such as a nitrogen-vacancy center in diamond, placed at a distance $z$ above of the 2D metal or spin liquid. Specifically the magnetic noise is given by $C\omega \mathcal{P}_{\rm FS}/z$, where $\mathcal{P}_{\rm FS}$ is the perimeter of the Fermi surface in momentum space and $C$ is a combination of fundamental constants of nature. Therefore these observables are controlled purely by the geometry of the Fermi surface and are independent of kinematic details of the quasi-particles, such as their effective mass and interactions. This behavior can be used as a new technique to measure the size of the Fermi surface of metals and as a smoking gun probe to pinpoint the presence of the elusive spinon Fermi surface in two-dimensional systems. We estimate that this universal regime is within reach of current nitrogen-vacancy center spectroscopic techniques for several spinon Fermi surface candidate materials.
\end{abstract}


\maketitle

\section{Introduction}

\begin{figure}
\includegraphics[scale=1.0]{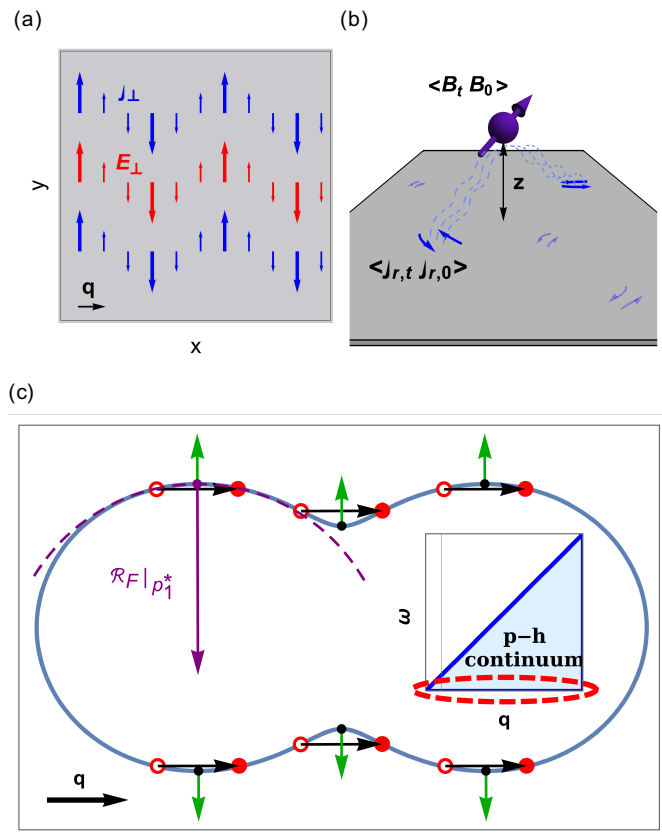}
\caption{\label{Fig.cartoon}
(a) Transverse currents (blue arrows) and electric fields (red arrows) with wavevector $\vec{q}$.
(b) Schematic of a spin qubit located at a distance $z$ above a 2D sample detecting the magnetic noise $\left\langle B_{t}B_{0}\right\rangle$ induced by current fluctuations $\left\langle j_{\vec{r},t}j_{\vec{r},0}\right\rangle$ in the sample.
(c) Depiction of the particle-hole excitations with small wavevector $\vec{q}$ that are tangential to the FS (solid blue line) and dominate the dissipative transverse conductivity.
For a given $\vec{q}$ these excitations are located near certain points, $\lbrace \vec{p} ^*_{i} \rbrace$, where the Fermi velocity is orthogonal to $\vec{q}$, and their contribution to the universal transverse conductivity depends only on the local radius of curvature
$\mathcal{R}_{\rm F} |_{\vec{p}_i^*}$, which we show explicitly in purple for one of these points $\vec{p}_1^*$.
The inset shows the region of the particle-hole continuum [cf. Fig.~\ref{Fig.condmetal}(b)] containing these $(\vec{q}, \omega \rightarrow 0)$ excitations.
}
\end{figure}

There are relatively few measurable properties of systems with a Fermi surface (FS) that remain unchanged by details of interactions and dispersion. One notable example is the invariance of the period of the quantum oscillations~\cite{Kohn1961}, which serves as tool to  measure of the cross sectional area of the FS. In this work we will demonstrate that a different quantity enjoys a similar degree of universality. This quantity is the quasi-static transverse or shear conductivity, denoted by $\sigma_{\perp, 0}(\vec{q})$ that measures the net current in response to a nearly static but spatially oscillating transverse or shear force (or equivalently a transverse electric field when the Fermi liquid is charged) with wavevector $\bf{q}$, as depicted in Fig.~\ref{Fig.cartoon}(a). As we will see in the ``quantum'' regime where the wavevector of the applied force satisfies $p_{\rm F} \gg q \gg l_{\rm mfp}^{-1}$, where $p_{\rm F}$ denotes the Fermi momentum and $l_{\rm mfp}$ the mean free path, and in the low-frequency quasi-static regime ($\omega \rightarrow 0$), the transverse conductivity takes the following universal form in two dimensions:
\begin{eqnarray}
\sigma _{\perp, 0} (\vec{q})
&=& (2S+1) \frac{e^2}{2 h q}\sum_i \mathcal{R}_i.\label{Eq.main1}
\end{eqnarray}
Here $e$ is the electron's charge, $h$ Planck's constant, $(2S + 1)$  is the spin degeneracy factor, and $\mathcal{R}_i$ is the absolute value of the local radius of curvature of the FS at points $i$ on the FS at which the Fermi velocity is orthogonal to the direction of the wavevector $\vec{q}$, as depicted in Fig.~\ref{Fig.cartoon}(c). Therefore, this limit is universal in the sense that it is independent of the quasi-particle mass and interactions, and only controlled by the local geometric shape of the FS. 

Remarkably, we have found that the exact same limit of Eq.~\eqref{Eq.main1} is also approached by the transverse electric conductivity of a strongly correlated state, namely, the U(1) spin liquid with a spinon FS (for reviews see~\cite{Savary2016,Zhou2017,Broholm2020}), which has been a ``holy grail'' of condensed matter research since the pioneering ideas of Anderson~\cite{ANDERSON1987}. This state features a form of spin-charge separation above one-dimension, in which the electron fractionalizes into a spinful fermion (the spinon) and a spinless boson (the chargon or holon). The chargon is gapped and the spinon remains in a gapless FS state, but both particles remain strongly coupled via an emergent photon field. This state displays electromagnetic responses that are a sort of blend of insulating and metallic behavior. On the one hand, while it has a vanishing electrical conductivity at $q=0$ in the DC zero temperature limit just like insulators, it also displays power-law subgap optical conductivity~\cite{Ng2007}, and even more strikingly, it can feature quantum oscillations under magnetic fields~\cite{Motrunich2006,Chowdhury2018,Sodemann2018} and cyclotron resonance~\cite{Rao2019}, in analogy to metals. Our findings therefore highlight that the electric transverse conductivity of the spinon FS not only behaves similar to a metal but, indeed, approaches the same universal limit at low frequencies given in Eq.~\eqref{Eq.main1}, although, as we will see, the crossover to such a regime occurs typically at much lower frequencies than in a metal.

While the universality of this limit of the transverse conductivity in ordinary Fermi liquids has been known since the early days of Landau Fermi liquid theory~\cite{Pines}, to our knowledge, its precise form for anisotropic FSs has not been derived previously in two-dimensions, although a related dependence of the quasi-static conductivity at finite wavevector on the FS curvature has also been discussed in the context of the anomalous skin effect in 3D metals~\cite{Pippard1954,Abrikosov}. More importantly, to this date there is no report of the experimental observation of this remarkable universal regime of the transverse conductivity even in ordinary two-dimensional Fermi liquids or metals. This is largely because it is experimentally challenging to probe the linear response regime, where Eq.~\eqref{Eq.main1} holds, by controllably applying external transverse electric fields (shear forces) with a finite wavevector that is not too small ($q \gg l_{\rm mfp}^{-1}$) and at very low frequencies. There is however, an alternative way to probe linear response functions that does not require actively applying external perturbations on the system, but instead to monitor its fluctuations since the fluctuation-dissipation theorem dictates that these are governed by the dissipative part of the same linear response susceptibilities. This is the key idea behind the technique of magnetic noise spectroscopy of nitrogen-vacancy center spin qubits~\cite{Casola2018}, which is emerging as a powerful tool to study current and spin correlations of diverse condensed matter systems~\cite{Andersen2019,Hsieh2019,Agarwal,Chatterjee2019,Ariyaratne2018,Kolkowitz1129}.

As we will demonstrate, the above regime of the transverse conductivity gives rise to a universal regime of the magnetic field noise when it is probed at a distance $z$ above the 2D sample, as depicted in Fig.~\ref{Fig.cartoon}(b), within the range $p_{\rm F}^{-1} \gg z \gg l_{\rm mfp}$, and at low temperatures and low frequencies. In this regime, the magnetic field autocorrelation function at a single point, takes the following universal form for both ordinary metals and spinon FS states:
\begin{equation}\label{Eq.main2}
\chi ''_{B_z B_z}  (z, \omega \rightarrow 0)
\simeq \frac{e^2 \mu _0 ^2}{16 \pi h} \frac{\omega}{z}\frac{(2S+1)}{2\pi}\mathcal{P}_{\rm FS} + \mathcal{O}(\omega^3).
\end{equation}
Here $\mathcal{P}_{\rm FS}$ is the perimeter of the FS in momentum space, and $\mu _0$ the permeability of free space.
This noise, which arises from orbital current fluctuations, dominates over the noise originating from the spin fluctuations in both the spinon FS state and metals. While this regime can be achieved for spinons only at much lower frequencies than for metals, we estimate that the required frequencies are of the order of MHz in organic spin liquid candidates and of order of GHz in transition metal dichalcogenide spin liquid candidates, placing them within experimental reach of current nitrogen-vacancy noise spectroscopic techniques~\cite{Casola2018}.

Our paper is organized as follows. 
In Sec.~\ref{Sec.condmetals}, we show that for a metal in the Landau Fermi liquid regime, the dissipative part of its conductivity tensor has only a single non-vanishing component in the quasi-static limit in the collisionless quantum regime, $\sigma _{\perp, 0} (\vec{q})$ [Eq.~\eqref{Eq.main1}]. The effects of collisions on the transverse conductivity are discussed in the isotropic system with a circular FS, where we establish the criteria on the frequency and wavevector to observe the universal value $\sigma _{\perp, 0} (\vec{q})$. 
We begin Sec.~\ref{Sec.condspinons} by introducing our treatment of the low-energy excitations of the spinon FS state, in which we replace its effective Lagrangian by the bosonized and linearized theory of the quantum Fermi liquid to obtain a bosonic bilinear theory in the Fermi radius operator and the internal gauge fields. We show that for an isotropic system, the conductivity obtained within this framework obeys the Ioffe-Larkin rule, and crucially, has the same quasi-static limit $\sigma _{\perp, 0} (\vec{q})$ in the collisionless quantum regime.  Analogous to the metallic case, we establish the criteria on the frequency and wavevector to observe the universal value. 
The physical intuition behind how an insulating spinon FS state exhibits the same quasi-static transverse conductivity as a metal is discussed in the last subsection.
In Sec.~\ref{Sec.noisespec}, we discuss how the low-frequency transverse conductivity can be probed by magnetic noise spectroscopy of nitrogen-vacancy center spin qubits.
In the collisionless regime, we show that the universal transverse conductivity gives rise to a corresponding universal quantum low-frequency noise [Eq.~\eqref{Eq.main2}]. 
We then proceed to discuss the effects of collisions on the noise and identify the relevant regimes for the frequency and the distance between sample and probe to detect the universal noise in both metals and spinon FS states experimentally. 
We conclude with a summary of our results and discuss their implications on the detection of U(1) spinon FS states in Sec.~\ref{Sec.summary}.
The derivations for the various results in each section are detailed in the Supplementary material~\cite{supplementary}.

\section{Transverse conductivity of electron Fermi surfaces}\label{Sec.condmetals}

\subsection{Universal transverse conductivity in the collisionless quantum regime}
At low temperatures, metals enter the quantum Landau Fermi liquid (LFL) regime. Its low-energy dynamics is described by the kinetic equation linearized to first order in the departure from the groundstate distribution $\delta n_{\vec{p}} (\vec{r}, t)$ and the 
amplitudes of the electromagnetic fields~\cite{Pines}

\begin{eqnarray}
\partial _t \delta n_{\vec{p}} &+& \vec{v}_{\vec{p}} \cdot \vec{\partial}_{\vec{r}}\delta \bar{n}_{\vec{p}} + \vec{E} \cdot \vec{v}_{\vec{p}} \delta (\epsilon _{\vec{p}} - \epsilon_{\rm F})
= I [\delta n_{\vec{p}}], \label{Eq.LKEfull} \\
\delta \bar{n}_{\vec{p}} &=& \delta n_{\vec{p}} + \sum _{\vec{p'}} f_{\vec{p}\vec{p'}}\delta (\epsilon _{\vec{p}} - \epsilon_{\rm F}) \delta n_{\vec{p'}}.
\end{eqnarray}
Here $\epsilon_{\vec{p}}$ and $\vec{v}_{\vec{p}} = \partial \epsilon_{\vec{p}}/\partial {\vec{p}}$ are the energy and velocity of a quasiparticle with momentum $\vec{p}$, $f_{\vec{p}\vec{p'}}$ the Landau interaction function, $\epsilon_{\rm F}$ the Fermi energy, $I$ the collision integral,
%
and $\vec{E}$ the net electric field that includes both the external field and the self-consistently induced field by the electronic liquid itself. Here and throughout most of this paper, we use the convention $e=\hbar=1$ for the electric charge and Planck's constant.
Notice that, at linear order in $\delta n_{\vec{p}}$ and the amplitudes of the electromagnetic fields,  the magnetic field $\vec{B}$ does not enter into the kinetic equation.
In this paper, we will focus on two-dimensional (2D) systems. We expect that certain qualitative aspects such as the universality of the results carry over to the three-dimensional (3D) case. 

The charge current density is given by~\cite{Pines}
\begin{eqnarray}\label{Eq.currdef}
\vec{J} &=& \frac{1}{\mathcal{A}} \sum _{\vec{p}}\vec{v}_{\vec{p}} \delta \bar{n}_{\vec{p}},
\end{eqnarray}
where $\mathcal{A}$ denotes the system area.
In the presence of a quasi-static ($\omega \rightarrow 0$), spatially modulated electric field with a finite wavevector ${\bf q}$, the solution of $\delta \bar{n}_{\vec{p}}$ to the transport equation Eq.~\eqref{Eq.LKEfull} in the absence of collisions [$I=0$ in Eq.~\eqref{Eq.LKEfull}] but in the presence of interactions encoded in the Landau parameters is given by~\cite{Pines}
\begin{eqnarray}\label{Eq.denseffdef}
\delta \bar{n}_{\vec{p}} = - i\frac{ \vec{E}(\vec{q}) \cdot \vec{v}_{\vec{p}} \delta (\epsilon _{\vec{p}} - \epsilon_{\rm F})}{\vec{q} \cdot \vec{v}_{\vec{p}} - i\eta},
\end{eqnarray}
where $i\eta \rightarrow i 0^+$.
From the above it can be shown~\cite{Pines,supplementary} that there is a finite, spatially oscillating quasi-static current that is purely transverse, i.e., $\vec{J}$ is perpendicular to ${\bf q}$, and therefore, the dissipative conductivity tensor in this limit has a single non-vanishing transverse component,
\begin{eqnarray}\label{Eq.anicondtensor}
{\rm Re~}\sigma  _{\alpha \beta = \|, \perp}(\vec{q}, \omega \rightarrow 0) &=& \left(
\begin{array}{cc}
0 & 0 \\
0 & \sigma _{\perp, 0} (\vec{q})
\end{array}
\right),
\end{eqnarray} 
where $\| (\perp)$ denote the components longitudinal (transverse) to the direction $\hat{q}=\vec{q}/q$ (see Appendix~\ref{Sec.suppcondperpderivation} of the Supplementary Material~\cite{supplementary} for details).
In writing this equation we have implicitly assumed time-reversal symmetry, which forces the Hall conductivity~\cite{Xiao2007,Haldane2004,Chen2017} to vanish. The quasi-static transverse conductivity in Eq.~\eqref{Eq.anicondtensor} is given by (momentarily restoring the electric charge and Planck's constant)
\begin{eqnarray}
\sigma _{\perp, 0} (\vec{q})
&=& (2S+1) \frac{e^2}{2 h q}\sum_i \mathcal{R}_{\rm F} |_{\vec{p}_i^* (\hat{q})},\label{Eq.anitranscond}
\end{eqnarray}
where $(2S+1)$ is the spin degeneracy factor, $\left\lbrace\vec{p}^*_i\right\rbrace$ is the set of points on the FS at which the Fermi velocity is orthogonal to $\hat{q}$ (or equivalently where $\hat{q}$ is tangential to the FS), and $\mathcal{R}_{\rm F} |_{\vec{p}_i^* (\hat{q})} $ the absolute value of the local radius of curvature of the FS at $\vec{p}_i^*$ (see Fig.~\ref{Fig.cartoon}(c) for illustration). 
%


We refer to the conductivity in Eq.~\eqref{Eq.anitranscond} as the ``universal transverse conductivity'', because it describes a physical response that only depends on fundamental constants of nature, the wavevector magnitude $q = |\vec{q}|$, and the pure geometry of the FS whereas information about interactions or even the electron mass disappears. Since the conductivities predicted within linearized Landau kinetic equation are identical to those obtained with the bosonization approach to FS~\cite{CastroNeto1995,Son2016,Shear}, and since the latter is believed to capture {\it exactly} the low-energy and long-wavelength properties of interacting LFLs (see e.g. Refs.~\onlinecite{Haldane2005,Houghton1993,Fradkin1994,Houghton2000} for a detailed review), Eqs.~\eqref{Eq.anicondtensor}-\eqref{Eq.anitranscond} are expected to describe the exact behavior of the conductivity of a collisionless interacting Fermi liquid in the precise ordering of limits $ \omega \ll v_{\rm F} q \ll E_{\rm F}$~\cite{Shearbd}). 

We would also like to emphasize that the off-diagonal components of the conductivity in Eq.~\eqref{Eq.anicondtensor} vanish generically only after imposing symmetries, for example when the wavevector ${\bf q}$ lies on a symmetric mirror plane~\cite{Shear,Shearbd}. However in the long-wavelength quasi-static limit $\omega \ll v_{\rm F} q \ll E_{\rm F}$, Eqs.~\eqref{Eq.anicondtensor}--\eqref{Eq.anitranscond} are expected hold regardless of spatial symmetries, namely, for any given $q$ along any direction the current is orthogonal to $q$ in the quasi-static limit. This is intimately related to the fact that longitudinal electric fields (those with vanishing curl) can be represented by a scalar electric potential, $\phi$, and any dissipative electric current response to these fields at finite $q$ (real part of the conductivity), should vanish in the quasi-static limit, because a static potential leads to a new time independent Hamiltonian with a well defined equilibrium state in which dissipative currents cannot exist~\cite{LLStat1}. Because of this reasoning, we also expect Eqs.~\eqref{Eq.anicondtensor}--\eqref{Eq.anitranscond} to hold even in systems with broken time reversal symmetry as a statement on the symmetrized real part of the conductivity tensor that controls dissipative currents.
%

\subsection{Effects of collisions on the transverse conductivity}\label{Sec.condperpcoll}

For simplicity, here we will consider a 2D system with time reversal and all the space symmetries of trivial vacuum, which leads to a circular FS.
In this special case, the transverse conductivity from Eq.~\eqref{Eq.anitranscond} reduces to
\begin{eqnarray}\label{Eq.sigfperp0}
\sigma _{\perp,0} (q)
&=& (2S+1)\frac{e^2}{h}\frac{p_{\rm F,0}}{q},
\end{eqnarray}
where $p_{\rm F,0}$ denotes the Fermi radius.
%
%
To include the effects of collisions, we proceed to solve the transport equation Eq.~\eqref{Eq.LKEfull}. We will restrict to analyzing spin symmetric fluctuations of the liquid, which can be done equivalently by considering a spinless model of fermions and restoring the spin degeneracy factor $(2S+1)$ at the end.
The Landau interaction function simplifies, $f(\theta, \theta') = f(\theta-\theta')$, where $\theta$ is an angle that parametrizes points along the FS.
%
By denoting the local deviation of the Fermi radius from its equilibrium value as $p_{\rm F}(\vec{r}, \theta)$, so that the local Fermi radius is $p_{\rm F,0} + p_{\rm F}(\vec{r}, \theta)$, the distribution deviation is given by
$\delta n_{\vec{p}} = \delta (p - p_{\rm F,0} ) p_{\rm F}(\vec{r}, \theta)$, which allows Eq.~\eqref{Eq.LKEfull} to simplify into the following form:
\begin{align}
\partial _t p_{\rm F} (\vec{r}, \theta) &+ \vec{v}_{\vec{p}} \cdot \vec{\partial}_{\vec{r}}\Bigl[ p_{\rm F} (\vec{r}, \theta)+ \int \frac{d \theta '}{2\pi} f(\theta - \theta ') p_{\rm F} (\vec{r}, \theta ') \Bigr]\nonumber \\
&= -  \vec{E} \cdot \vec{v}_{\vec{p}} + I [p_{\rm F}]. \label{Eq.LKE}
\end{align}
The collision integral can be modeled as~\cite{LevGregNat,LevGregPNAS,Alekseev2018}
\begin{eqnarray}
I [p_{\rm F}] &=& - \Gamma _1 (p_{\rm F} - P_0[p_{\rm F}]) \nonumber \\
&&- \Gamma _2 (p_{\rm F} - P_0[p_{\rm F}] - P_1[p_{\rm F}] - P_{-1} [p_{\rm F}]),\\
P_m [p_{\rm F}] &=& e^{im\theta} \int \frac{d\theta'}{2\pi} p_{\rm F} (\vec{r}, \theta') e^{-im\theta'},
\end{eqnarray}
which captures momentum-relaxing processes such as electron-impurity collisions, as well as momentum-preserving processes originating from electron-electron collisions, respectively characterized by collision rates $\Gamma _1$ and $\Gamma _2$.
Here $P_m[p_{\rm F}]$ projects the Fermi radius onto the m-th harmonic $e^{i m \theta}$.

\begin{figure}
\includegraphics[scale=1.0]{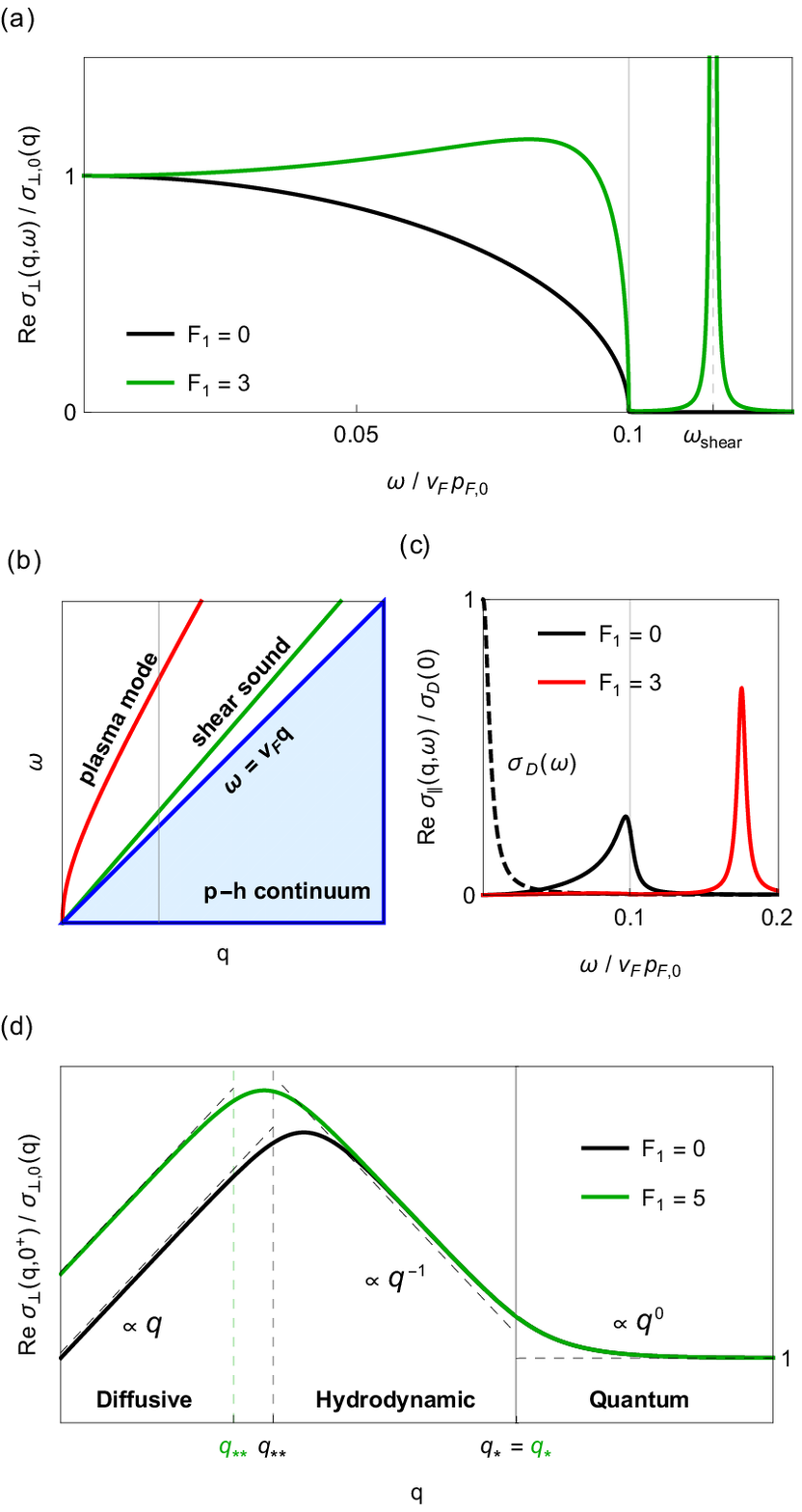}
\caption{
(a) Conductivity Re~$\sigma _\perp (q, \omega)$ in the quantum regime at $q = 0.1 p_{\rm F,0}$ [vertical line cut in (b)] for the non-interacting case (black) and in the presence of interactions $F_1 = 3$ (green).
For nonzero interactions the shear sound resonance at $\omega _{\rm shear}$ emerges from the particle-hole continuum. Regardless of interactions, Re~$\sigma _\perp (q, \omega)$ approaches the universal transverse conductivity $\sigma _{\perp,0} (q)$ in the $\omega \rightarrow 0$ limit.
(b) Dispersion of collective modes and particle-hole excitations in a LFL. 
(c) Solid black and red lines show Re~$\sigma _\| (q, \omega)$ corresponding to those shown in  (a) but with a finite collision rate $\Gamma _1 = 5 \times 10^{-3} v_{\rm F}p_{\rm F,0}$ and $ \Gamma _2 = 0$.
For reference, the Drude conductivity (black dashed) is shown at the same scale.
(d) The different transport regimes in LFLs in the presence of collisions for the case of $q_C > q_D$. The quasi-static transverse conductivity scales differently with $q$ in each of the regimes, separated by the momentum scales $q_{*}$ and $q_{**}$.
}
\label{Fig.condmetal}
\end{figure}

Following the approach in Ref.~\onlinecite{Shearbd}, we obtain exact analytic expressions of the response functions by solving Eq.~\eqref{Eq.LKE} with finite Landau parameters $\left\lbrace F_0 ,F_1\right\rbrace$ (see Appendix~\ref{Sec.suppfullsolve} of the Supplementary Material~\cite{supplementary} for details). 
When the driving field is spatially modulated along an arbitrary direction $\hat{q}$, due to the mirror symmetries of the system, 
$\sigma $ can be decoupled into a longitudinal ($\sigma _\|$) and transverse ($\sigma _\perp$) component, corresponding respectively to the response to the net field component parallel $\vec{E}_{\|}(\vec{q}) = \hat{q}\cdot \vec{E}(\vec{q})\hat{q}$ and orthogonal $\vec{E}_{\perp}(\vec{q}) = \vec{E}(\vec{q}) - \hat{q}\cdot \vec{E}(\vec{q})$ to the direction of modulation $\hat{q}$,
\begin{eqnarray}
J_\| (\vec{q}, \omega) &=& \sigma _\| (\vec{q},\omega) E _\| (\vec{q}, \omega), \\
J_\perp (\vec{q}, \omega) &=& \sigma _\perp (\vec{q},\omega) E _\perp (\vec{q}, \omega). \label{Eq.transcond}
\end{eqnarray}
From the solutions to Eq.~\eqref{Eq.LKE}, we obtain the longitudinal and transverse conductivities
\begin{align}
&\sigma _\| (q, \omega) = \frac{n}{m}\frac{2 i}{\frac{2in}{m} \rho _*(q,\omega) + F_1 \omega _- - \omega _+ - 2 i \Gamma _2 }, \label{Eq.condfpll} \\
&\sigma _\perp (q, \omega) = \frac{n }{m}\frac{2i}{ F_1 \omega _-  - \omega _+  - 2i \Gamma _2 }, \label{Eq.condfperp}\\
&\rho_* (q, \omega) =
- i\frac{1}{ n^2 \kappa} \frac{q^2}{\omega }, \quad \kappa = \frac{1}{n E_{\rm F}} \frac{1}{1 + F_0},\\
&\omega _\pm = \omega-i (\Gamma_1 +\Gamma_2) \pm \sqrt{\left[\omega-i (\Gamma_1 +\Gamma_2)\right] ^2 -(v_{\rm F}q)^2}, \\
&\omega _+ \omega _- = (v_{\rm F}q)^2,\label{Eq.zpm}
\end{align}
where $n = p_{\rm F,0}^2/ 4\pi $ denotes the carrier density, $m = m^*/(1+F_1)$ the transport mass, $\kappa$ the compressibility and $E_{\rm F} = p_{F,0}^2 /2m^*$ the Fermi energy.
The difference between transverse and longitudinal conductivities can be more conveniently expressed in terms of resistivities $\rho _i = \sigma _i ^{-1}$,
\begin{eqnarray}
\rho_\| (q, \omega) &=& \rho_\perp (q, \omega) + \rho_* (q, \omega), \label{Eq.condpllperprel}
\end{eqnarray}
%
%
In the limit of $v_{\rm F} q \ll \omega$  both conductivities approach the familiar Drude conductivity,
\begin{eqnarray}
\sigma  _D(\omega) = \sigma _\| (0, \omega) = \sigma _\perp (0, \omega) =\frac{ne^2}{m\left( i \omega + \Gamma _1 \right)}.\label{Eq.Drudecond}
\end{eqnarray}
The distinction between the transverse and longitudinal conductivities become evident at finite $q$, as 
illustrated in Fig.~\ref{Fig.condmetal}(a) and Fig.~\ref{Fig.condmetal}(c).
Most notably, while the real part of the longitudinal conductivity vanishes in the quasi-static limit, the real part of the transverse conductivity approaches a finite value given by 
\begin{align}
&\sigma _\perp (q, 0^+) = (2S+1)\frac{e^2}{h}\frac{p_{\rm F,0}}{Q(q)}, \label{Eq.sigfperp0scatt}\\
&Q(q) =  q_D + \sqrt{q^2 + q_C^2} -q_C, \label{Eq.condrescale}\\
&q_D = \frac{2}{1+F_1}\frac{\Gamma _1}{v_{\rm F}}, \quad q_C = \frac{1}{v_{\rm F}}(\Gamma _1 + \Gamma _2),
\end{align}
%
%
where we have restored the spin degeneracy factor, electric charge, and Planck's constant.
This remarkable difference remains even in the presence of sufficiently strong interactions, where the shear sound mode emerges out of the particle-hole continuum $\omega _{\rm shear} > v_{\rm F} q$ and carries along with it a substantial weight of the transverse current fluctuations~\cite{Shear,Shearbd,Conti,Gao2010,valentinis2020optical,valentinis2020observing} as illustrated in Fig.~\ref{Fig.condmetal}.

As shown in Fig.~\ref{Fig.condmetal}(d) and summarized in Table~\ref{Table.transcond}, the presence of collisions gives rise to two additional transport regimes -- the hydrodynamic and diffusive regimes -- when $q_C > q_D$. These are separated by momentum scales $q_{**} = \sqrt{q_C q_D}$ and $q_* = q_C$ respectively. 
Interestingly, the existence of a well defined window of hydrodynamic behavior for the transverse conductivity can be achieved above some temperature $T$ in clean samples with mean free path $l_{\rm mfp}$, when $\Gamma _2 \sim (E_{\rm F}/2\pi)(k_B T/E_{\rm F})^2 \gg \Gamma _1 \sim \hbar v_{\rm F} /l_{\rm mfp}$ (up to logarithmic corrections)~\cite{eecolrate},
but apparently also in dirty samples $\Gamma _2 \ll \Gamma _1$, with large $F_1 \gg 1$.
When $q_C < q_D$, the hydrodynamic regime is absent and $q_* = q_D$ sets the momentum scale separating the diffusive and quantum regimes~\cite{Shearbd}.
In the absence of Landau interaction parameters, these results are in qualitative agreement with those obtained in Ref.~\onlinecite{Agarwal}, where the quantum regime is referred to as the ballistic regime. 
In summary, probing the universal transverse conductivity in the quantum regime requires a momentum $q \gg q_*$, where the quantum momentum scale $q_*$ is defined as
\begin{eqnarray}\label{Eq.qstar}
q_* = {\rm max} (q_C, q_D).
\end{eqnarray}
where $q_C$ and $q_D$ were defined in Eq.~\eqref{Eq.condrescale}. 

As a consistency check, we remark that using the conductivity in the absence of collisions $\sigma^0_{\|,\perp}(q,\omega)$ as a starting point, the effect of collisions on the conductivity can be taken into account by a simple rule discovered in Ref.~\onlinecite{Conti} (see Appendix~\ref{sec:collisons} for details),
 \begin{align}
 [\sigma_{\|}(q,\omega)]^{-1}=&[\sigma^0_{\|}(q,\omega-i\Gamma_{12})]^{-1}- \Gamma_2\frac{m}{n}\notag\\
 &-
 \frac{i\Gamma_{12}}{\omega(\omega-i\Gamma_{12})} \lim_{\omega'\to0}\frac{\omega'}{\sigma^0_{\|}(q,\omega')}\label{sigma_par_conti_main}
  \\
 [\sigma_{\perp}(q,\omega)]^{-1}=&[\sigma^0_{\perp}(q,\omega-i\Gamma_{12})]^{-1}- \Gamma_2\frac{m}{n}\label{sigma_perp_conti_main}
  \end{align} 
with the short-hand notation $\Gamma_{12}=\Gamma_1+\Gamma_2$. It is straightforward to show that our solutions in Eqs.~\eqref{Eq.condfpll}--\eqref{Eq.zpm} indeed satisfy these relations.


\begin{table}[t]
\begin{tabular}[b]{cccc}
\hline\\[-1.em]
\hline\\[-1.em]
& \quad Diffusive \quad \quad & \quad Hydrodynamic \quad \quad & Quantum \\ [.1em]
& $q \ll q_{**}$ & $q_{**} \ll q \ll q_{*}$ & $q_{*} \ll q$\\ [.5em]
\hline\\[-1.em]
$\sigma  _\perp (q, 0^+)$ & {\large $\frac{ne^2}{m}\frac{\hbar}{\Gamma _1}$} & $g_S${\large $\frac{e^2}{h}\frac{2 m^*\Gamma _2}{ q^2}$} & $g_S${\large $\frac{e^2}{h}\frac{p_{\rm F,0}}{q}$} \\ [1.em]
\hline\\[-1.em]
\hline\\[-1.em]
\end{tabular}
\caption{Transport regimes of the Landau Fermi liquid when $q_C > q_D$ in which case $q_{*} = q_C$ and $q_{**} = \sqrt{q_C q_D}$. Here, $m^* = p_{\rm F,0}/v_{\rm F}$ denotes the quasiparticle mass while $m = m^*/(1+F_1)$ denotes the transport mass, and $g_S = 2S+1$ denotes the spin degeneracy factor. 
When $q_D > q_C$, the hydrodynamic regime is absent and $q_* = q_D$ sets the momentum scale between the diffusive ($q \ll q_*$) and quantum ($q \gg q_*$) regimes.
}\label{Table.transcond}
\end{table}

\section{Conductivity of U(1) spinon Fermi surface states}\label{Sec.condspinons}

Given the geometric and universal nature of the transverse conductivity in metals established in the previous section, it is natural to ask if this behavior extends to other phases of matter that exhibit FSs. In this section, we show that gapless quantum spin liquids with a spinon FS also exhibit the same universal transverse conductivity.

\subsection{Spinon Fermi surface low-energy theory}~\label{Sec.spinonlowETh}

We begin by discussing the low-energy effective description of the spinon FS state (SFSS), in which electrons fractionalize into a spinless boson (the chargon) and spinful fermion (the spinon). For the underlying microscopic lattice model, the electron creation operator, $\psi^\dagger _{r \sigma}$, at lattice site $r$ and spin $\sigma$, can be written as
\begin{eqnarray}
\psi^\dagger _{r \sigma} = f^\dagger _{r \sigma} b^\dagger _{r},
\end{eqnarray}
where $f^\dagger _{r \sigma}$ and $b^\dagger _{r}$ are respectively the spinon and chargon creation operators. The electron operator is the only physical operator out of which every microscopic Hamiltonian is constructed and also the elementary object 
whose action allows to construct physical states.
For example, only electrons hop between lattice sites, and not isolated spinons or chargons. As a consequence the lattice partons are always forced to hop together. Therefore, the spinon, chargon, and electron occupations are identical in every physical state
\begin{align}\label{Eq.escdens}
\rho _{r} = \sum _\sigma \psi^\dagger _{r \sigma} \psi_{r \sigma} = \rho _{s,r} = \sum _\sigma f^\dagger _{r \sigma} f_{r \sigma} = \rho _{c,r} = b^\dagger _{r} b_{r}.
\end{align}
Similarly, the lattice particle current density operators satisfy
\begin{align}\label{Eq.escj}
\vec{j} _{r} = \vec{j}_{s,r} = \vec{j} _{c,r}.
\end{align}

The traditional scenario in which the SFSS can be realized requires the electrons to be at half-filling of the lattice (one electron per site), allowing the chargons to form a trivial Mott insulator, and the spinons to form a Fermi sea with a volume equal to half of the Brillouin zone, although other variants such as the composite exciton Fermi sea, 
which have different filling constraints due to the coexistence of spinon-like particle and hole pockets,
are possible as well~\cite{Chowdhury2018,Sodemann2018}. Hence, the ground state can be taken to be a product state of chargons in a trivial Mott insulator and the spinons in the Fermi sea projected to satisfy the constraints from Eq.~\eqref{Eq.escdens}. An effective field theory capturing this can be formally written as
\begin{align}\label{Eq.Leff}
\mathcal{L} = \mathcal{L}^{\rm FS}_{\rm spinon} (\vec{p}-\vec{a}) + \mathcal{L}^{\rm Mott} _{\rm chargon} (\vec{p} + \vec{a} - \vec{A}) + \cdots.
\end{align}
Here $\vec{a}$ is an internal gauge field and $\vec{A}$ is the physical electromagnetic field. In this notation $\mathcal{L}^{\rm FS}_{\rm spinon} (\vec{p}-\vec{a})$ would be a Lagrangian of Free spinons in a FS state if we momentarily imagine the internal gauge field $\vec{a}$ to be a non-dynamical probe field (an external parameter) and similarly $\mathcal{L}^{\rm Mott} _{\rm chargon} (\vec{p} + \vec{a} - \vec{A})$ would be the one of bosonic Mott insulator if we imagine the field $(\vec{A} - \vec{a})$ to be a probe non-dynamical gauge field. 
At this level in which the role of the gauge field $\vec{a}$ is simply to enforce the constraints of Eq.~\eqref{Eq.escdens} and Eq.~\eqref{Eq.escj},
the ``$\cdots$'' in Eq.~\eqref{Eq.Leff} denotes a list of other terms that only involve spinon and chargon interactions, but not pure gauge field terms such as the Maxwell action for $\vec{a}$. 
Such Maxwell terms arise upon integrating out the chargon degrees of freedom, which is legitimate within the low-energy description since they form a gapped Mott insulating state. Before doing so, we would like to emphasize some important conceptual points, which will prove useful when interpreting our results later on. Notice that the Lagrangian in Eq.~\eqref{Eq.Leff} implies that the self-consistent electromotive forces that spinon, chargon and electron experience are
\begin{eqnarray}
\vec{F}_s &=& \vec{\mathsf{e}} + \vec{v} \times \vec{b}, \label{Eq.forcespinon}\\
\vec{F}_c &=& (\vec{E} - \vec{\mathsf{e}}) + \vec{v} \times (\vec{B} - \vec{b}), \\
\vec{F}_e &=& \vec{E} + \vec{v} \times \vec{B},
\end{eqnarray}
where
\begin{align}
&\vec{\mathsf{e}} = -\vec{\partial}_{\vec{r}} \phi - \partial _t \vec{\mathsf{a}}, \quad
\vec{\mathsf{b}} = \vec{\partial}_{\vec{r}} \times \vec{\mathsf{a}}, \\
&\vec{E} = -\vec{\partial}_{\vec{r}} \Phi - \partial _t \vec{A} , \quad
\vec{B} = \vec{\partial}_{\vec{r}} \times \vec{A},
\end{align}
Here we are implicitly replacing the underlying compact lattice gauge fields by non-compact ones appealing to the idea that this is legitimate at sufficiently low energies in the SFSS where the compactness has been argued to be irrelevant even in two spatial dimensions~\cite{Lee2008}. Even though the SFSS is nominally an insulating state to DC transport, the net electromotive force that the spinons can experience in response to external electromagnetic fields can ultimately drive a variety of remarkable electromagnetic responses of this state, such as the power law sub-gap optical conductivity~\cite{Ng2007}, quantum oscillations~\cite{Motrunich2006,Chowdhury2018,Sodemann2018}, and sub-gap cyclotron resonance~\cite{Rao2019}. Notice that the velocity, $\vec{v}$, of the electron, chargon, and spinon are identical as it follows from the constraint in Eq.~\eqref{Eq.escj}. Thus even though they all move at the same speed they actually experience different electromotive forces. This is at the heart of the very useful Ioffe-Larkin rule~\cite{Ioffe1989,LNW2006}, which can be obtained by imagining that each of their currents are driven by their respective effective net electromotive forces as follows~\footnote{notice that we are using a convention in which $\vec{j}_s$ and $\vec{j}_c$ are the chargon and spinon particle number currents which are equal since they measure the change in spinon and chargon densities, as opposed to their respective currents under the gauge field $\vec{a}$ which would be opposite, since the spinon and chargon have opposite gauge charges which are for example employed in Ref.~\onlinecite{LNW2006}}:
\begin{eqnarray}
\vec{j}_s &=& \vec{j}_e = \sigma _s \vec{e}, \label{Eq.js}\\
\vec{j}_c &=& \vec{j}_e = \sigma _c (\vec{E} - \vec{e}),\label{Eq.jc}
\end{eqnarray}
which leads to the following physical electron conductivity (Ioffe-Larkin rule):
\begin{eqnarray}\label{Eq.IoffeLarkin}
\sigma  ^{-1} = \sigma _s ^{-1} + \sigma _c ^{-1}.
\end{eqnarray}
The above formula is valid even when the conductivities are viewed as frequency and wavevector
dependent tensors. 

For purposes of understanding low-energy properties, we can consider the Lagrangian that results from that in Eq.~\eqref{Eq.Leff} after integrating out the chargons, which to the leading order in the effective gauge field experienced by the chargons, $\vec{A}-\vec{a}$, is a Maxwell action because they are assumed to form a trivial time-reversal-invariant Mott insulator. The effective Lagrangian density can be formally written as
\begin{align}\label{Eq.Leffintb}
\mathcal{L} = \mathcal{L}^{\rm FS}_{\rm spinon} (\vec{p}-\vec{a}) + \frac{\epsilon}{2}( \vec{\mathsf{e}} - \vec{E})^2  - \frac{1}{2\mu} (\vec{\mathsf{b}} - \vec{B})^2 + \cdots.
\end{align}
Here ``$\cdots$'' would include higher order, gauge-invariant terms of the field $\vec{A}-\vec{a}$ and other spinon interactions, and $\epsilon$ and $\mu$ are effective dielectric and magnetic susceptibilities of the Chargons, which we will explicitly relate to the Mott scale, or more precisely the optical pseudo-gap of the SFSS later on.
Even after all these simplifications, the effective field theory written above remains strongly coupled~\cite{Lee2009}, featuring corrections to the spinon single-particle propagator that are in fact singular to leading order. It has been long known that such singularities disappear to leading order in the gauge neutral spinon particle-hole propagators, whose forms resemble those of the particle-hole propagators in LFLs~\cite{Kim1994}.
When the range of the gauge-field propagator is extended, which is for example physically justified in the related problem of composite Fermi seas in half-filled Landau levels with Coulomb interactions, a controlled double expansion approach has been devised to show that, indeed, such leading RPA LFL-like results are exact in the limit of a large number of Fermion flavors and for small deviations of the range of the gauge propagator from the Coulomb point~\cite{Mross2010}. We will employ a treatment that is ultimately able to reproduce these RPA results at small $(\omega, \vec{q})$, while also allowing for spinon interaction effects in the form of Landau parameters to be included, similar in spirit to the quantum Boltzmann approach employed in the pioneering work of Ref.~\onlinecite{Kim1995}.
In particular, we are interested in computing the spinon conductivity which involves only the spinon particle-hole propagator, and not the more singular spinon single-particle propagator.

Our approach is to replace the Lagrangian of the SFSS by the bosonized and linearized theory of the quantum Fermi liquid~\cite{Haldane2005,Houghton1993,Fradkin1994}, which is a second quantized version of ordinary LFL theory~\cite{Son2016,Shear}. The resulting theory is a bosonic bilinear theory in the Fermi radius operator and the gauge fields, which can therefore be solved exactly. Moreover because the theory is bosonic bilinear, the quantum and classical equation of motions of its operators are identical~\cite{Shear} and therefore response functions, such as the conductivities, can be obtained, without loss of generality, by simply solving the classical kinetic equation describing the distribution function of spinons experiencing the $\vec{\mathsf{e}}$ and $\vec{\mathsf{b}}$ fields together with the Maxwell equations describing the dynamics of these fields. Therefore our effective Lagrangian can then be written as
\begin{align}
&\mathfrak{L} = \int d^2 \vec{r} \mathcal{L}\\
&\mathcal{L} = \frac{\epsilon}{2}( \vec{\mathsf{e}} - \vec{E})^2  - \frac{1}{2\mu} (\vec{\mathsf{b}} - \vec{B})^2 - \rho _s \phi + \vec{j}_s\cdot \vec{\mathsf{a}} + \mathcal{L}^{\rm FS}_s
\end{align}
where $\mathcal{L}^{\rm FS}_s$ is a short-hand notation for the linearized and bosonized spinon FS Lagrangian. It is straightforward to verify that the physical electron density and currents in the above Lagrangian are identical to the spinon density and currents, namely, that
\begin{equation}
(\rho , \vec{j}) = \left( \frac{\delta \mathfrak{L}}{\delta \Phi}, \frac{\delta \mathfrak{L}}{\delta \vec{A}} \right) = (\rho _s, \vec{j}_s) = \left( \frac{\delta \mathfrak{L}}{\delta \phi}, \frac{\delta \mathfrak{L}}{\delta \vec{\mathsf{a}}} \right),
\end{equation}
which is the low-energy version of the microscopic identities in the underlying model stated in
Eqs.~\eqref{Eq.escj} and~\eqref{Eq.Leff}. 
Therefore, from here on, we will drop the ``$s$'' when referring to spinon densities and currents. 
The equations of motion of the internal gauge fields that follow from the above Lagrangian are
\begin{eqnarray}
\rho &=& \epsilon \vec{\partial}_{\vec{r}} \cdot ( \vec{\mathsf{e}} - \vec{E}), \label{Eq.spinoneom1}\\
\vec{j} &=& \frac{1}{\mu} \vec{\partial}_{\vec{r}} \times (\vec{\mathsf{b}} - \vec{B}) - \epsilon \partial _t ( \vec{\mathsf{e}} - \vec{E}).\label{Eq.spinoneom1b}
\end{eqnarray}
The spinon densities and currents are then related to the deviation of the spinon distribution
function $\delta n_{\vec{p}}$ as follows:
\begin{eqnarray}
\rho &=& \frac{1}{\mathcal{A}} \sum _{\vec{p}}\delta n_{\vec{p}}, \\
\vec{j} &=& \frac{1}{\mathcal{A}} \sum _{\vec{p}}\vec{v}_{\vec{p}} \delta \bar{n}_{\vec{p}}.
\end{eqnarray}
The above equations need to be complemented by a kinetic equation for the spinon’s FS describing the response of low-energy, interacting spinon particle-hole excitations to the electromotive forces they experience in Eq.~\eqref{Eq.forcespinon}. This equation is identical to Eq.~\eqref{Eq.LKEfull} that we wrote in the case of electrons in ordinary metals, except that the physical electric field $\vec{E}$ is replaced by the internal electric field $\vec{\mathsf{e}}$ experienced by the spinons,
\begin{align}
\partial _t \delta n_{\vec{p}} &+ \vec{v}_{\vec{p}} \cdot \vec{\partial}_{\vec{r}}\delta \bar{n}_{\vec{p}} +   \vec{\mathsf{e}} \cdot \vec{v}_{\vec{p}} \delta (\epsilon _{\vec{p}} - \epsilon_{\rm F})
= I [\delta n_{\vec{p}}].\label{Eq.spinoneom2}
\end{align}
Just as for electrons, notice that at the linearized level the effective magnetic field, $\vec{\mathsf{b}}$, does not enter into the spinon kinetic equation. 
This provides a complete set of coupled equations that will be used to compute the spinon electrical conductivity in the next section.
Also, it is important to bear in mind that the collision terms, $I[\delta n_{\vec{p}}]$, in the case of spinons can be very different from those of electrons due to gauge field fluctuations. Impurities lead to a momentum-relaxing collisions with a simple Drude-like form $\Gamma_1 \sim v_{\rm F}/l_{\rm mfp}$, with $l_{\rm mfp}$ the spinon mean free path, but gauge field fluctuations lead to momentum relaxation rates that have been argued to scale as $\Gamma_1 \sim \omega^{4/3}$ in 2D~\cite{Lee1992}, although the scaling with temperature and frequency can be non-trivially affected by umklapp processes~\cite{lee2020}.

\subsection{Conductivities of the Spinon Fermi surface state}~\label{Sec.spinonFStranscond}

We will now consider the response of the spinon FS to a space-time oscillating field $\vec{E}(\vec{q}, \omega)$. For simplicity, we consider an isotropic system with a circular FS, for which the mirror symmetries guarantee a decomposition into transverse and longitudinal responses,
\begin{eqnarray}
j_\| (\vec{q}, \omega) &=& \sigma  _\| (\vec{q},\omega) E _\| (\vec{q}, \omega), \\
j_\perp (\vec{q}, \omega) &=& \sigma  _\perp (\vec{q},\omega) E _\perp (\vec{q}, \omega).
\end{eqnarray}

Using the Faraday's laws $\vec{\partial} _r \times \vec{E} = - \partial _t \vec{B}$ and $\vec{\partial} _r \times \vec{e} = - \partial _t \vec{b}$, we arrive at the following set of coupled equations of motion for the combined spinon-gauge field system,
\begin{align}
j_\| (\vec{q}, \omega)  &= i \epsilon \omega (E_\|(\vec{q}, \omega) - \mathsf{e}_\|(\vec{q}, \omega)),\label{Eq.jspll} \\
j_\perp (\vec{q}, \omega) 
&= i \epsilon \omega\left(1 - \frac{q^2}{\epsilon \mu\omega ^2} \right)  (E_\perp (\vec{q}, \omega) - \mathsf{e}_\perp (\vec{q}, \omega)), \label{Eq.sglin1m}\\
j_\| (\vec{q}, \omega) &= \sigma _{s\|} (\vec{q},\omega) \mathsf{e} _\| (\vec{q}, \omega), \\
j_\perp (\vec{q}, \omega) &= \sigma _{s\perp} (\vec{q},\omega) \mathsf{e} _\perp (\vec{q}, \omega).\label{Eq.jsperp}
\end{align}
Solving these equations (see Appendix~\ref{Sec.suppfullsolve} of the Supplementary Material \cite{supplementary}), we arrive at the longitudinal and transverse conductivities of the SFSS,
\begin{align}
\sigma  _\| ^{-1}(q,\omega) &= \sigma _{c\|}^{-1} (\omega) + \sigma _{s\|}^{-1}(q,\omega),\label{Eq.condsppll}\\
\sigma  _\perp ^{-1}(q,\omega) &= \sigma _{c\perp}^{-1} (q, \omega) + \sigma _{s\perp}^{-1}(q,\omega),\label{Eq.condspperp}\\
\sigma _{s\|} (q, \omega) &= \frac{n }{m}\frac{2 i}{\frac{2in}{m} \rho _*(q,\omega) + F_1 \omega _- - \omega _+ - 2 i \Gamma _2 },\\
\sigma _{s\perp} (q, \omega) &= \frac{n }{m}\frac{2i}{ F_1 \omega _-  - \omega _+  - 2i \Gamma _2 }, \\
\sigma _{c\|} (\omega) &= i \frac{n }{m} \frac{\omega }{\omega _p^2}, 
\quad \omega _p = \sqrt{\frac{n }{m \epsilon}}, \\
\sigma _{c\perp} (q, \omega) &= i \frac{n }{m} \frac{\omega }{\omega _p^2}\left( 1  - \frac{c^2 q^2}{\omega ^2}\right), \quad c = \frac{1}{\sqrt{\epsilon \mu}}, \label{Eq.condchperp}
\end{align}
where we introduce the plasma-like frequency $\omega _p$ and the velocity of the emergent photon $c$. Notice that the spinon longitudinal and transverse conductivities $\sigma _{s\|}$ and $\sigma _{s\perp}$ above are identical to those of the electron conductivities Eqs.~\eqref{Eq.condfpll}--\eqref{Eq.condpllperprel}.
As we will see, the plasma frequency controls the Mott scale, or more specifically, it measures the optical pseudo-gap of the SFSS.

\begin{figure}
\includegraphics[scale=1]{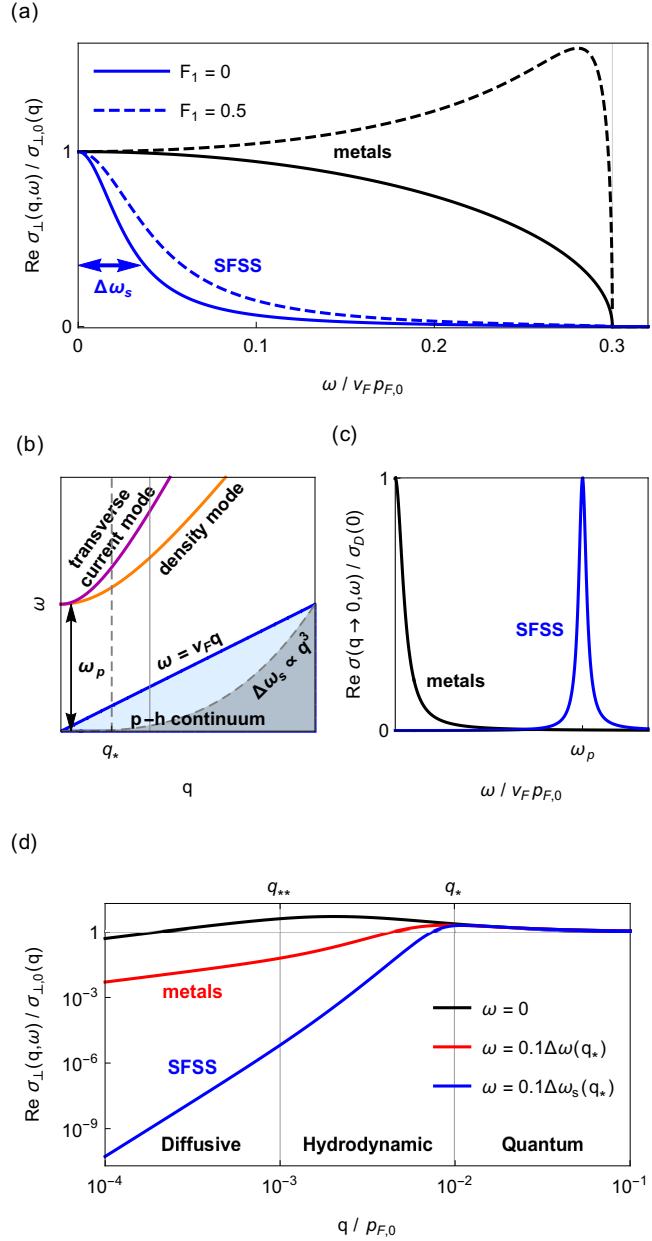}
\caption{
(a) Transverse conductivity Re~$\sigma  _\perp (q, \omega)$ of the SFSS in the collisionless limit at $q = 0.3 p_{\rm F,0}$ [along the vertical line cut in (b)], with $F_1=0$ (blue solid curve) and for $F_1 = 0.5$ (blue dashed curve). The corresponding conductivities in metals are shown in black. 
Conductivities in both systems converge to $\sigma _{\perp,0} (q)$ in the $\omega \rightarrow 0$ limit. The SFSS conductivity becomes rapidly suppressed for frequency above $\Delta \omega _{\rm s}$.
(b) Dispersion of various collective modes and particle-hole excitations in the SFSS ($F_l = 0$). 
(c) Optical conductivity Re~$\sigma _\| (q \rightarrow 0, \omega)$ for a metal and a SFSS in the presence of weak collisions $\Gamma _1 = \Gamma _2 = 0.1 v_{\rm F} p_{\rm F,0}$. 
The Drude peak in metals is shifted to the plasma frequency $\omega _p > E_{\rm F}$ for the SFSS.
(d) Plots of Re~$\sigma _\perp (q, \omega)$ for metals (red) and SFSS (blue) away from the quasi-static $\omega \rightarrow 0$ limit (black). 
A smaller frequency $0.1 \Delta \omega _{\rm s}(q_*)$ is required in SFSSs compared to $0.1 \Delta \omega(q_*)$ in metals to probe the quasi-static limit in the quantum regime $q > q_*$.
}
\label{Fig.condspinon}
\end{figure}

For the case where only $F_0$ and $F_1$ are non-zero, one can solve the full spectrum of collective and particle-hole excitations of the SFSS exactly from the coupled Eqs.~\eqref{Eq.spinoneom1}--\eqref{Eq.spinoneom2}.
What we find is that the spectrum separates into a particle-hole continuum and two isolated collective modes, as shown in Fig.~\ref{Fig.condspinon}(b). One of these collective modes is purely longitudinal while the other is purely transverse, the dispersions of which can be directly obtained from the poles of longitudinal and transverse conductivities respectively (see Appendix~\ref{Sec.suppcolldisp} of the Supplementary Material~\cite{supplementary} for full expression).
To leading order in $q$, their dispersions are
\begin{eqnarray}
\omega _L &\simeq& \omega _p + \frac{1}{2\omega _p}\left(\frac{3 + 2 F_0}{4}\right)v_{\rm F}^2 q^2  + \mathcal{O}(q^4), \label{Eq.wLdisp}\\
\omega _T &\simeq& \omega _p + \frac{1}{2\omega _p}\left(\frac{1}{4} + \frac{c^2}{v_{\rm F}^2} \right)v_{\rm F}^2q^2  + \mathcal{O}(q^4).\label{Eq.wTdisp}
\end{eqnarray}
At long wavelengths the longitudinal mode has all its weight on the charge density and longitudinal charge currents, and it is the analogue of a plasma oscillation of a charged fluid.  On the other hand, at long wavelengths the transverse mode is a coherent mixture of the emergent photon and the transverse electric currents, and its gapping is analogous to the plasma gap of photons in metals~\cite{Pines}. These modes are gapped here even in 2D because the emergent photon propagates strictly within the sample, unlike the physical photon.


In addition, there is a continuum of spinon particle-hole excitations that remains gapless, as shown in Fig.~\ref{Fig.condspinon}(a). However, the physical properties of the SFSS can in general be very distinct from that of metals. 
This can be clearly seen, for example, in the conductivity in response to spatially uniform electric fields, obtained in the limit $\bf{q}\rightarrow 0$, which governs optical and transport properties, and is given by
\begin{align}
\sigma (\omega) = \sigma _\| (0, \omega) = \sigma _\perp (0, \omega)  
&= i\frac{n  \omega }{m\omega _p^2 - m \omega (\omega - i \Gamma _1)}. \label{Eq.condq0} 
\end{align}
As shown in Fig.~\ref{Fig.condspinon}(a), this conductivity features a peak at $\omega = \omega _p$. 
This peak can be viewed as the optical pseudo-gap or Mott optical lobe of these correlated states. We caution that our approach is aimed at capturing only long-wavelength and low-energy properties of the SFSS, and, therefore this peak should be taken as a caricature of these high-frequency optical phenomena. Notably, at low frequencies the conductivity lacks the distinctive Drude peak and displays a power law, first obtained in~\cite{Ng2007}, given by
\begin{eqnarray}
{\rm Re}~\sigma (\omega) \simeq \frac{n}{m} \frac{\Gamma _1 \omega ^2}{\omega_p^4} + \mathcal{O}(\omega ^4).
\end{eqnarray}

The aforementioned vanishing of the DC conductivity and optical pseudo-gap are often emphasized by referring to the SFSS as an ``insulator''. This state, however, can under some probes resemble a metal~\cite{Chowdhury2018,Sodemann2018,Motrunich2006,Rao2019}. Remarkably, in the long-wavelength quasi-static limit $(\omega \ll v_{\rm F} q \ll E_{\rm F})$, it follows from Eqs.~\eqref{Eq.condsppll}--\eqref{Eq.condchperp} that the transverse conductivity of this state is {\it identical} to that of a metal, and given by
\begin{eqnarray}\label{Eq.sigperpscatt}
\sigma _{\perp} (q,0^+) &=&  (2S+1)\frac{e^2}{h}\frac{p_{\rm F,0}}{Q(q)}
\end{eqnarray}
with $Q(q)$ defined in Eq.~\eqref{Eq.condrescale}. 
Likewise, the dissipative longitudinal conductivity vanishes, $\sigma _{\|} (q, \omega \rightarrow 0) \rightarrow 0$, as expected for a quasi-static longitudinal perturbation.
Importantly, the above implies that the transverse quasi-static conductivity in the collisionless quantum regime $q \gg q_*$ [see Eq.~\eqref{Eq.qstar}] is also a universal number controlled only by the geometry of the spinon FS, and it is given by [cf. Eq.~\eqref{Eq.sigfperp0}] 
\begin{equation}\label{Eq.sigperp0scatt}
\sigma _{\perp,0} (q) = (2S+1)\frac{e^2}{h}\frac{p_{\rm F,0}}{q},
\end{equation}
%
and more generally for FSs of arbitrary shapes [cf. Eq.~\eqref{Eq.anitranscond}] by
\begin{eqnarray}
\sigma _{\perp,0} (\vec{q}) &=& (2S+1)
\frac{e^2}{2 h q}\sum_i \mathcal{R}_{\rm F} |_{\vec{p}_i^* (\hat{q})} .\label{Eq.anisigperp0scatt}
\end{eqnarray}

Although the transverse conductivities of metals and SFSSs approach the same value in the quasi-static limit, their behavior is markedly different at finite frequencies, as shown in Figs.~\ref{Fig.condspinon}(c)--(d). 
In the case of metals, the transverse conductivity vanishes for frequencies exceeding the particle-hole continuum threshold, $\omega \gtrsim v_{\rm F} q$. In the case of SFSSs, however, the transverse conductivity vanishes rapidly over a much narrower frequency window as illustrated in Fig.~\ref{Fig.condspinon}(a), which can be characterized by its half-peak frequency
\begin{eqnarray}\label{Eq.spinonpeakwidth}
\Delta \omega  _{\rm s} (q)= \frac{ v_{\rm F} c^2 q^2}{2\omega _p^2}Q(q),
\end{eqnarray}
where $Q(q)$ is defined in Eq.~\eqref{Eq.condrescale} (see Appendix~\ref{Sec.suppcondperpweffect} of the Supplementary Material~\cite{supplementary} for details). Therefore to reach the quasi-static regime, the transverse conductivity of the SFSS should be measured at $\omega < \Delta \omega_{\rm s}(q)$. 
In particular, to probe the universal transverse conductivity in the quantum regime, the transverse conductivity needs to be measured for wavevectors above $q> q_C$ [see Eq.\eqref{Eq.qstar}] and for frequencies below scales $\Delta \omega$ and $\Delta \omega _{\rm s}$ respectively for metals and SFSSs, given by 
%
\begin{eqnarray}
\Delta \omega (q)&=& v_{\rm F} q,\label{Eq.balpeakwidthmetal}\\
\Delta \omega _{\rm s} (q)&=& \frac{ v_{\rm F} c^2}{2\omega _p^2} q^3.\label{Eq.balpeakwidthspinon}
\end{eqnarray}
%
The shaded region of the particle-hole continuum in Fig.~\ref{Fig.condspinon}(b) is where the universal quantum behavior appears.
These frequencies are estimated for some spinon FS candidate 2D materials in Table~\ref{Table.spinon} for $q = 0.1p_{\rm F,0}$. 
More details on the organic material candidates d-mit and $\kappa$-ET can be found in Ref.~\onlinecite{Zhou2017}. The suggestion that 1T-TaS$_2$ might harbor a U(1) spin liquid is more recent~\cite{Law2017,He2018}, and more specifically, the case for a spinon FS state has been argued based on heat transport measurements~\cite{Murayama2020}. The possibility that a state like the spinon FS might be realized in monolayer WTe$_2$ has been highlighted by the recent remarkable observation of clear quantum oscillations of resistivity in a strongly insulating regime~\cite{Wang2021}.

Notice that the additional frequency-dependent scattering rate of the spinons $\Gamma _\omega \sim\omega^{4/3}$ induced by gauge field fluctuations~\cite{Lee1992} does not affect the quasi-static limit of the transverse conductivity in Eq.~\eqref{Eq.sigperpscatt}, and also it will not change the width of the transverse conductivity frequency dependence from Eq.~\eqref{Eq.spinonpeakwidth}, provided that $\Delta \omega_{\rm s} \ll E_{\rm F}$, which is a criterion easily satisfied as seen in Table~\ref{Table.spinon}.
The precise scaling of the low-frequency scattering rate in the spinon FS is still a subject of debate (see e.g.~\cite{lee2020}), but provided this rate is much smaller than $E_{\rm F}$, we expect the quasi-static limit of the transverse conductivity to be given by Eq.~\eqref{Eq.sigperpscatt}, and will feature a window of wavevectors $\Gamma _\omega < v_{\rm F} q \ll E_{\rm F}$ over which it will be governed by the universal quantum limit of Eqs.~\eqref{Eq.sigperp0scatt}--\eqref{Eq.anisigperp0scatt}.

\begin{table}[t]
\begin{tabular}{lcccc}
\hline \hline
Spinon FS candidate & $\epsilon_{\rm F}$
& $\omega _p$
& $\Delta \omega _{\rm s}/ 2\pi$ & $\Delta \omega _{\rm s} / 2\pi$ \\ [-.2em]
& (meV) & (meV) & $(c = 0.5v_{\rm F})$ & $(c = 2v_{\rm F})$ \\ \hline 
EtMe$_3$Sb[Pd(dmit)$_2$]$_2$ & 59 & 80~\cite{Rao2019} & 7.8 GHz & 120 GHz \\ 
$\kappa$-(ET)$_2$Cu$_2$(CN)$_3$ & 98 & 87~\cite{Rao2019} & 30 GHz & 480 GHz \\ 
1T-TaS$_2$ & 1753 & 200~\cite{Rao2019} & 33 THz  & 520 THz \\ 
monolayer WTe$_2$ & 29 & 60~\cite{Wang2021} & 1.6 GHz  & 26 GHz \\ 
\hline \hline
\end{tabular}
\caption{Order-of-magnitude estimate of the frequency width scale of transverse conductivity [see Fig.~\ref{Fig.condspinon}(a)] for various U(1) spinon FS candidate systems at $q = 0.1 p_{\rm F,0}$ estimated from Eq.~\eqref{Eq.balpeakwidthspinon}.
}\label{Table.spinon}
\end{table}


\subsection{Physical picture for the transverse metallic conductivity of spinons}

\begin{figure}[h]
\includegraphics[scale=1.0]{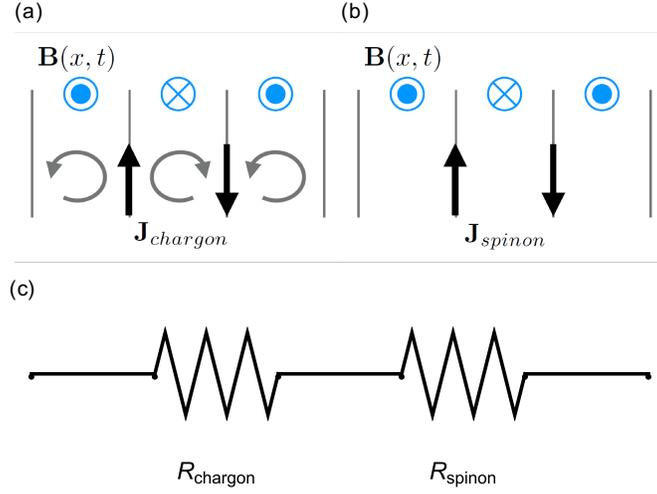}
\caption{Illustration of the difference between the chargon current (a), which corresponds to the non-dissipative magnetization current, to the spinon current (b) that is solely responsible for the dissipative DC transverse conductivity. (c) Analogous to resistors connected in series, the least conductive of the two will dominate the response.}
\label{Fig.s1}
\end{figure}

We would like to give an intuitive explanation for the apparent contradiction that at low frequencies the longitudinal transport properties of the SFSS are characteristic of an insulator, whereas its transverse conductivity is identical to a metallic Fermi liquid. From Eqs.~\eqref{Eq.js}--\eqref{Eq.jc} or equivalently Eqs.~\eqref{Eq.jspll}--\eqref{Eq.jsperp} it follows that the electric fields experienced by the spinons can be written as
\begin{eqnarray}
\mathsf{e}_{\alpha = \|,\perp} = \frac{\sigma _{c, \alpha}}{\sigma _{c, \alpha}+\sigma _{s, \alpha}}E_\alpha,
\end{eqnarray}
where the frequency and wavevector dependences are implicit. Now, the key to the dramatic difference of responses lies in the different behavior of the chargon's longitudinal and transverse conductivity, which are simply those of an ordinary dielectric diamagnetic insulator with permittivities $(\epsilon, \mu)$. Insulators can support non-zero transverse currents, because these have zero divergence and encode the spatial variation of the magnetization without leading to charge density fluctuations, while long-wavelength currents are suppressed at low frequencies because of the incompressibility of insulators. In fact, while the longitudinal conductivity of the chargon insulator vanishes analytically at low frequencies, 
\begin{align}
 \sigma _{c,\|} (q, \omega) &= i \epsilon \omega ,
\end{align}
the transverse conductivity diverges as 
\begin{eqnarray}\label{Eq.transcondchargon}
\sigma _{c, \perp} (q, \omega \rightarrow 0) &=& -i  \frac{q^2}{\mu\omega}.
\end{eqnarray}
The above forms follow simply from the classic relations between polarization, electric fields and longitudinal currents on one hand, $\vec{P} = \epsilon \vec{E}_\| $, $\vec{j}_\| = \partial _t \vec{P}$, and magnetization, magnetic fields, and transverse currents on the other, $\vec{B} = \mu \vec{M}$, $\vec{j}_\perp = \vec{\partial}_{\vec{r}} \vec{\times} \vec{M}$, combined with Faraday's law $\vec{\partial}_{\vec{r}} \vec{\times} \vec{E}=-\partial _t \vec{B}$. These results are therefore expected to be exact at low frequencies and long wavelengths in a trivial ordinary insulating ground state. It follows that the effective longitudinal electric field, $\mathsf{e}_\|$, that the spinons experience in the regime 
$v_{\rm F} q \ll \omega \ll \Gamma _1 \ll \omega _p$ is
\begin{align}
\mathsf{e}_\| \approx i\frac{\omega \Gamma _1}{\omega_p^2}\left( 1 - i \frac{\omega \Gamma _1}{\omega _p^2} \right)  E_\| ,
\end{align}
%
where we used the Drude form for the spinon conductivity. Therefore, we see that at low frequencies an external longitudinal electric field will induce a vanishingly small longitudinal effective electric field on the spinons, and this is the reason for the absence of their electrical Drude weight. In other words, one cannot apply an effectively DC longitudinal electric field $\mathsf{e}_\|$ on the spinons, because we do not have external sources outside the sample for this field and because its effective coupling to the physical external field $E_\|$ vanishes at small frequencies. This can also be intuitively pictured from the Ioffe-Larkin rule, by noting that the most insulating resistor dominates (see Fig.~\ref{Fig.s1}), which in this case is the chargon, leading to an essentially insulating response.

Remarkably, on the other hand, in the transverse quasi-static limit [$q \ll p_{\rm F}$, $\omega \ll {\rm min}( v_{\rm F} q, c q)$],
the transverse electric field experienced by the spinons approaches the full externally applied transverse electric field,
\begin{eqnarray}
\mathsf{e}_\perp \approx \left( 1 - i \frac{\mu \omega}{q^2} \frac{2\pi (2S+1)p_{\rm F}}{Q(q)}\right) E_\perp.
\end{eqnarray}
This behavior can be understood again by appealing to the Ioffe-Larkin rule and noting that in this case the diverging transverse conductivity of the chargons in Eq.\eqref{Eq.transcondchargon}, leads to the spinons dominating the transverse resistivity in this case. Notably, Eq.~\eqref{Eq.IoffeLarkin} combined with Faraday's laws for both the emergent and the physical electromagnetic fields imply that within linear response theory, the effective magnetic field experienced by the spinons, $\vec{\mathsf{b}}$, approaches the external physical magnetic field, $\vec{B}$, in this limit, $\vec{\mathsf{b}} \approx \vec{B}$.
This property is in line with the curious mean-field finding in Ref.~\onlinecite{Sodemann2018} that the effective field experienced by spinons with a parabolic dispersion equals the external magnetic field in 2D at zero temperature. More broadly, this is intimately related to the distinction between applying static magnetic fields and electric fields to spinons, where the former are known to induce an average emergent magnetic field that the spinons experience in resemblance to metals, while remaining largely unresponsive to DC electric fields.

\section{Spin qubit noise spectroscopy of metals and spinon Fermi surface states}\label{Sec.noisespec}

Nitrogen vacancy (NV) center spin qubits are emerging as powerful and versatile  detectors of magnetic properties of condensed matter systems~\cite{Casola2018}. In particular, the NV center allows to measure the two-time autocorrelation function of the magnetic field, at a point located at a distance $z>0$ above the sample, $\chi _{B_\mu B_\nu} (z, t) = \langle B_\mu (\vec{r}+z \hat{z}, t) B_\nu (\vec{r}+z \hat{z}, 0) \rangle$, where translational invariance in the 2D sample coordinate $\vec{r}$ is assumed~\cite{Casola2018,Langsjoen2012}. This magnetic field fluctuations, are related to the dissipative part of the causal retarded correlation function of magnetic fields \cite{GiuVig,Casola2018,Agarwal,Chatterjee2019}
\begin{align}
\mathcal{N}_{\mu \nu} (z, \omega) = -\frac{2\pi}{\hbar} \coth \Bigl(\frac{\beta \hbar \omega }{2}\Bigr) {\rm Im}\chi_{B_\mu B_\nu} (z, \omega) . \label{Eq.noise}
\end{align}
The magnetic field has contributions from the orbital magnetic moments caused by electric currents in the sample and the spin magnetic moments. We will first discuss the contributions arising from electric currents and demonstrate later on that the spin fluctuation contributions are subdominant at low frequencies.

\subsection{Universal quantum low-frequency noise}\label{Sec.noiseclean}

The contribution from currents can be obtained by using the Biot-Savart law. As detailed in Appendix~\ref{Sec.suppnoisederivation} of the Supplementary Material~\cite{supplementary} (see also Ref.~\onlinecite{Agarwal}), we have found that the low-frequency noise is controlled by the transverse quasi-static conductivity alone, and it is given by
\begin{align}
\chi ''_{B_z B_z}  (z, \omega \rightarrow 0)
&\simeq \frac{\mu _0 ^2 \omega}{4}\int \frac{d^2\vec{q}}{4\pi^2} e^{-2qz} \sigma _{\perp ,0}  (\vec{q}) + \mathcal{O}(\omega^3),\label{Eq.chiBzBzw0}\\
\chi ''_{B_i B_j}  (z, \omega \rightarrow 0)
&\simeq \frac{\mu _0 ^2 \omega}{4}\int \frac{d^2\vec{q}}{4\pi^2} e^{-2qz} \sigma _{\perp ,0}  (\vec{q}) \hat{q}\cdot \hat{e}_i \hat{q}\cdot \hat{e}_j \nonumber \\
& \quad + \mathcal{O}(\omega^3)\label{Eq.chiBiBjw0},
\end{align}
where $\mu _0$ is the permeability of free space, and $i,j = x, y$ denote the directions in the plane of the 2D sample with corresponding basis vectors $\hat{e}_{i,j}$. Here and in the following, we write  $\mathcal{F}'$ and $\mathcal{F}''$  to denote the real and imaginary parts of the complex function $\mathcal{F} = \mathcal{F}' + i \mathcal{F}''$.
The remaining components of the noise $\chi ''_{B_i B_z}  (z, \omega \rightarrow 0)$ vanish in the presence of a symmetry that enforces the quasiparticle dispersion to satisfy $\epsilon_{\bf p}= \epsilon_{-{\bf p}}$, such as time reversal or space inversion (see discussion in Appendix~\ref{Sec.suppnoisederivation} of the Supplementary Material~\cite{supplementary}).
An interesting property that can serve as a consistency check of the low-frequency regime in which only transverse currents control the noise, is that the trace of the in-plane noise tensor equals the out-of-plane noise in this regime,
\begin{align}\label{Eq.ImchiBBxyzsum}
\chi ''_{B_x B_x} (z, \omega) + \chi ''_{B_y B_y} (z, \omega) = \chi ''_{B_z B_z} (z, \omega) + \mathcal{O}(\omega^3).
\end{align}
When the NV center is located at distances so as to probe the quantum collisionless regime (also referred to as ballistic in Ref.~\onlinecite{Agarwal}), the above formula together with Eq.~\eqref{Eq.anitranscond} leads to a remarkable geometric expression for the low-frequency noise 
\begin{equation}
\chi ''_{B_z B_z}  (z, \omega \rightarrow 0)
\simeq \frac{e^2 \mu _0 ^2}{16 \pi h} \frac{\omega}{z}\frac{(2S+1)}{2\pi}\mathcal{P}_{\rm FS} + \mathcal{O}(\omega^3). \label{Eq.ImchiBzBz}
\end{equation}
See Appendix~\ref{Sec.suppnoisederivation} of the Supplementary Material~\cite{supplementary} for derivation.
Here $\mathcal{P}_{\rm FS}$ is the perimeter of the FS in momentum space, which in the special case of a circular FS equals $2\pi p_{\rm F,0}$. Therefore, in the quantum regime, the low-frequency noise only depends on the perimeter of the FS, and not on dynamical properties such as the quasiparticle mass or interactions, and following the discussion in Sec.~\ref{Sec.spinonFStranscond}, we see that it is identical for the SFSS and metallic Fermi liquids.

Both the frequency and distance dependence of Eq.~\eqref{Eq.ImchiBzBz} are fingerprints of the regime of universal quantum low-frequency noise, which can serve as consistency checks in experiments. 
In particular, the linear in $\omega$-dependence of the noise is a hall-mark of systems with finite density of 
gapless states that contribute to magnetic noise either via current or spin fluctuations. Therefore, in addition to SFSSs and metals, it would also be present in idealized magnets with perfect SU(2) spin rotational symmetry leading to a quadratic magnon dispersion, and in $\mathbb{Z}_2$ spin liquids with a FS, as is shown in Ref.~\onlinecite{Chatterjee2019} for the noise from spin contributions for some of these systems.
The dependence on distance appears to be even more special, as it distinguishes between SFSSs with U(1) gauge fields, where $\chi''_{B_z B_z}\propto 1/z$, and gapless $\mathbb{Z}_2$ SFSSs. In the latter case, orbital current fluctuations are suppressed due to a gapped gauge field. Consequently, the spin fluctuations, $\chi ''_{{\rm spin},B_i B_j} (z,\omega)\propto 1/z^3$ (see below), dominate the noise and the distance dependence could serve as a smoking gun to detect the elusive U(1) SFSS in correlated materials.
We discuss the detailed distance dependence in the next subsection.

\subsection{Collisions and spin contributions to low-frequency noise}\label{Sec.noisecoll}

For simplicity, we focus here on the case of isotropic systems with circular FS. In this case, as we detail in Appendix~\ref{Sec.suppnoisewcoll} of the Supplementary Material~\cite{supplementary}, the only non-trivial components of the magnetic noise can be written as
\begin{align}
\chi ''_{B_i B_i}  (z, \omega) 
&= \frac{\mu _0 ^2 \omega}{16 \pi}\int dq q e^{-2qz} \left( \sigma '_{\|} (q, \omega) + \sigma '_{\perp} (q, \omega)  \right), \label{Eq.chiBBiso} \\
\chi ''_{B_z B_z}  (z, \omega)
&= \frac{\mu _0 ^2 \omega}{8 \pi}\int dq q e^{-2qz} \sigma '_{\perp}  (q, \omega), \label{Eq.BBresp}
\end{align}
where $i \in {x,y}$ in the above. The $q e^{-2qz}$ factor in the integrand acts like a filtering function that is peaked around $q \sim z^{-1}$ that allows the noise from current fluctuations to be probed at different wavevectors $q$. This wavevector selection is analogous to the noise from spin fluctuations, which is facilitated instead by a $q^3 e^{-2qz}$ factor in the integrand~\cite{Chatterjee2019} (see also Appendix~\ref{Sec.suppspinnoise} of the Supplementary Material~\cite{supplementary}). 
For purposes of probing the finite but low-frequency regime, in which the noise approximates its quasi-static quantum response $\chi ''_{B_z B_z}  (z, \omega) \simeq \chi ''_{B_z B_z}  (z, \omega \rightarrow 0)$, 
\begin{eqnarray}\label{Eq.ImchiBzBziso}
\chi ''_{B_z B_z}  (z, \omega \rightarrow 0)
\simeq (2S+1)\frac{e^2 \mu _0 ^2}{16 \pi h} \frac{\omega p_{\rm F,0}}{z}, 
\end{eqnarray} 
we focus in the following discussion on the out-of-plane noise $\chi ''_{B_z B_z}  (z, \omega)$. At low frequencies $\omega \ll v_{\rm F} /z$, in which case $\sigma '_{\|}  (q, \omega) \ll \sigma '_{\perp}  (q, \omega)$ and therefore $\chi ''_{B_i B_i}  (z, \omega)  \simeq \chi ''_{B_z B_z}  (z, \omega)/2$ .

To reach the quantum regime of noise given in Eq.~\eqref{Eq.ImchiBzBz} and Eq.~\eqref{Eq.ImchiBzBziso} in both metals and SFSSs, experiments must measure the noise at distances that are much larger than the typical Fermi wavelength $p_{\rm F,0}^{-1}$, but much smaller than the classical collision length $z_* \sim q_*^{-1}$, with $q_*$ defined in Eq.~\eqref{Eq.qstar}, namely,
\begin{equation}\label{Eq.distcriteria}
p_{\rm F,0}^{-1} \ll z \ll z_* = {\rm min} \left( \frac{(1+F_1)v_{\rm F}}{2\Gamma _1}, \frac{v_{\rm F}}{(\Gamma _1 + \Gamma _2)}\right).
\end{equation}
At low temperatures, where the momentum preserving collision rate $\Gamma _2$ vanishes, and the momentum relaxing collisions rates are dominated by elastic impurities $\Gamma _1 \sim v_{\rm F} l _{\rm mfp}^{-1}$ (assuming $F_1$ is not large), the above criterion can be simply expressed as
\begin{equation}
p_{\rm F,0}^{-1} \ll z \ll z_* \sim l_{\rm mfp}.
\end{equation}
In general, the noise at longer distances will deviate from the behavior of Eq.~\eqref{Eq.ImchiBzBz} and Eq.~\eqref{Eq.ImchiBzBziso} as shown in Fig.~\ref{Fig.Bperpnoise} and summarized in Table~\ref{Table.chiBzBzscaleQS}. The behavior of the noise in the classical hydrodynamic or diffusive regimes was discussed in Ref.~\onlinecite{Agarwal} and we review it and extend it to SFSSs in Appendix~\ref{Sec.suppnoisewcoll} of the Supplementary Material~\cite{supplementary}. While the aforementioned requirement on the distance is the same for metals and SFSSs, the requirement in terms of the frequency are more stringent for the SFSS than for the metal. This is a consequence of the behavior of the transverse conductivity illustrated in Fig.~\ref{Fig.condspinon}(a), and discussed in the text surrounding Eqs.~\eqref{Eq.spinonpeakwidth}--\eqref{Eq.balpeakwidthspinon}. More specifically, provided the noise is measured at a distance satisfying Eq.~\eqref{Eq.distcriteria}, in order to reach the quantum regime in metals the frequency must be in the regime
\begin{equation}\label{Eq.Deltawmetal}
\omega \ll \frac{v_{\rm F}}{z} \quad \text{for metals}.
\end{equation}
However, in the case of SFSSs within a distance satisfying Eq.~\eqref{Eq.distcriteria}, the noise must be measured at frequencies
\begin{equation}\label{Eq.Deltawspinon}
\omega \ll \Delta \omega _s ( q = z^{-1}) = \frac{v_{\rm F} c^2}{2\omega _p^2} \frac{1}{z^3} \quad \text{for SFSSs}.
\end{equation}
Typical values of $\Delta \omega _s$ are shown in Table~\ref{Table.spinon} for several spinon FS candidates. The behavior of the noise at different finite frequencies for metals and SFSSs as a function of distance is shown in Fig.~\ref{Fig.Bperpnoise}. We can see in this figure that satisfying the frequency requirement of Eq.~\eqref{Eq.Deltawmetal}--~\eqref{Eq.Deltawspinon}, which guarantees that the noise is probing the quasi-static behavior of the transverse conductivity,  is actually easier in the quantum regime than in the classical regimes for both metals and SFSSs. This is because the classical regimes are accessed at larger distances from the sample and both frequency scales $v_{\rm F}/z$ (for metals) and $\Delta \omega _s $ (for SFSSs) decrease as the distance from the sample increases (the probed wavevector decreases).

In addition to current fluctuations, spin fluctuations also contribute to the noise and may therefore affect its frequency and distance dependence. However, it can be shown that (see Appendix~\ref{Sec.suppspinnoise})
the low-frequency limit of the magnetic noise originating from spin fluctuations is given by
\begin{align}\label{Eq.spinnoise}
\chi ''_{{\rm spin},B_i B_j} (z,\omega)
&\simeq  -\delta _{ij} \frac{e^2 \mu ^2 _0}{8 \pi ^2} \left(\frac{g_s m^*}{4m_0}\right)^2
\frac{\omega}{p_{\rm F,0} z^3},
\end{align}
where $m_0$ is the electron rest mass in vacuum.
In the low-frequency limit from Eqs.~\eqref{Eq.Deltawmetal}--\eqref{Eq.Deltawspinon} where the noise is dictated by the quasistatic behavior of the transverse conductivity, the ratio of the noise originating from spin fluctuations to the noise from current fluctuations from Eq.~\eqref{Eq.ImchiBzBziso} is given by
\begin{eqnarray}\label{Eq.spinnoiseratio}
\left|\frac{\chi ''_{{\rm spin}, B_{z} B_{z}} (z,\omega \rightarrow 0)}{\chi ''_{B_{z} B_{z}} (z,\omega \rightarrow 0)}\right|
&\simeq & \left(\frac{g_s m^*}{4m_0}\right)^2 
\frac{2}{p_{\rm F,0}^2z^2},
\end{eqnarray}
Therefore we see that the spin noise is suppressed by a factor $(p_{\rm F,0} z)^{-2}$ relative to the current noise, in the quantum regime and therefore it is highly subdominant once the criterion Eq.~\eqref{Eq.distcriteria} is satisfied. More details of the behavior of the spin noise in the quantum limit are presented in Appendix~\ref{Sec.suppspinnoise} of the Supplementary Material~\cite{supplementary}.

Before closing this section we would like to contrast our work with and make a few comments on a recent and very interesting, closely related study of magnetic noise of spin liquid states in Ref.~\onlinecite{Chatterjee2019}. We begin by noting that Ref.~\onlinecite{Chatterjee2019} focused only on the contributions to magnetic noise originating from spin fluctuations such as those we describe in Appendix~\ref{Sec.suppspinnoise} of the Supplementary Material~\cite{supplementary}, but did not consider the possibility of orbital current fluctuations. The latter can be justified in strongly insulating spin liquid states such as those with $\mathbb{Z}_2$ gauge fields and electrically neutral spinons, but it is not necessarily justified in U(1) spin liquids, especially when they feature a gapless FS as we have demonstrated in our study. 
At the distances relevant for probing the universal noise, $p_{\rm F,0}^{-1} \ll z \ll l_{\rm mfp}$, the low-frequency noise [$\omega \ll \Delta \omega_s$ with $\Delta\omega_s$ given in Eq.~(\ref{Eq.Deltawspinon})] is dominated by the universal result $\propto \omega/z$ in Eq.~(\ref{Eq.ImchiBzBziso}). For frequencies above $\Delta \omega_s$ the noise decreases with frequency due to the suppression of orbital currents and the much weaker noise contribution from spin fluctuations $\propto \omega/z^3$ dominates only at a parametrically larger frequency scale, namely, in the regime
\begin{equation}
 p_F z\Delta \omega_s (q = 1/z)\ll \omega \ll \frac{v_{\rm F}}{z}.
\end{equation}
The noise in this regime is suppressed by a factor $\sim (p_Fz)^{-2}$ with respect to the zero-frequency noise.

Another discrepancy between our results and those of Ref.~\onlinecite{Chatterjee2019} concerns the comparison of $\mathbb{Z}_2$ and U(1) spin liquids with a FS. In the presence of time reversal symmetry and neglecting spin-orbit coupling, the spin anti-symmetric fluctuations of the SFSS decouple from the spin-symmetric fluctuations and also from the U(1) gauge fields, and hence they behave as in ordinary metals, irrespective of the gauge structure. In particular this means, the low-frequency spin noise, given by Eqs.~(\ref{Eq.noise}) and (\ref{Eq.spinnoise}), obeys the same scaling $\coth(\beta\omega/2)\omega/z^{3}$ with frequency, temperature and distance for U(1) or $\mathbb{Z}_2$ gauge fields  (assuming temperature remains low enough compared to the Fermi energy that collisions can be ignored). This is in contrast to Tables I and II of Ref.~\onlinecite{Chatterjee2019}, which predict different distance and temperature dependences of the noise from spin fluctuations in $\mathbb{Z}_2$ and U(1) spin liquids with a FS in certain limiting cases. The origin of these discrepancies appear to be mistakes in Ref.~\onlinecite{Chatterjee2019} in the calculation of the spin correlation functions in the case of the $\mathbb{Z}_2$ spin liquid, which we highlight explicitly in Appendix~\ref{Sec.suppspinnoise} of the Supplementary Material~\cite{supplementary}.

Moreover, the equivalence of the scaling with distance and frequency of the spin fluctuations in $\mathbb{Z}_2$ and U(1) spin liquids with a FS remains true at elevated frequencies, which implies that the range of validity of our results for the noise arising from spin fluctuations in Appendix~\ref{Sec.suppspinnoise} is identical to the one expected for an ordinary LFL, namely,  $\omega \ll v_{\rm F} /z$. In contrast, Eq.~(B28) of Ref.~\onlinecite{Chatterjee2019} predicts a different range of validity. The origin of this issue is that Ref.~\onlinecite{Chatterjee2019} computed the spin correlator from a single bubble diagram with dressed Green's functions containing an imaginary self-energy scaling as $\sim \omega^{2/3}$, but without any vertex corrections. As has been shown before~\cite{Kim1994,Mross2010,Kim1995}, if one tries to go beyond the RPA-like treatmeant with bare bubble diagrams, the vertex corrections are crucial, because they cancel divergences from the self-energy in gauge invariant particle-hole correlation functions in the U(1) SFSS restoring the behavior of LFL theory at small $\omega$ and $q$.

We conclude that SFSSs with U(1) or $\mathbb{Z}_2$ gauge fields can be distinguished in noise spectroscopy because the former has a dominant orbital current contribution with a different distance dependence at low frequencies. In contrast, the contribution from spin fluctuations to the noise does not depend on the gauge field and is the same in both cases to the behavior of spin fluctuations in ordinary metals.

\begin{table}[t]
\begin{tabular}[b]{cccc}
\hline\\[-1.em]
\hline\\[-1.em]
& \quad Diffusive \quad \quad & \quad Hydrodynamic \quad \quad & Quantum \\ [.1em]
& $z \gg z_{**}$ & $z_{**} \gg z \gg z_{*}$ & $z_{*} \gg z$\\ [.5em]
\hline\\[-1.em]
$\chi ''_{B_z B_z} (z) $ & {\large $\frac{\omega}{z^2}$} & $\omega f(z)$ & {\large $\frac{\omega}{z}$} \\ [1.em]
\hline\\[-1.em]
\hline\\[-1.em]
\end{tabular}
\caption{Frequency and distance dependence of low-frequency magnetic noise from current fluctuations in various transport regimes which is the same in both metals and SFSSs. 
For the hydrodynamic regime, the function $f(z) \sim {\rm Ei}(-z/z_*) - {\rm Ei}(-z/z_{**})$, where Ei$(x)$ denotes the exponential integral function (See Appendix~\ref{Sec.suppnoisewcoll} of the Supplementary Material for details~\cite{supplementary}).
}\label{Table.chiBzBzscaleQS}
\end{table}


\begin{figure}[h]
\includegraphics[scale=1.0]{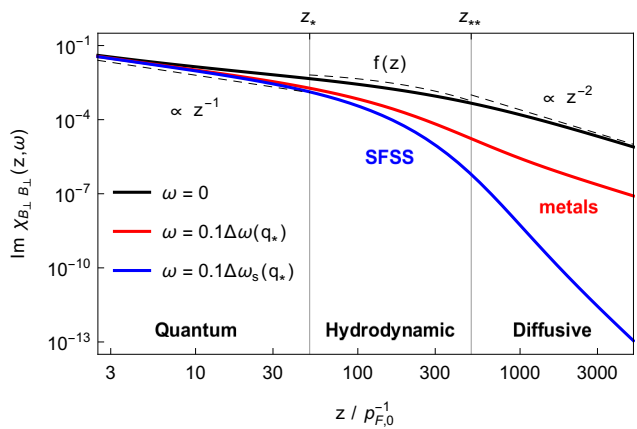}
\caption{
Plots showing the deviation of the magnetic noise due to current fluctuations from the quasi-static noise (black) for metals at frequency $\omega = 0.1\Delta \omega (q_*)$ (red) and SFSSs at frequency $\omega = 0.1\Delta \omega _s(q_*)$ (blue) corresponding to those shown in Fig.~\ref{Fig.condspinon}.
}
\label{Fig.Bperpnoise}
\end{figure}

\section{Summary and discussion}\label{Sec.summary}

In this paper, we have derived several remarkable results that may each have profound consequences for the study of material properties: 

First, we have extended previous work to show that the quasi-static transverse electrical conductivity of two-dimensional metals in the quantum (or ballistic) limit $p_{\rm F} \gg q \gg l^{-1}_{\rm mfp}$ takes a universal form, given in Eq.~\eqref{Eq.main1}, which is only controlled by the FS geometry and is independent of kinematic details or interactions. This is a direct consequence of the quantum degeneracy of electrons in the FS and markedly different from the classical Drude-type transport at zero wavevector.

Second, we have demonstrated that the universal transverse conductivity manifests itself as a universal low-frequency magnetic noise above the 2D sample, given by Eq.~\eqref{Eq.main2}, which can be directly measured by spin qubit noise spectroscopy. In experiments, the dependence of the noise on the qubit-sample distance can be used to identify the quantum regime, in which the noise is fully determined by the FS perimeter.

Third, we have found that the quasi-static transverse electrical conductivity and the corresponding magnetic noise of two-dimensional U(1) spin liquids with a spinon FS are identical to those of a metal. In a sense, we have shown that in a certain limit the SFSS behaves in the same way as an ordinary metal. This is in striking contrast to the insulating behavior of SFSSs in standard transport experiments at zero wavevector. We have also pointed out how noise spectroscopy can be used to distinguish between spinon FS states with $\mathbb{Z}_2$ and U(1) gauge fields.

At first glance, one might feel uneasy about the metallic behavior of a U(1) SFSS, especially because it is a common jargon in the community to refer to the spinon as a ``neutral fermion''. How can then a ``neutral fermion'' respond to a long-wavelength transverse quasi-static electric field in the same way as an ordinary  electron? This viewpoint is based on traditional parton descriptions of the SFSS, where the spinon is viewed as carrying zero electric charge and the chargon carrying the full charge of the electron. However, this charge assignment is essentially a book-keeping convention that is motivated by viewing this state as an ``electrical insulator''. Only the net charge assignment to the combined spinon and chargon bound state is physical, and if calculations are done in a consistent and gauge-invariant fashion, it is equally legitimate to assign the electrical charge to the spinon while viewing the chargon as electrically neutral.

Nevertheless, the above concern about the metallic response of a SFSS might seem particularly worrysome, if one takes the standpoint that such a state can in principle emerge out of a pure-spin Heisenberg-like model, in which there are no microscopic electrons in the Hilbert space, as has sometimes been emphasized by referring to the U(1) SFSS as a spin Bose metal~\cite{Sheng2009,Block2011}. 
Our results can be reconciled with this point of view by noticing that, according to Eq.~\eqref{Eq.spinonpeakwidth}, the metallic behavior of the transverse conductivity only occurs in a frequency window $\Delta \omega_s$ that decreases as $\sim 1/\omega_p^2$,
where $\omega_p$ is the effective plasma scale of the problem. This plasma scale is essentially the Mott scale, or more precisely the scale controlling the pseudo-gap for optical absorption at $q=0$. To put it another way, if we take the Fermi energy to be controlled by the Heisenberg-like exchange scale of the problem $v_{\rm F} p_{\rm F,0} \sim J$ (since this scale often controls the spinon energy bandwidth in lattice models~\cite{Zhou2017}),  take the effective speed of light to be comparable to the Fermi velocity $c \sim v_{\rm F}$, and take the optical gap to be controlled by a Hubbard-like scale $\omega_P \sim U$ (see e.g. Ref.~\onlinecite{Florens2004}), the edge of the particle-hole continuum of the spinon excitations would be at frequencies
\begin{equation}
\frac{\omega}{J} \sim \frac{q}{p_{\rm F,0}},
\end{equation}
while the metallic behavior of the transverse conductivity would emerge at a much lower frequency scale $\Delta \omega_s$ of the order
\begin{equation}
\frac{\Delta \omega_s}{J} \sim \left(\frac{J}{U}\right)^2 \left(\frac{q}{p_{\rm F,0}}\right)^3.
\end{equation}
Therefore this window of low-frequency transverse metallicity is expected to disappear if one takes $U \rightarrow \infty$ while keeping the Heisenberg scale $J$ constant. Our result implies that even though the SFSS can emerge and be understood in the limit of $U \rightarrow \infty$ where the electron might disappear from the relevant Hilbert space, a large but finite $U$ scale remains in a sense a relevant perturbation that alters the low-frequency and long-wavelength responses of this phase of matter in a non-perturbative fashion. It also emphasizes the complex and non-trivial behavior of physical observables near the corner of small $\omega$ and $q$ of the spinon particle-hole continuum whose behavior changes even more dramatically than in an ordinary metal, especially the transverse conductivity that resembles that of an insulator when first taking $q\to 0$ and then $\omega\to 0$, while behaving like a metal in the reverse order of limits.

We would like to close by making a strong case for viability of the observation of the universal low-frequency noise as a powerful experimental ``smoking gun'' for the presence of the U(1) SFSS in correlated materials. The key to be able to confirm this regime in materials lies in observing the predicted scaling with frequency $\omega$ and distance $z$ from the sample $\propto \omega/z$, which emerges at low temperatures ($T\ll E_F$) and in the clean long wavelength regime $(p_{\rm F,0}^{-1} \ll z \ll l_{\rm mfp}^{-1})$. The observation of linear in $\omega$ scaling of the noise signals that the system has a finite density of states contributing, such as it is the case for the spin noise of a different SFSS with $\mathbb{Z}_2$ gauge field. Therefore, while this linear $\omega$ scaling of the noise would be a non-trivial indicator the presence of a finite density of states in the system of interest, it alone is not sufficient to distinguish the SFSS from other states. The $1/z$ dependence is however much more special, as this is directly an indication of the $1/q$ scaling of the transverse conductivity. To our knowledge the only other states of matter in 2D that can produce the same $1/z$ scaling are metallic states with a FS in the clean long-wavelength regime $(p_{\rm F,0}^{-1} \ll z \ll l_{\rm mfp})$. The additional fact that the prefactor of the noise in this regime is controlled only by the perimeter of the FS in momentum space and universal constants of nature [see Eq.~\eqref{Eq.main2}], makes this a highly robust indication of the presence of the SFSS. A possible metallic FS state can be easily ruled out by ordinary DC transport measurements, since the DC conductivity of the spinon is expected to vanish in the ideal zero temperature limit in analogy to an insulator.

Finally, we would like to contrast the metallic universal transverse conductivity and the universal magnetic field noise, with other remarkable and non-trivial behaviors that can be used to advocate for the presence of the SFSS, namely, the presence of quantum oscillations~\cite{Motrunich2006,Chowdhury2018,Sodemann2018} and cyclotron resonance~\cite{Rao2019}. One aspect that makes our proposal conceptually advantageous over the above is that the transverse conductivity and the magnetic field noise are determined by the linear response correlation functions of the SFSS, which do not require active modification of the state of interest. This is especially true for the noise, which is ideally a non-invasive probe that simply monitors the fluctuations of the equilibrium state~\cite{Casola2018}. On the other hand, observing quantum oscillations and cyclotron resonance requires exerting an external magnetic field, which induces non-perturbative modifications to the FS state and, strictly speaking, demands a description beyond the linear-response regime. In the clean limit, such non-perturbative modifications occur ideally at arbitrarily low magnetic fields and are therefore finger-prints of the FS, but in practice they require the application of sizable magnetic fields to overcome disorder and temperature fluctuations, making it more delicate to differentiate from competing scenarios such as the quantum oscillations expected in inverted band insulators with small gaps~\cite{Knolle2015,Zhang2016,Knolle2017}. Also, it is worth emphasizing that even though the oscillations and cyclotron resonance of SFSS resemble those of metals, the detailed features can be different and depend on hard to estimate microscopic constants such as the ratio of effective magnetic field to  applied physical magnetic field that the spinons experience~\footnote{see however Ref.~\onlinecite{Sodemann2018} for an example where the period of de Haas-van Alphen oscillations of 2D SFSS was found to be identical to that of electrons.}. 
The robustness of the expected universal magnetic noise against microscopic details summarized in Eq.~\eqref{Eq.main2} is therefore a highly appealing feature of this probe. The above prompts us to advocate that the observation of the regime of universal low-frequency noise from Eq.~\eqref{Eq.main2}, would constitute a ``smoking gun'' evidence to finally pin-point the presence of the long-sought-after U(1) SFSS in correlated materials.

\section{Acknowledgments}
We would like to thank Suraj Hegde for valuable discussions.
%


%

\clearpage

\appendix

\onecolumngrid

\section{Derivation of the conductivities of isotropic metals and spinon Fermi surface states}\label{Sec.suppfullsolve}

In this appendix, we outline the derivation of the longitudinal and transverse conductivities of isotropic SFSSs, Eqs.~\eqref{Eq.condsppll}--\eqref{Eq.condchperp} in the main text, following the discussion in Sections~\ref{Sec.spinonlowETh}--\ref{Sec.spinonFStranscond}. The corresponding conductivities for metals, Eqs.~\eqref{Eq.condfpll}--\eqref{Eq.condfperp} in the main text, are obtained in a similar fashion and will not be shown here explicitly.
The linearized kinetic equation with collisions reads,
\begin{eqnarray}
\partial _t p_\theta + v_{\rm F} \hat{p}_{\theta} \cdot \vec{\partial} _{r} \left(p_\theta + \int d\theta'  F_{\theta, \theta '} p_{\theta'} \right) 
= \vec{\mathsf{e}} \cdot \hat{p}_{\theta} - \Gamma _1 (p_\theta - P_0[p_\theta]) 
- \Gamma _2 (p_\theta - P_0[p_\theta] - P_1[p_\theta] - P_{-1} [p_\theta]),
\end{eqnarray}
where $P_l [p_\theta] = e^{i l \theta} \int  \frac{d\theta}{2\pi} e^{-i l \theta} p_\theta$, while the linearized coupled spinon-gauge field equations of motion read,
\begin{eqnarray}
\epsilon \vec{\partial} _r \cdot (\vec{\mathsf{e}} - \vec{E}) &=& \int \frac{d^2 \vec{k}}{(2\pi)^2} \delta (p_{\rm F,0} - k) p_\theta,\\
\vec{\partial} _r \times (\vec{\mathsf{b}} - \vec{B}) - \epsilon \mu \partial _t (\vec{\mathsf{e}} - \vec{E}) &=& \mu \frac{p_{\rm F,0}}{m} \hat{p}_\theta \int \frac{d^2 \vec{k}}{(2\pi)^2} \delta (p_{\rm F,0} - k) p_\theta,
\end{eqnarray}
where $\epsilon$ and $\mu$ are the gauge dielectric constant and magnetic permeability respectively and $m$ is the transport mass.
Using ansatz of the form $p_\theta = p(\vec{q},\theta, \omega) e^{i(\omega t - \vec{q} \cdot \vec{r})}$, $\vec{E} = \vec{E}(\vec{q}, \omega) e^{i(\omega t - \vec{q} \cdot \vec{r})}$ and $\mathsf{e} = \mathsf{e}(\vec{q}, \omega) e^{i(\omega t - \vec{q} \cdot \vec{r})}$ (similarly for $\vec{B}$ and $\vec{b}$), for $F_{l >1} = 0$, we have the following coupled equations (henceforth suppressing the $\omega$ label):
\begin{eqnarray}
\left( i \omega - i v_{\rm F} q \cos \theta + \Gamma _1 + \Gamma _2 \right) p(\vec{q},\theta) 
&=& \vec{\mathsf{e}} (\vec{q}) \cdot \hat{p}_{\theta} + \left(i F_0 v_{\rm F} q \cos \theta + \Gamma _1 + \Gamma _2 \right) P_0 (\vec{q}) + \left(i F_1 v_{\rm F} q \cos \theta + \Gamma _2 \right) p_1 (\vec{q}, \theta), \label{Eq.LLKE} \\
- i \epsilon \vec{q} \cdot (\vec{\mathsf{e}}(\vec{q}) - \vec{E}(\vec{q})) &=& \frac{p_{\rm F,0}}{2\pi} P_0 (\vec{q}), \\
- i \vec{q} \times (\vec{\mathsf{b}}(\vec{q}) - \vec{B}(\vec{q})) &-& i \epsilon \mu \omega (\vec{\mathsf{e}}(\vec{q}) - \vec{E}(\vec{q})) = \mu \frac{(p_{\rm F,0})^2}{2\pi m} \left(P^+_1 (\vec{q}) \hat{q} + P^-_1 (\vec{q}) \hat{q}_\perp \right),
\end{eqnarray}
 where henceforth $\theta$ is taken relative to $\vec{q}$, $\hat{p}_\theta = \cos \theta \hat{q} + \sin \theta \hat{q}_\perp$  and
\begin{eqnarray}
p (\vec{q}, \theta) &=& \sum_{l=-\infty}^{\infty} P_l (\vec{q}) e^{i l \theta} = P_0 (\vec{q}) +  \sum_{l=1}^{\infty} p_l (\vec{q},\theta), \\
p _l (\vec{q}, \theta) &=&  2\left(P_l ^+ (\vec{q}) \cos \left(l \theta \right) + P_l ^- (\vec{q}) \sin \left( l \theta \right) \right),  \\
P_l ^{+} (\vec{q}) &=& \int \frac{d\theta}{2\pi} \cos (l \theta) p  (\vec{q},\theta), \quad P_l ^{-} (\vec{q}) = \int \frac{d\theta}{2\pi} \sin(l \theta) p (\vec{q}, \theta).\label{Eq.Chebydef}
\end{eqnarray}
Rewriting the LKE,
\begin{eqnarray}\label{Eq.collsoln}
&&p(\vec{q},\theta) = \frac{-i \tilde{\mathsf{e}}(\vec{q}) \cdot \hat{p}_\theta + \left(F_0 \cos \theta - i \tilde{\Gamma} _1- i \tilde{\Gamma} _2 \right) P_0 (\vec{q}) + \left(F_1 \cos \theta - i \tilde{\Gamma} _2 \right) p_1 (\vec{q}, \theta) }{\left( s - \cos \theta - i \tilde{\Gamma} _1 - i \tilde{\Gamma} _2 \right) }, \\
&& s = \frac{\omega}{v_{\rm F} q}, \quad \tilde{\Gamma} _i = \frac{\Gamma _i}{v_{\rm F} q}, \quad \tilde{\mathsf{e}}(\vec{q}) = \frac{\mathsf{e}(\vec{q})}{v_{\rm F} q}.
\end{eqnarray}

Anticipating the decoupling between the even and odd solutions, we express the gauge and EM fields in terms of their transverse and longitudinal components relative to $\hat{q}$, $\vec{f} = f_{\||}\hat{q} + f_\perp \hat{q}_\perp + f_z \hat{z}$ ($\vec{f} = \vec{e}, \vec{b}, \vec{E}, \vec{B}$ and $\hat{z} = \hat{q} \times \hat{q}_\perp$) and project Eq.~\eqref{Eq.collsoln} to the 0th and 1st components,
\begin{eqnarray}
P_0 (\vec{q}) &=& -i q \tilde{\mathsf{e}}_\|(\vec{q})\Omega _1 (q)+  q\left(F_0 \Omega _1 (q) - i \left(\tilde{\Gamma} _1+ \tilde{\Gamma} _2\right)\Omega _0  (q) \right) P_0 (\vec{q}) +  2q \left(F_1 \Omega _2  (q)- i \tilde{\Gamma} _2 \Omega _1  (q) \right)  P_1^+ (\vec{q}), \label{Eq.p0}\\ 
P_1^+ (\vec{q}) &=& -i q \tilde{\mathsf{e}}_\|(\vec{q})\Omega _2 (q) + q\left(F_0 \Omega _2  (q) - i \left(\tilde{\Gamma} _1+ \tilde{\Gamma} _2 \right) \Omega _1  (q) \right) P_0 (\vec{q}) +  2q \left(F_1 \Omega _3  (q) - i \tilde{\Gamma} _2 \Omega _2  (q) \right)  P_1^+ (\vec{q}), \label{Eq.p1p} \\ 
P_1^- (\vec{q}) &=& q\Big(\Omega _0  (q)  - \Omega _2  (q) \Big) \left( -i \tilde{\mathsf{e}}_\perp(\vec{q}) -2i \tilde{\Gamma} _2 P_1^- (\vec{q})\right)  +  2  q\Big( \Omega _1  (q)  - \Omega _3 (q)\Big) F_1 P_1^- (\vec{q}),  \label{Eq.p1m}  \\
\Omega _l  (q)&=& \frac{1}{q}\int \frac{d\theta}{2\pi} \frac{(\cos \theta )^l}{\left(s - \cos \theta - i \tilde{\Gamma} _1 - i \tilde{\Gamma} _2 \right)} = v_{\rm F} \int \frac{d\theta}{2\pi} \frac{(\cos \theta )^l}{\left(\omega  - v_{\rm F} q \cos \theta - i \Gamma _1 - i \Gamma _2 \right)},
\end{eqnarray}
with the linearized coupled spinon-gauge field equations of motion,
\begin{eqnarray}
\frac{p_{\rm F,0}}{2\pi} P_0 (\vec{q}) &=& - i \epsilon q (\mathsf{e}_{\|}(\vec{q}) - E_{\|}(\vec{q})),\label{Eq.sglin0} \\
\frac{(p_{\rm F,0})^2}{2\pi m} P^+_1 (\vec{q}) &=& - i \epsilon \omega (\mathsf{e}_\|(\vec{q}) - E_\|(\vec{q})) , \label{Eq.sglin1p} \\
\frac{(p_{\rm F,0})^2}{2\pi m} P^-_1 (\vec{q}) &=& i \mu ^{-1} q(\mathsf{b}_z(\vec{q}) - B_z(\vec{q})) - i \epsilon \omega (\mathsf{e}_\perp(\vec{q}) - E_\perp(\vec{q})) \\
&=& -i \epsilon \left(\frac{\omega ^2 - c^2 q^2}{\omega} \right)  (\mathsf{e}_\perp (\vec{q}) - E_\perp (\vec{q})) \label{Eq.sglin1m}
\end{eqnarray}
where in the last line we used the respective Faraday's laws $\vec{\partial} _r \times \vec{E} = - \partial _t \vec{B}$ and $\vec{\partial} _r \times \vec{e} = - \partial _t \vec{b}$ and defined the velocity of the gauge boson in the medium $c = 1/\sqrt{\epsilon \mu}$.

For the case $ \tilde{\Gamma}_{1,2} > 0$, we extend $s$ into the complex plane ($z$) and evaluate these integrals by a change of variables $z = e^{i \theta}$, 
\begin{eqnarray}
\Omega _{l} (q) &=& \frac{1}{q} \frac{i}{2\pi} \frac{1}{2^{l-1}} \oint _{C} dz \frac{(z^2 + 1)^l}{z^l (z-z_+)(z-z_-)}, \\
&=& - \frac{1}{q}\frac{1}{2^{l-1}} \left({\rm Res}(0) (1-\delta_{l,0}) + {\rm Res}(z_\xi) \right), \\
&=& - \frac{1}{q}\frac{1}{2^{l-1}} \left(\frac{(1-\delta_{l,0})}{(l-1)!}\frac{d^{l-1}}{dz^{l-1}}\left.\left( \frac{(z^2 + 1)^l}{(z-z_-)(z-z_+)}\right)\right|_{z=0} + \frac{(z_\xi + z_{-\xi})^l}{z_\xi -z_{-\xi}}\right),
\end{eqnarray}
where in the above $C$ denotes the unit circle and
\begin{eqnarray}
z_{\pm} &=& \zeta \pm \sqrt{\zeta ^2-1}, \quad z_+z_- = 1, \\
\zeta &=& s - i( \tilde{\Gamma}_1 + \tilde{\Gamma}_2), \\
\xi &=& -{\rm sgn}(s).
\end{eqnarray}
For $|z_+| \neq |z_-| \neq 1$, corresponding to solutions outside the particle-hole continuum, we have explicitly
\begin{eqnarray}
\Omega _{0} (q) &=& -\frac{2}{q} \frac{z_\xi}{z_\xi^2 -1} = \frac{{\rm sgn}(s)}{q} \frac{1}{\sqrt{\zeta ^2 - 1}},\\
\Omega _{1} (q) &=& -\frac{1}{q} \left( 1 + \frac{z_\xi^2 + 1}{z_\xi^2 - 1} \right) = \frac{{\rm sgn}(s)}{q} \frac{z_\xi}{\sqrt{\zeta ^2-1}} = z_\xi \Omega _{0} (q), \\
\Omega _{2} (q) &=& -\frac{1}{2q} \left( 2\zeta + \frac{(2\zeta)^2 }{z_\xi-z_\xi ^{-1}}\right) = \frac{{\rm sgn}(s)}{q} \frac{\zeta z_\xi}{\sqrt{\zeta ^2-1}} = \zeta \Omega _{1} (q), \\
\Omega _{3} (q) &=& -\frac{1}{4q} \left( 2 + (2\zeta)^2+ \frac{(2\zeta)^3 }{z_\xi-z_\xi ^{-1}} \right) = \frac{1}{q} \left(-\frac{1}{2} + {\rm sgn}(s) \frac{\zeta ^2 z_\xi}{\sqrt{\zeta ^2-1}}\right) = -\frac{1}{2q} + \zeta \Omega _{2} (q), \\
\Omega _{4} (q) &=& -\frac{1}{8q} \left( 2 (2\zeta ) + (2\zeta )^3+ \frac{(2\zeta)^4 }{z_\xi-z_\xi ^{-1}} \right) = \frac{\zeta}{q} \left(-\frac{1}{2} + {\rm sgn}(s) \frac{\zeta ^2 z_\xi}{\sqrt{\zeta ^2-1}}\right) = \zeta \Omega _{3} (q), \\
\Omega _{0} (q) &-& \Omega _{2} (q) = \frac{1}{q}z_\xi, \quad \Omega _{1} (q) - \Omega _{3} (q) = \frac{1}{2q}z_\xi ^2,
\end{eqnarray}

As a consistency check, let us rewrite Eq.~\eqref{Eq.p1p},
\begin{eqnarray}
P_1^+ (\vec{q}) &=& -i q \tilde{\mathsf{e}}_\|(\vec{q})\Omega _1 (q) \zeta + q\left(F_0 \Omega _1 (q) \zeta - i \left(\tilde{\Gamma} _1+ \tilde{\Gamma} _2 \right) \Omega _0  (q) z_\xi \right) P_0 (\vec{q}) +  2q \left(F_1 \left(-\frac{1}{2q}+\zeta \Omega _2  (q)\right) - i \tilde{\Gamma} _2 \Omega _1  (q) \zeta \right)  P_1^+ (\vec{q}) \nonumber \\ 
&=& \zeta P_0 (\vec{q}) - i q \left(\tilde{\Gamma} _1+ \tilde{\Gamma} _2 \right) \Omega _0  (q) (z_\xi - \zeta)  P_0 (\vec{q}) - F_1 P_1^+ (\vec{q}) \nonumber \\
P_1^+ (\vec{q}) &=& \frac{s}{1+F_1} P_0 (\vec{q}), \label{Eq.ChargeconsvP0}
\end{eqnarray}
which is consistent with the relation obtained from Eq.~\eqref{Eq.sglin0} and Eq.~\eqref{Eq.sglin1p} with the quasi-particle mass $m^* = \frac{p_{\rm F,0}}{v_{\rm F}} = (1+F_1)m$ as is expected since the relaxation terms introduced do not violate charge conservation.

This allows us to solve for $P_0 (\vec{q})$ in terms of $\vec{\mathsf{e}}_\|(\vec{q})$,
\begin{eqnarray}
P_0 (\vec{q}) &=&\frac{-i q \Omega _1 (q)}{1- q\left(F_0 \Omega _1 (q) - i \left(\tilde{\Gamma} _1+ \tilde{\Gamma} _2\right)\Omega _0  (q) \right) - 2\frac{s q}{1+F_1} \left(F_1 \Omega _2  (q)- i \tilde{\Gamma} _2 \Omega _1  (q) \right) } \tilde{\mathsf{e}}_\|(\vec{q}) = -i \Pi ^{\mathsf{e}} _\| (q) \mathsf{e}_\|(\vec{q}) \\
\Pi ^{\mathsf{e}} _\| (q) &=& \frac{1}{v_{\rm F}q} \frac{ q \Omega _1 (q)}{1- q \Omega _0 (q)\left(F_0 z_\xi- i \left(\tilde{\Gamma} _1+ \tilde{\Gamma} _2\right) \right) - 2\frac{s}{1+F_1} q \Omega _1  (q) \left(F_1 \zeta - i \tilde{\Gamma} _2  \right) }  \\
&=& \frac{1}{v_{\rm F}q} \frac{1}{{\rm sgn}(s) z_{-\xi}\sqrt{\zeta ^2 - 1}- \left(F_0 - i \left(\tilde{\Gamma} _1+ \tilde{\Gamma} _2\right) z_{-\xi} \right) - 2\frac{s}{1+F_1} \left(F_1 \zeta - i \tilde{\Gamma} _2  \right) } 
\end{eqnarray}
and using Eq.~\eqref{Eq.sglin0}, we have
\begin{eqnarray}
i \epsilon q  E_{\|}(\vec{q}) &=& \frac{p_{\rm F,0}}{2\pi} P_0 (\vec{q}) \left(1 - \epsilon q \frac{2\pi}{p_{\rm F,0}}\frac{1}{\Pi ^{\mathsf{e}} _\| (q)}\right), \\
\frac{p_{\rm F,0}}{2\pi} P_0 (\vec{q})  &=& i \epsilon q \left(1 - \epsilon q \frac{2\pi}{p_{\rm F,0}}\frac{1}{\Pi ^{\mathsf{e}} _\| (q)}\right)^{-1} E_{\|}(\vec{q}) 
\end{eqnarray}
and from Eq.~\eqref{Eq.sglin1p}, we obtain the longitudinal conductivity $\sigma _\| (q)$,
\begin{eqnarray}
j_\| (\vec{q}) &=& \frac{(p_{\rm F,0})^2}{2\pi m} P^+_1 (\vec{q}) =\frac{p_{\rm F,0}}{m} \frac{p_{\rm F,0}}{2\pi} \frac{s}{1+F_1} P_0 (\vec{q}) = \sigma _\| (q) E_{\|}(\vec{q}), \\
\sigma _\| (q) &=& i \epsilon \omega \left(1 - \epsilon q \frac{2\pi}{p_{\rm F,0}}\frac{1}{\Pi ^{\mathsf{e}} _\| (q)}\right)^{-1}.
\end{eqnarray}

The longitudinal resistivity can be written as a linear sum of the bosonic (chargon) $ \rho_{c\|} $ and fermionic (spinon) $ \rho_{s\|}  (q) $ contributions,
\begin{eqnarray}
\rho_\| (q) &=& \sigma ^{-1} _\| (q)  = \rho _{c\|} + \rho_{s\|} (q), \label{Eq.rhosppll}\\
\rho _{c\|} &=& \frac{1}{i \epsilon \omega}, \quad \rho_{s\|} (q) =  - \frac{q}{i \omega	} \frac{2\pi}{p_{\rm F,0}}\frac{1}{\Pi ^{\mathsf{e}} _\| (q)}.
\end{eqnarray}

Similarly, we can solve for the transverse component Eq.~\eqref{Eq.p1m},
\begin{eqnarray}
P_1^- (\vec{q}) &=& \frac{-iq\Big(\Omega _0  (q)  - \Omega _2  (q) \Big)}{1+2i q\Big(\Omega _0  (q)  - \Omega _2  (q) \Big) \tilde{\Gamma} _2 - 2  q\Big( \Omega _1  (q)  - \Omega _3 (q)\Big) F_1} \tilde{\mathsf{e}}_\perp(\vec{q})  = -i \Pi ^{\mathsf{e}} _\perp (q) \mathsf{e}_\perp(\vec{q}),  \\
\Pi ^{\mathsf{e}} _\perp (q) &=& \frac{1}{v_{\rm F}q} \frac{q\Big(\Omega _0  (q)  - \Omega _2  (q) \Big)}{1+2i q\Big(\Omega _0  (q)  - \Omega _2  (q) \Big) \tilde{\Gamma} _2 - 2  q\Big( \Omega _1  (q)  - \Omega _3 (q)\Big) F_1}  \\
&=& \frac{1}{v_{\rm F}q} \frac{z_\xi}{1+2i z_\xi \tilde{\Gamma} _2 -  F_1 z_\xi ^2}  \\
\end{eqnarray}
The transverse conductivity $\sigma _\perp (q)$ is obtained from Eq.~\eqref{Eq.sglin1m}, 
\begin{eqnarray}
j_\perp (\vec{q}) &=& \frac{(p_{\rm F,0})^2}{2\pi m} P^-_1 (\vec{q}) = \sigma _\perp (q) E_{\perp}(\vec{q}), \\
\sigma _\perp (q) &=& i \epsilon \left(\frac{\omega ^2 - c^2 q^2}{\omega} \right) \left( 1 - \epsilon \left(\frac{\omega ^2 - c^2 q^2}{\omega} \right)\frac{m}{p_{\rm F,0}}\frac{2\pi}{p_{\rm F,0}}\frac{1}{\Pi ^{\mathsf{e}} _\perp (q)}\right)^{-1}
\end{eqnarray}
The transverse resistivity can be written as a linear sum of the bosonic $\rho_{c\perp} (q)$ and fermionic contributions $\rho_{s\perp} (q) $,
\begin{eqnarray}
\rho_\perp (q) &=& \sigma ^{-1} _\perp (q)  = \rho_{c\perp} (q) + \rho_{s\perp} (q), \label{Eq.rhospperp} \\
\rho_{c\perp} (q) &=& \frac{1}{i \epsilon \omega} \left(\frac{\omega ^2}{\omega ^2-c^2 q^2}\right), \quad \rho_{s\perp} (q) =  - \frac{m}{i p_{\rm F,0}} \frac{2\pi}{p_{\rm F,0}}\frac{1}{\Pi ^{\mathsf{e}} _\perp (q)},
\end{eqnarray}
implying a series-stacking of these resistances, i.e. the Ioffe-Larkin rule.

\section{Alternative derivation of the effect of collions on the conductivity}\label{sec:collisons}

A useful consistency check can be done by comparing our calculation in the presence of collisions to a method proposed in Ref.~\onlinecite{Conti} by Conti and Vignale. In Ref.~\onlinecite{Conti} the authors demonstrate that the current-current response functions of a Fermi liquid  $\chi_{\|,\perp}(q,\omega)$ in the presence of collisions can be obtained from the same response functions $\chi^0_{\|,\perp}(q,\omega)$ in the absence of collisions by a simple set of rules [see Eqs.~(5.10)--(5.13) of Ref.~\onlinecite{Conti}, note the different convention with an opposite sign of $\omega$]:
\begin{itemize}
 \item For momentum relaxing collisions with rate $\Gamma_1$:
 \begin{align}
  \frac{1}{\chi_{\|}(q,\omega)}=&\frac{\omega-i\Gamma_1}{\omega} \frac{1}{\chi^0_{\|}(q,\omega-i\Gamma_1)}-
 \frac{i\Gamma_1}{\omega^2(\omega-i\Gamma_1)} \lim_{\omega'\to0}\frac{\omega'^2}{\chi^0_{\|}(q,\omega')}\\
  \frac{1}{\chi_\perp(q,\omega)}=&\frac{\omega-i\Gamma_1}{\omega}  \frac{1}{\chi^0_\perp(q,\omega-i\Gamma_1)}.
 \end{align}
\item For momentum conserving collisions with rate $\Gamma_2$:
 \begin{align}
 \frac{1}{\chi_{\|}(q,\omega)}=&\frac{\omega-i\Gamma_2}{\omega} \frac{1}{\chi^0_{\|}(q,\omega-i\Gamma_2)}+\frac{i \Gamma_2}{\omega}\frac{m}{n}-\frac{i\Gamma_2}{\omega^2(\omega-i\Gamma_2)}\Bigl(\lim_{\omega'\to 0}\frac{\omega'^2}{\chi^0_\|(q,\omega')}\Bigr)\\
 \frac{1}{\chi_{\perp}(q,\omega)}=&\frac{\omega-i\Gamma_2}{\omega} \frac{1}{\chi^0_{\perp}(q,\omega-i\Gamma_2)}+\frac{i \Gamma_2}{\omega}\frac{m}{n}.
  \end{align} 
\end{itemize}
If both types of collisions are present, we can simply concatenate the two relations. After some simple manipulations, we obtain the following expressions for the conductivity $\sigma(\omega)=-i\chi(q,\omega)/\omega$ (note that $\chi$ already contains the diamagnetic contribution in Ref.~\onlinecite{Conti}):
 \begin{align}
 \frac{1}{\sigma_{\|}(q,\omega)}=&\frac{1}{\sigma^0_{\|}(q,\omega-i\Gamma_{12})}- \Gamma_2\frac{m}{n}-
 \frac{i\Gamma_{12}}{\omega(\omega-i\Gamma_{12})} \lim_{\omega'\to0}\frac{\omega'}{\sigma^0_{\|}(q,\omega')}\label{sigma_par_conti}
  \\
 \frac{1}{\sigma_{\perp}(q,\omega)}=&\frac{1}{\sigma^0_{\perp}(q,\omega-i\Gamma_{12})}- \Gamma_2\frac{m}{n}\label{sigma_perp_conti}
  \end{align} 
with the short-hand notation $\Gamma_{12}=\Gamma_1+\Gamma_2$.
In our case, we have from Eq.~(15) of the main text 
\begin{align}
  \lim_{\omega'\to0}\frac{\omega'}{\sigma^0_{\|}(q,\omega')}=  \lim_{\omega'\to0}\omega'\rho^*(q.\omega)=\frac{-iq^2}{n^2\kappa}.\label{sigma0_conti}
\end{align}
Using Eqs.~\eqref{sigma_par_conti}--\eqref{sigma0_conti}, it is straightforward to verify that our results in Eqs.~\eqref{Eq.condfpll}--\eqref{Eq.zpm} of the main text, can be obtained from the same results in the absence of collisions by the substitutions above.

\section{Dispersion of collective modes in metals and spinon Fermi surface states}\label{Sec.suppcolldisp}

In this appendix, we show explicitly the dispersion relations of the collective modes of the SFSS for the case $F_{l \geq 1} = 0$.
These can be obtained by solving for the poles of the longitudinal and transverse conductivities of the SFSS, or equivalently, the zeroes of their respective resistivities, Eq.~\eqref{Eq.rhosppll} and Eq.~\eqref{Eq.rhospperp}.
%
%
\begin{equation}
\omega _L = \frac{(1+F_0)q^2+ 2\omega _p^2}{\sqrt{(1+2F_0)q^2 + 4\omega _p^2}},
\end{equation}
and the transverse collective mode dispersion,
\begin{align}
\omega _T &= \omega _p \left(2- \frac{v_{\rm F}^2 q^2}{2 \omega _p^2} \right)^{-1/2}\sqrt{1 + \frac{c^2 q^2}{\omega _p^2}\left( 1 - \frac{v_{\rm F} ^2 q^2}{2\omega _p^2} \right) + \sqrt{1 + 2 \frac{c^2 q^2}{\omega _p^2}\left( 1 - \frac{v_{\rm F} ^2 q^2}{2\omega _p^2} + \frac{c^2 q^2}{2\omega _p^2} \right)} }.
\end{align}
To leading order in $q$, we find
\begin{eqnarray}
\omega _L &\simeq& \omega _p + \frac{1}{2\omega _p}\left(\frac{3 + 2 F_0}{4}\right)v_{\rm F}^2 q^2  + \mathcal{O}(q^4), \\
\omega _T &\simeq& \omega _p + \frac{1}{2\omega _p}\left(\frac{1}{4} + \frac{c^2}{v_{\rm F}^2} \right)v_{\rm F}^2q^2  + \mathcal{O}(q^4),
\end{eqnarray}
i.e. Eqs.~\eqref{Eq.wLdisp}--\eqref{Eq.wTdisp} in the main text.

\section{Derivation of the quasi-static transverse conductivity}\label{Sec.suppcondperpderivation}

In this appendix, we derive the result of the universal transverse conductivity and the magnetic noise spectrum for anisotropic FSs given in Eqs.~\eqref{Eq.anitranscond}
of the main text.
%
%
The charge current of a Fermi liquid in the presence of a quasi-static $(\omega \rightarrow 0)$ electrical field is given by substituting Eq.~\eqref{Eq.denseffdef} into Eq.~\eqref{Eq.currdef} of the main text [see Eq.~(3.121) from Ref.~\onlinecite{Pines}, in units of $\hbar = 1$]
\begin{equation}
\vec{J}_{\vec{q}} = - i \frac{e^2}{\mathcal{A}} \sum _{\vec{p}} \delta (\epsilon _{\vec{p}} - \epsilon _{\rm F}) (\vec{E} \cdot \vec{v}_{\vec{p}}) \vec{v}_{\vec{p}} \left\lbrace P \left( \frac{1}{\vec{q} \cdot \vec{v}_{\vec{p}}} \right) + i \pi \delta (\vec{q} \cdot \vec{v}_{\vec{p}})\right\rbrace,
\end{equation}
where $\mathcal{A}$, $\vec{v}_{\vec{p}} = \frac{\partial \epsilon _{\vec{p}}}{\partial \vec{p}}$, $\epsilon _{\vec{p}}$ and $ \epsilon _{\rm F}$ are respectively the system area, quasiparticle velocity, energy dispersion and the Fermi energy. 
For an anisotropic FS, the Fermi momentum $p_{\rm F}(\theta) $ as well as the Fermi velocity $v_{\rm F}(\theta)$ varies with angle in momentum space.
Nonetheless, independent of the symmetries of the FS, it follows that the real part of the conductivity is nonzero only when $\vec{q}$ and $\vec{v}_p$ are orthogonal. Hence the only non-trivial component of the real conductivity tensor is the transverse-transverse component Re~$\sigma _{\perp \perp} (\vec{q}, \omega \rightarrow 0) = \sigma _{\perp, 0} (\vec{q})$, i.e. Eq.~\eqref{Eq.anicondtensor} of the main text.
Without loss of generality, let us consider the case with $\vec{q} = q \vec{\hat{x}}$ is along the positive $x$-direction so that $\vec{E} = E_{\perp} \vec{\hat{y}}$ and $\vec{J}_{\vec{q}} = J_{\perp, \vec{q}} \vec{\hat{y}}$. We find,
%
%
\begin{eqnarray}
\sigma _{\perp, 0} (\vec{q})
&=&  \pi \frac{e^2}{\mathcal{A}} \sum _{\vec{p}} \delta (\epsilon _{\vec{p}} - \epsilon _{\rm F}) (\vec{\hat{y}} \cdot \vec{v}_{\vec{p}})^2 \delta (\vec{q} \cdot \vec{v}_{\vec{p}}) \nonumber \\
&=& \pi e^2\int \frac{d^2 \vec{p}}{(2\pi)^2} 
\sum_i (v_{\vec{p}}^y)^2 \frac{\delta ^2 (\vec{p} - \vec{p}^*_i)}{q \mathcal{J}_{\vec{p}}}, \quad \mathcal{J}_{\vec{p}} = \rm{det} \left( \begin{array}{cc}
\partial _{p_x} \epsilon _{\vec{p}} & \partial _{p_y} \epsilon _{\vec{p}}\\
\partial _{p_x} v_{\vec{p}}^x & \partial _{p_y} v_{\vec{p}}^x
\end{array}\right),
\nonumber \\
&=& \frac{e^2}{4\pi q} \sum_i \int d^2 \vec{p}
(v_{\vec{p}}^y)^2 \frac{\delta ^2 (\vec{p} - \vec{p}^*_i)}{\left| \cancel{v_{\vec{p}}^x} m_{yx}^{-1} (\vec{p})- v_{\vec{p}}^y m_{xx}^{-1} (\vec{p}) \right|}, \quad v_{\vec{p}}^j = \partial _{p_j} \epsilon _{\vec{p}}, \quad m_{ij}^{-1} (\vec{p}) = \partial _{p_i} \partial _{p_j} \epsilon _{\vec{p}}, \\
&=& \frac{e^2}{4\pi q}\sum_i |m_{xx} (\vec{p}_i^*) v_{\vec{p}_i^*}^y| = \frac{e^2}{4\pi q}\sum_i \left| \frac{\partial _{p_y}\epsilon _{\vec{p}^*_i}}{\partial _{p_x}^2\epsilon _{\vec{p}^*_i}} \right|,
\end{eqnarray}
%
%
where $\left\lbrace\vec{p}^*_i\right\rbrace$ denote the set of points on the FS at which the Fermi velocity is orthogonal to $\vec{q}$ or equivalently, where the tangents are parallel to $\vec{q}$. For general $\vec{q}$, the above can be written, restoring $\hbar = h/2\pi$ and spin degeneracy $g_S = 2S +1$, as
\begin{eqnarray}
\sigma _{\perp, 0} (\vec{q})
&=& (2S+1)\frac{e^2}{2 h q}\sum_i \left| \mathcal{R}_{\rm F} |_{\vec{p}_i^* (\hat{q})} \right|,
\end{eqnarray}
where $\mathcal{R}_{\rm F} |_{\vec{p}_i^* (\hat{q})} $ denotes the radius of curvature (or equivalently the inverse curvature) of the FS at $\vec{p}_i^*$.
For a circular FS, the quasiparticle mass is constant so that $\mathcal{R}_{\rm F} = p_{\rm F,0} = m^* v_{\rm F}$, and for any given $\hat{q}$ there are two points giving rise to a factor of two so that
\begin{eqnarray}
\sigma _{\perp, 0} (\vec{q})
&=& (2S+1)\frac{e^2}{h}\frac{p_{\rm F,0}}{q}, \quad \text{isotropic FS.}
\end{eqnarray}
%

We derive also the imaginary transverse-transverse conductivity Im~$\sigma _{\perp \perp}(\vec{q}, \omega \rightarrow 0)$, a result we will use in a subsequent section. We find,
\begin{eqnarray}
{\rm Im}~\sigma _{\perp \perp} (\vec{q}, \omega \rightarrow 0) 
&=&  - \frac{e^2}{\mathcal{A}} \sum _{\vec{p}} \delta (\epsilon _{\vec{p}} - \epsilon _{\rm F}) (\vec{\hat{y}} \cdot \vec{v}_{\vec{p}})^2 P \left( \frac{1}{\vec{q} \cdot \vec{v}_{\vec{p}}} \right) \nonumber \\
&=&  - \frac{e^2}{4\pi^2 q} \int d\theta \int p dp \frac{\delta \big(p - r_{\rm F} (\theta)\big)}{|\vec{v}_{\vec{p}}|} 
\frac{(v^y_{\vec{p}})^2}{v^x_{\vec{p}}}, \quad r_{\rm F} (\theta) = p_{\rm F,0} + p_{\rm F}(\theta) \nonumber \\
&=&  - \frac{e^2}{4\pi^2 q} \int d\theta ~r_{\rm F} (\theta)
\frac{\big(v^y_{\rm F}(\theta)\big)^2}{|\vec{v}_{\rm F}(\theta)| v^x_{\rm F}(\theta)}, \quad \vec{v}_{\rm F}(\theta) = \big(v^x_{\rm F}(\theta), v^y_{\rm F}(\theta)\big)= |\vec{v}_{\rm F}(\theta)|\big(\cos \theta_{\rm F}(\theta), \sin \theta_{\rm F}(\theta)\big)  \nonumber \\
&=&  - \frac{e^2}{4\pi^2 q} \int d\theta ~\mathcal{I}(\theta), \\
\mathcal{I}(\theta) = 
&=& \frac{r_{\rm F} (\theta)}{\sqrt{r^2_{\rm F} (\theta) + \dot{r}^{2}_{\rm F} (\theta)}} \frac{\Big(\dot{r}_{\rm F} (\theta)\cos \theta - r_{\rm F} (\theta) \sin \theta \Big)^2}{\dot{r}_{\rm F} (\theta) \sin \theta + r_{\rm F} (\theta) \cos \theta}, \quad \dot{r}_{\rm F} (\theta ) = \frac{d}{d\theta} r_{\rm F} (\theta ),
\end{eqnarray}
where $r_{\rm F} (\theta)$ denotes the Fermi radius, $\vec{v}_{\rm F}(\theta)$ the Fermi velocity, and the last line was obtained from the gradient of the normal to $r_{\rm F} (\theta)$,
\begin{eqnarray}
\tan \theta _{\rm F}(\theta) = \frac{r_{\rm F} (\theta)\sin \theta - \dot{r}_{\rm F} (\theta)\cos \theta}{\dot{r}_{\rm F} (\theta)\sin \theta + r_{\rm F} (\theta) \cos \theta}.
\end{eqnarray}

For a system with spatial inversion symmetry, only the diagonal components of $\sigma _{\alpha \beta}$ are non-trivial, $\sigma _{\| \|} = \sigma _{\|}$ and $\sigma _{\perp \perp} = \sigma _{\perp}$. In particular, $r_{\rm F} (\theta) = r_{\rm F} (\theta + \pi)$, so that the integrand $\mathcal{I} (\theta + \pi) = -\mathcal{I} (\theta)$ and ${\rm Im}~\sigma _{\perp} (\vec{q}, \omega \rightarrow 0) = 0$. In this case, the quasi-static transverse-transverse conductivity is purely real~\cite{Pines},
\begin{eqnarray}\label{Eq.invsymmcondtensor}
\sigma _{\perp} (\vec{q}, \omega \rightarrow 0) = \sigma _{\perp, 0} (\vec{q}), \quad \text{time reversal or space inversion symmetry.}
\end{eqnarray}

\section{Effect of finite frequency on momentum-dependence of transverse conductivity in the different transport regimes}\label{Sec.suppcondperpweffect}

In this appendix, we analyze the effects of finite frequency and collision on the transverse conductivity leading to the discussion in Sec.~\ref{Sec.condperpcoll} of the main text.
The transverse conductivity of the SFSS can be expressed in terms of the momentum scales $q_C$ and $q_D$ as
\begin{eqnarray}
\sigma _\perp (q, q_\omega, q_p) &=& \frac{n e^2}{m}\frac{2i/v_{\rm F}}{ F_1 q_-  - q_+  - 2i (q_C - (1+F_1)q_D/2) + 2q_\omega\left(\frac{q _p^2}{q_\omega ^2 - (c/v_{\rm F})^2 q^2}\right)} 
= g_S\frac{e^2}{h}\frac{p_{\rm F,0}}{Q(q, q_\omega, q_p)},\\
q_\pm &=& q_\omega-i q_C \pm \sqrt{(q_\omega-i q_C) ^2 -q^2}, \quad q_+ q_- = q^2, \\
q_D &=& \frac{2}{1+F_1}\frac{\Gamma _1}{v_{\rm F}}, \quad q_C = \frac{1}{v_{\rm F}}(\Gamma _1 + \Gamma _2), \quad m^* = m(1+F_1),
\end{eqnarray}
where we recast frequencies $\omega$ and $\omega _p$ in terms of their associated momenta $q_\omega = \omega / v_{\rm F}$ and $q_p = \omega _p / v_{\rm F}$, and $g_S = 2S+1$ denotes the spin degeneracy factor. In the following, it is implicit that $q_p$ is always much larger than every other momentum scale. The transverse conductivity for metals is therefore $\sigma _\perp (q, q_\omega) \sim \sigma _\perp (q, q_\omega, q_p = 0)$.
We consider the case $q_C \gg q_D$ for which the hydrodynamic regime exists. 
Let us first study the effect of small frequency $q_\omega \ll q_C$ on the momentum dependence in the quantum regime $q \gg q_C$, for which case we expand $Q(q, q_\omega, q_p)$ to leading order in $q_\omega$:
%
%
\begin{eqnarray}
Q (q, q_\omega, q_p) &\simeq& Q(q) - \frac{q_\omega ^2 q^2}{2(q^2+q_C^2)^{3/2}}+ i q_\omega \xi (q, q_p) + \mathcal{O}(q_\omega^3),\label{Eq.Qqwexp}\\
Q(q) &=&  q_D + \sqrt{q^2 + q_C^2} -q_C, \\
\xi (q, q_p)&=&\left(a_1 + \frac{q_C}{\sqrt{q^2 + q_C^2}} + \frac{2q _p^2}{(1+F_1)(c/v_{\rm F})^2 q^2}\right), \quad a_1 =  \frac{1 - F_1}{F_1 + 1},\label{Eq.xiqwexp}
\end{eqnarray}
where we restrict to $F_1 >0$ so that $|a_1| \leq 1$. The real part of the transverse conductivity takes a simple form in this limit,
\begin{eqnarray}
{\rm Re}~ \sigma _\perp (q, q_\omega, q_p) &\simeq& 
g_S\frac{e^2}{h}\frac{p_{\rm F,0} Q(q)}{Q^2(q)+q_\omega^2 \left[\xi ^2 (q, q_p) - \frac{Q(q)q^2}{(q^2+q_C^2)^{3/2}}\right]} + \mathcal{O}(q_\omega^3).
\end{eqnarray}
In the quantum regime, $Q(q) \sim q \gg q_C \gg q_D$, the above expressions further simplify such that the frequency scale at which ${\rm Re}~ \sigma _\perp$ deviates from its $q^{-1}$ dependence in the quasi-static limit is given by $q_\omega \sim q/\sqrt{|\xi^2 - Qq^2/(q^2+q_C^2)^{3/2}|}$. For metals and spinon FSs respectively, we have (for simplicity we consider $F_1 \ll 1$),
\begin{eqnarray}
\Delta \omega (q) &\sim & \frac{v_{\rm F} q}{\sqrt{\left|\xi ^2(q, 0) - \frac{Q(q)q^2}{(q^2 + q_C^2)^{3/2}}\right|}} \sim v_{\rm F}\left( \frac{F_1 + 1}{3- F_1} \frac{q^3}{q_C}\right)^{1/2}, \\
\Delta \omega _{\rm s} (q) &\sim & \frac{v_{\rm F} q}{|\xi (q, q_p)|} \sim (1+F_1)\frac{v_{\rm F}c^2}{2 \omega_p^2} q^3,
\end{eqnarray}
i.e. Eqs.~\eqref{Eq.balpeakwidthmetal} and~\eqref{Eq.balpeakwidthspinon} of the main text when $F_1 =0$. At much larger frequencies $\omega \gg \Delta \omega _{(\rm s)} (q)$, the $q$-dependence of the transverse conductivities for 
the SFSS is given by
\begin{eqnarray}
{\rm Re}~ \sigma _\perp (q, q_\omega, q_p) &\simeq & 
g_S\frac{e^2}{h}\frac{p_{\rm F,0} q}{q_\omega^2 \xi ^2 (q, q_p)} \simeq g_S\frac{e^2}{h}(1+F_1)^2 \left(\frac{c}{v_{\rm F}}\right)^4\frac{p_{\rm F,0} q^5}{q_\omega^2 q _p^4},
\end{eqnarray}
while for the metallic case, the expansion in small $q_\omega$ is no longer valid and a large $q_\omega$ expansion is required instead, from which one finds,
\begin{eqnarray}
{\rm Re}~ \sigma _\perp (q, q_\omega) &\simeq & 
g_S\frac{e^2}{h} \frac{(1+F_1)^2 p_{\rm F,0}}{4q_\omega ^2} \left( q_D + \frac{q_C q^2}{2q_\omega ^2} \right) + \mathcal{O}(q_\omega ^{-5}).
\end{eqnarray}

\begin{figure}[h]
\includegraphics[scale=1.0]{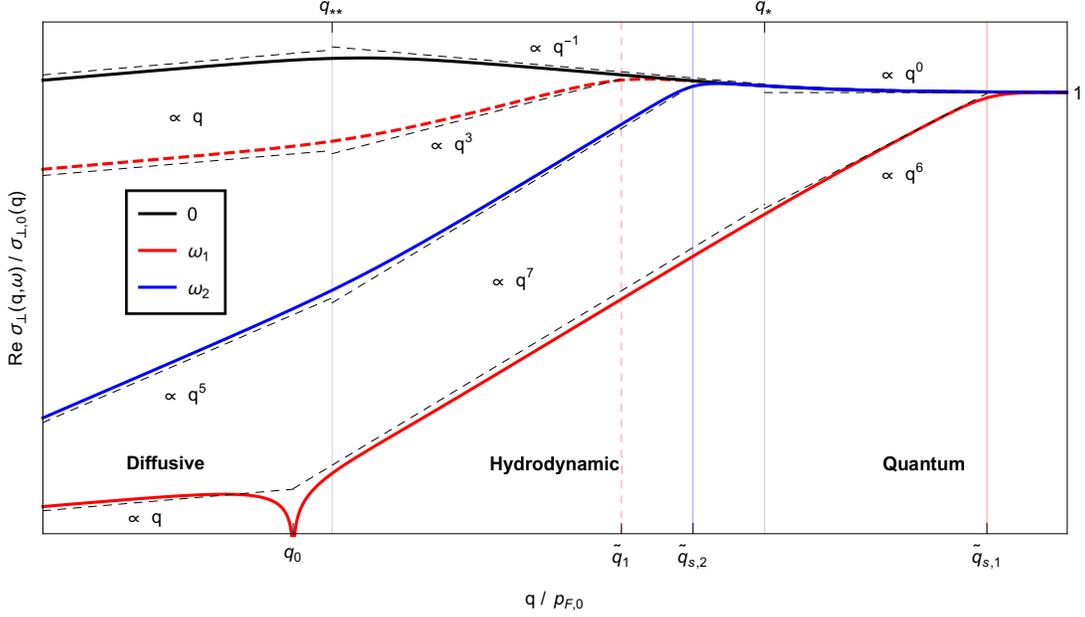}
\caption{
The effect of finite frequency on Re~$\sigma _\perp (q, \omega)$ of metals (red dashed curve) and SFSS (blue solid curve) in the different transport regimes with $F_1 = 0$. Both systems have the same conductivity in the strict quasi-static $\omega \rightarrow 0$ limit (black). 
At finite frequencies, the respective conductivities approach this quasi-static limit only at momenta much larger than a frequency-dependent momentum cut-off scale, which is much larger in SFSSs ($\tilde{q}_s$) than in metals ($\tilde{q}$).
Plots are shown for frequencies $\omega _1 = 0.01 \Delta \omega (q_*)$ in red, and $\omega _2 = 0.01 \Delta \omega _{\rm s}(q_*)$ in blue, where $\omega _2 \ll \omega _{HQs} \ll \omega _1 \ll \omega _{HQ}$.
}
\label{Fig.condspinonsupp}
\end{figure}

For completeness, we perform the same analysis for the diffusive and hydrodynamic transport regimes, which can be studied concurrently by expanding $Q(q, q_\omega, q_p)$ to leading order in $q_C^{-1}$:
\begin{eqnarray}
Q (q, q_\omega, q_p) &\simeq& Q(q) + i q_\omega \xi (q, q_\omega, q_p) + \mathcal{O}(q_C^{-2}),\label{Eq.Qclassical}\\
\xi (q, q_\omega, q_p)&=&\left(\frac{2}{1+F_1} - \frac{2q _p^2}{(1+F_1)(q_\omega ^2 - (c/v_{\rm F})^2 q^2)}\right),\\
{\rm Re}~\sigma _\perp (q, q_\omega, q_p) &\simeq& 
g_S\frac{e^2}{h}\frac{p_{\rm F,0}Q(q)}{Q^2(q)+q_\omega^2 \xi ^2 (q, q_\omega, q_p)}.
\end{eqnarray}
Similarly, the $q$-dependence of ${\rm Re}~ \sigma _\perp$ in each regime deviates from its quasi-static limit [see Table~\ref{Table.transcond} of the main text] at frequencies larger than the scale set by $Q(q) \sim q_\omega |\xi (q, q_\omega, q_p)|$, where $Q(q) \simeq q_D$ in the diffusive regime and $Q(q) \simeq q^2/2q_C$ in the hydrodynamic regime. These results are summarized in Table~\ref{Table.transcondwmetal} and~\ref{Table.transcondwspinon} for metals and spinon FSs respectively for the case of $F_1 \ll 1$, the $q$-dependence of which are illustrated in Fig.~\ref{Fig.condspinonsupp}.

\begin{table}[h]
\begin{tabular}[b]{cccc}
\hline\\[-1.em]
\hline\\[-1.em]
&\quad \quad \quad \quad \quad Diffusive \quad \quad \quad \quad & \quad \quad \quad \quad Hydrodynamic \quad \quad \quad \quad \quad & Quantum \\ [.1em]
& $q \ll q_{**}$ & $q_{**} \ll q \ll q_*$ & $q_* \ll q$\\ [.5em]
\hline\\[-1.em]
$\Delta \omega (q)$ & {\large $\frac{(1+F_1)v_{\rm F}q_D}{2}$} & {\large $\frac{1+F_1}{4q_C}$}$v_{\rm F}q^2$ & {\large $\sqrt{ \frac{F_1 + 1}{3- F_1} \frac{1}{q_C}}$}$v_{\rm F}q^{3/2}$ \\ [1.em]
\hline\\[-1.em]
Re~$\sigma  _\perp (q, \omega \gg \Delta \omega (q))$ & $g_S${\large $\frac{e^2}{h} \frac{(1+F_1)^2}{4}\frac{v_{\rm F}^2 p_{\rm F,0}q_D}{\omega ^2}$} & $g_S${\large $\frac{e^2}{h}\frac{(1+F_1)^2}{8} \frac{v_{\rm F}^2 p_{\rm F,0}}{q_C \omega ^2}$}$q^2$ & 
$g_S${\large $\frac{e^2}{h} \frac{(1+F_1)^2v_{\rm F}^2 p_{\rm F,0}}{4\omega ^2}$} $\Big(q_D +${\large $\frac{q_C v_{\rm F}^2}{2\omega^2}$}$q^2\Big)$ \\ [1.em]
\hline\\[-1.em]
\hline\\[-1.em]
\end{tabular}
\caption{The $q$-dependence of the real part of the transverse conductivity in metals, Re~$\sigma  _\perp (q, \omega \gg \Delta \omega (q))$, for the various transport regimes ($q_C \gg q_D$) at frequencies larger than the respective cutoff frequency scales $\Delta \omega (q)$ required for quasi-static approximation for the case of $F_1 \ll 1$.
}
\label{Table.transcondwmetal}
\end{table}

\begin{table}[h]
\begin{tabular}[b]{cccc}
\hline\\[-1.em]
\hline\\[-1.em]
&\quad \quad \quad \quad \quad Diffusive \quad\quad \quad \quad \quad & \quad \quad \quad \quad Hydrodynamic \quad \quad \quad \quad \quad & Quantum \\ [.1em]
& $q \ll q_{**}$ & $q_{**} \ll q \ll q_*$ & $q_* \ll q$\\ [.5em]
\hline\\[-1.em]
$\Delta \omega _{\rm s}(q \gg q_0)$ & {\large $\frac{(1+F_1)v_{\rm F}q_D}{2}\frac{c^2}{\omega _p^2}$}$q^2$ & {\large $\frac{(1+F_1)v_{\rm F}}{4q_C }\frac{c^2}{\omega _p^2}$}$q^4$ & {\large $\frac{(1+F_1)v_{\rm F}}{2}\frac{c^2}{\omega_p^2}$}$q^3$ \\ [1.em]
\hline\\[-1.em]
Re~$\sigma  _\perp (q \gg q_0, \omega \gg \Delta \omega _{\rm s}(q))$ & $g_S${\large $\frac{e^2}{h} \frac{(1+F_1)^2}{4}\frac{v_{\rm F}^2 p_{\rm F,0}q_D}{\omega ^2}\frac{c^4}{\omega _p^4}$}$q^4$ & $g_S${\large $\frac{e^2}{h}\frac{(1+F_1)^2}{8} \frac{v_{\rm F}^2 p_{\rm F,0}}{q_C \omega ^2}\frac{c^4}{\omega _p^4}$}$q^6$ & $g_S${\large $\frac{e^2}{h} \frac{(1+F_1)^2}{4}\frac{v_{\rm F}^2 p_{\rm F,0}}{\omega ^2}\frac{c^4}{\omega _p^4}$}$q^5$ \\ [1.em]
\hline\\[-1.em]
Re~$\sigma  _\perp (q \ll q_0, \omega)$ & $g_S${\large $\frac{e^2}{h} \frac{(1+F_1)^2}{4}\frac{v_{\rm F}^2 p_{\rm F,0}q_D \omega ^2}{\omega _p^4}$} & $g_S${\large $\frac{e^2}{h}\frac{(1+F_1)^2}{8} \frac{v_{\rm F}^2 p_{\rm F,0}}{q_C} \frac{\omega ^2}{\omega _p^4}$}$q^2$ & 
$g_S${\large $\frac{e^2}{h}\frac{1+F_1}{\omega _p^4} \left(\frac{p_{\rm F,0}^2 v_{\rm F}^{2} \omega ^{10}}{4}\right)^{\frac{1}{3}}$}
 \\ [1.em]
\hline\\[-1.em]
\hline\\[-1.em]
\end{tabular}
\caption{The $q$-dependence of the real part of the transverse conductivity in spinon FSs, Re~$\sigma _\perp (q, \omega \gg \Delta \omega _{\rm s}(q))$, for the various transport regimes ($q_C \gg q_D$) at frequencies larger than the respective cutoff frequency scales $\Delta \omega _{\rm s}(q)$ required for quasi-static approximation for the case of $F_1 \ll 1$. At small frequencies $\omega \ll v_{\rm F} q_C$ of interest, its behavior is always different from the quasi-static limit when $q \ll q_0$.
}
\label{Table.transcondwspinon}
\end{table}

Conversely, a given frequency determines a momentum scale, $\tilde{q}(\omega)$ and $\tilde{q}_{s}(\omega)$ for the metal and spinon FS respectively, such that the corresponding transverse conductivity can be approximated by its quasi-static limit when $q \gg \tilde{q}_{(s)}(\omega)$. This momentum scale is obtained by inverting the appropriate cutoff frequency scale $\Delta \omega _{(s)}(q)$ that is consistent with the transport regime the momentum scale lies in, and is therefore determined by two threshold frequencies $\omega_{DH(s)}$ and $\omega_{HQ(s)}$,
\begin{eqnarray}
\omega_{DH} &=& \frac{1+F_1}{2} v_{\rm F} q_D, \hspace{3.5em}\omega_{HQ} = \sqrt{ \frac{F_1 + 1}{3- F_1}} v_{\rm F} q_C, \\
\omega_{DHs} &=& \frac{1+F_1}{2} \frac{c^2}{\omega _p^2} v_{\rm F} q_C q_D^2, \quad \omega_{HQs} = \frac{1+F_1}{2} \frac{c^2}{\omega _p^2}v_{\rm F} q_C^3.
\end{eqnarray}
The momentum scales for metals and SFSSs are summarized in Table~\ref{Table.momcutoff}.

\begin{table}[h]
\begin{tabular}[b]{cccc}
\hline\\[-1.em]
\hline\\[-1.em]
&\quad \quad \quad \quad \quad Diffusive \quad\quad \quad \quad \quad & \quad \quad \quad \quad Hydrodynamic \quad \quad \quad \quad \quad & Quantum \\ [.1em]
& $\omega < \omega _{DH(s)}$ & $\omega _{DH(s)} < \omega < \omega _{HQ(s)}$ & $\omega _{HQ(s)} < \omega$\\ [.5em]
\hline\\[-1.em]
$\tilde{q} (\omega)$ & $\simeq 0$ & {\large $\left(\frac{4q_C}{1+F_1}\frac{\omega}{v_{\rm F}}\right)^{1/2}$} & {\large $\left(\frac{(3-F_1)q_C}{F_1+1}\frac{\omega ^2}{v_{\rm F}^2}\right)^{1/3}$} \\ [1.em]
\hline\\[-1.em]
$\tilde{q} _{\rm s}(\omega)$ & {\large $\left(\frac{2}{(1+F_1)q_D}\frac{\omega _p^2}{c^2}\frac{\omega}{v_{\rm F}}\right)^{1/2}$} & {\large $\left(\frac{4q_C }{1+F_1}\frac{\omega _p^2}{c^2}\frac{\omega}{v_{\rm F}}\right)^{1/4}$} & {\large $\left(\frac{2}{(1+F_1)}\frac{\omega_p^2}{c^2}\frac{\omega}{v_{\rm F}}\right)^{1/3}$} \\ [1.em]
\hline\\[-1.em]
\hline\\[-1.em]
\end{tabular}
\caption{Expressions for the momentum scales $\tilde{q} (\omega)$ and $\tilde{q}_{\rm s} (\omega)$, in metals and SFSSs respectively, such that Re~$\sigma _\perp (q \gg \tilde{q}_{\rm s}(\omega), \omega) \simeq \sigma _\perp (q, 0^+) = g_S (e^2/h) p_{\rm F,0}/Q(q)$ and Re~$\sigma ^{f}_\perp (q \gg \tilde{q}(\omega), \omega) \simeq \sigma ^{f}_\perp (q, 0^+) = g_S (e^2/h) p_{\rm F,0}/Q(q)$. 
}
\label{Table.momcutoff}
\end{table}

Particular care has to be taken for SFSSs. The presence of the emergent photon with dispersion $\omega = c q$ gives rise to a divergent $Q(q,q_\omega, q_p)$ on resonance, and consequently, a vanishing transverse conductivity. 
Unlike in the quantum regime, the momentum at which resonance occurs for a given frequency, $q_0 = \omega / c$, becomes relevant in the hydrodynamic and diffusive regimes even when $\omega \ll v_{\rm F} q_C$. In particular, when $q \ll q_0$, frequencies larger than the Fermi energy are required to satisfy $Q(q) > q_\omega |\xi (q, q_\omega, q_p)|$,
\begin{eqnarray}
\omega > \frac{2\omega _p^2}{(1+F_1) v_{\rm F} Q(q)} > E_{\rm F}.
\end{eqnarray}
Consequently, at low frequencies of interest, the transverse conductivity of the SFSS never approaches its quasi-static behavior when $q \ll q_0$.

For completeness, let us consider the case $q_D \gtrsim q_C$, in which the hydrodynamic regime is absent. In this case, as pointed out in the main text, $q_* = q_D$ sets the momentum scale that separates the diffusive from quantum transport regimes. For $F_1 \geq 0$, $q_D$ takes a maximum value $2q_C$ when $F_1 = 0$ and $\Gamma _2 = 0$. By proceeding with an analogous analysis, one finds the same results as in the case of $q_C \gg q_D$ from Tables~\ref{Table.transcondwmetal},~\ref{Table.transcondwspinon} and~\ref{Table.momcutoff}, but with the following threshold frequencies that determine the appropriate form of the momentum scale $\tilde{q}_{(s)}(\omega)$,
\begin{eqnarray}
\omega_{DQ} &=& 
v_{\rm F} q_D, \\
\omega_{DQs} &=& \frac{1+F_1}{2} \frac{c^2}{\omega _p^2}v_{\rm F} q_D^3,
\end{eqnarray}
the metallic frequency and momentum scales required in the quantum regime,
\begin{eqnarray}
\Delta \omega (q \gg q_D) &\sim& v_{\rm F} \left(\frac{2}{5-F_1}\frac{q^3}{q_D} \right)^{1/2},\\
\tilde{q}(\omega > \omega _{DQ}) &\sim& \left(\frac{5-F_1}{2}q_D \frac{\omega ^2}{v_{\rm F}^2}\right)^{1/3},
\end{eqnarray}
and the asymptotic $q$-dependence of the real part of the transverse conductivity in metals,
\begin{eqnarray}
{\rm Re}~\sigma  _\perp (q, \omega \gg \Delta \omega (q)) 
&\simeq&g_S\frac{e^2}{h} \frac{(1+F_1)^2 p_{\rm F,0}}{4q_\omega ^2} q_D.
\end{eqnarray}

\begin{figure}[h]
\includegraphics[scale=1.0]{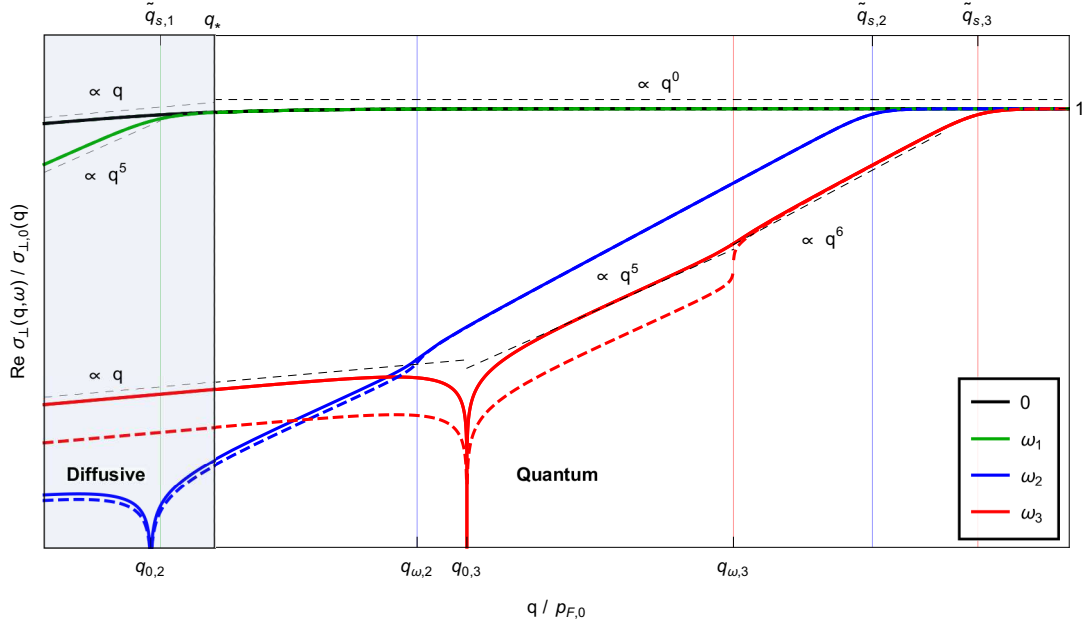}
\caption{
Plots of ${\rm Re}~\sigma  _\perp (q, \omega)$ of the SFSS for various frequencies, showing that the effect of including (solid plots) or omitting (dashed plots) the frequency-dependent momentum relaxation starts becoming noticeable only at frequencies for which $\Gamma _\omega \simeq \Gamma _1$ (blue) and significant at larger frequencies for which $\Gamma _\omega \gg \Gamma _1$ (red). Frequencies plotted here are $\omega _1 = 0.01 \omega _{DQs}$, $\Gamma_{\omega_2} = \Gamma _1$ and $\Gamma_{\omega_3} = 35\Gamma _1$, with $\omega _1 \ll \omega _{DQs} \ll \omega _2 \ll \omega _3$.
}
\label{Fig.transcondspsupp}
\end{figure}

Finally, we analyze the case of spinons with frequency-dependent momentum relaxation rate induced by gauge field fluctuations, $\Gamma _\omega \sim  E_{\rm F}^{-1/3} \xi ^{4/3}$ with $\xi = {\rm max}(\omega, T)$, which directly alters the transverse conductivity when $q \ll \Gamma _\omega /v_{\rm F}$~\cite{Lee1992}. A consistent treatment within our framework is to add this term to the impurity scattering rate $\Gamma _1 \rightarrow \Gamma _1 (\omega) = \Gamma _1 + E_{\rm F}^{-1/3} \omega ^{4/3}$, where we consider the more interesting low-temperature limit.
We further consider for simplicity the case with $\Gamma _2 = 0$, so that the transverse conductivity is expressed in terms of $q_1 (\omega) = \Gamma _1 (\omega) /v_{\rm F}$, with $q_D = 2q_1 (\omega)/(1+F_1)$ and $q_C = q_1 (\omega)$. 
The analysis proceeds largely as per the case with $q_D \gtrsim q_C$ before and is identical at frequencies $\Gamma _1 \gg \Gamma _\omega$ as well as momenta $q \gg q_\omega, q_1$. We therefore focus on the frequency range, $\Gamma _\omega \gg \Gamma _1$, or equivalently, the ultra-clean limit $\Gamma _1 \rightarrow 0$. Expanding in large $q_\omega$, one finds
\begin{eqnarray}
Q (q, q_\omega, q_p) &\simeq& 
\frac{2q_\omega}{1+F_1} \left( \alpha _\omega - i  \frac{q _p^2}{q_\omega ^2 - (c/v_{\rm F})^2 q^2}\right), \quad \alpha _\omega = \left(\frac{2q_\omega}{p_{\rm F,0}}\right)^{1/3} < 1,
\end{eqnarray}
and thus the transverse conductivity,
\begin{eqnarray}
{\rm Re}~\sigma  _\perp (\omega \gg v_{\rm F} q) \simeq g_S \frac{e^2}{h} \left(\frac{p_{\rm F,0}}{2}\right)^{2/3} \frac{1+F_1}{q_p^4} \times \left\lbrace
\begin{array}{cc}
q_\omega ^{10/3}, & q \ll q_0 \\
\frac{c^4}{v_{\rm F}^4} q_\omega ^{-2/3}q^4, & q_0 \ll q \ll q_\omega 
\end{array} \right. ,
\end{eqnarray}
where the second regime is significant only when $c \gg v_{\rm F}$. These scaling regimes are shown in Fig.~\ref{Fig.transcondspsupp}, showing that the effect of $\Gamma _\omega$ only becomes significant at large frequencies for which $\Gamma _\omega \gg \Gamma _1$, and only for the range of $q$ for which ${\rm Re}~\sigma  _\perp (\omega \gg v_{\rm F} q)$. Therefore, even in ultra-clean samples in which the diffusive regime vanishes, effect of $\Gamma _\omega$ is negligible when probing the universal quasi-static transverse conductivity at frequencies and momenta satisfying $\omega \ll \Delta \omega _s (q) \ll v_{\rm F} q$.

We close this appendix by showing that the various classical (i.e. non-quantum) regimes admits a hydrodynamical interpretation by rederiving the above asymptotic transverse conductivities starting from the Navier-Stokes equation with an additional per unit area external force, $\vec{f}$, and friction $\vec{f}_{\rm fr} = -n m \gamma \vec{v}$, 
\begin{equation}
\eta \vec{\partial}_{\vec{r}}^2 \vec{v} = n m (\partial _t + \vec{v} \cdot \vec{\partial}_{\vec{r}} )\vec{v} + \vec{\partial}_{\vec{r}} p - \vec{f} - \vec{f}_{\rm fr}
\end{equation}
For metals, $\vec{f} = ne\vec{E}$, so that linearizing the above equation in $\vec{v}$, one finds that the transverse current $j_\perp  = nev_\perp$ has an associated transverse conductivity
\begin{eqnarray}
\sigma _{\perp} &=& \frac{ne^2}{m \left(i \omega  + \gamma + \frac{\eta}{nm} q^2\right)} = g_S \frac{e^2}{h}\frac{p_{\rm F,0}}{Q'(q, q_\omega)},\\
Q'(q, q_\omega) 
&=& q'_D + \frac{q^2}{2q'_C} + i q_\omega \xi_0,\\
q'_D &=& \frac{2}{1+F_1}\frac{\gamma}{v_{\rm F}}, \quad q'_C = \frac{n p_{\rm F,0}}{4\eta}, \quad \xi_0 = \frac{2}{1+F_1},
\end{eqnarray}
i.e. Eq.~\eqref{Eq.Qclassical} with redefined parameters $q'_C$ and $q'_D$. For the SFSS, we consider the force per unit area $\vec{f} = ne\mathsf{e}$, and consider the relation between the physical current, spinon and chargon currents, respectively Eqs.~\eqref{Eq.escj},~\eqref{Eq.js} and \eqref{Eq.jc} in the main text, the SFSS conductivity can be obtained straightforwardly by applying the Ioffe-Larkin rule Eq.~\eqref{Eq.IoffeLarkin} using the expression for the chargon conductivity, Eq.~\eqref{Eq.condchperp} in the main text, which follows from Maxwell's equations. Doing so recovers Eq.~\eqref{Eq.Qclassical} in its entirety with redefined parameters $q'_C$ and $q'_D$,
\begin{eqnarray}
Q'(q, q_\omega, q_p) &=& q'_D + \frac{q^2}{2q'_C} + i q_\omega \xi (q, q_\omega, q_p).
\end{eqnarray}
The $q$- and $\omega$-dependences in the various cases immediately follow from the previous analysis.

\section{Derivation of low-frequency magnetic noise from conductivity}\label{Sec.suppnoisederivation}

In this appendix, we derive the two-time time magnetic field correlations associated to the current correlations of the system in the collisionless regime $\Gamma _{1,2} = 0$ leading to the discussion in Sec.~\ref{Sec.noiseclean} of the main text, and in particular, the expression for the low-frequency noise Eq.~\eqref{Eq.main2}. We begin by decomposing the density and current fluctuations into longitudinal ($\|$) and transverse ($\perp$) Fourier modes. Treating each mode as an independent source, we solve Maxwell's equations to obtain the associated electromagnetic field distributions $\vec{E}^{\|,\perp} (\vec{x},z,t) = \vec{E}^{\|,\perp}_{\vec{q}}(\omega ,z) e^{i(\omega t - \vec{q}\cdot \vec{x})}$ and $\vec{B}^{\|,\perp} (\vec{x},z,t) = \vec{B}^{\|,\perp}_{\vec{q}}(\omega ,z) e^{i(\omega t - \vec{q}\cdot \vec{x})}$, where $\vec{x}$ is a 2D-vector denoting the coordinate parallel to the plane of the system while $z$ denotes the out-of-plane coordinate. For a longitudinal mode,
\begin{eqnarray}
\rho (\vec{x},z,t) &=& \rho _{\vec{q}} \delta (z) e^{i (\omega t - \vec{q}\cdot \vec{x})} ,\label{Eq.rhoL}\\
\vec{j} _\| (\vec{x},z,t) &=& j_{\| ,\vec{q}} \delta (z) e^{i (\omega t - \vec{q}\cdot \vec{x})}\hat{q} \quad j_{\| ,\vec{q}} = v \rho _{\vec{q}}, \label{Eq.jL}
\end{eqnarray}
where $v = \omega /q$ is the velocity of the traveling wave moving along the modulation direction $\hat{q}$. 
In contrast, a transverse mode has no density fluctuations,
\begin{eqnarray}
\rho (\vec{x},z,t) &=& 0, \label{Eq.rhoT}\\
\vec{j} _\perp (\vec{x},z,t) &=& j_{\perp ,\vec{q}} \delta (z) e^{i (\omega t - \vec{q}\cdot \vec{x})}\hat{q}_{\perp}, \quad \hat{q} \times \hat{q}_{\perp} = \hat{z},\label{Eq.jT}
\end{eqnarray}
where we have assumed a charge neutral background. The corresponding magnetic fields generated by these source modes are
\begin{align}
\vec{B}^{\|}_{\vec{q}}(\omega,z)
&= \frac{\mu _0}{2} {\rm sgn} (z) e^{-\frac{q}{\gamma }|z|} \vec{j}_{\| ,\vec{q}} \times \hat{z}, \\
\vec{B}^{\perp}_{\vec{q}}(\omega,z)
&= \frac{\mu _0}{2} e^{-\frac{q}{\gamma }|z|}  \vec{j}_{\perp ,\vec{q}} \times  \left( {\rm sgn} (z) \hat{z} + i \gamma  \hat{q}\right), 
\end{align}
where $\gamma = 1/\sqrt{1- \omega^2/c^2 q^2}$ is the Lorentz factor and $c$ the speed of light.

In the non-relativistic limit, $\gamma \rightarrow 1$ and the frequency dependence of $\vec{B}^{\|,\perp}_{\vec{q}}$ drop out. 
Consequently, these expressions can be directly quantized, so that the time-evolution of the corresponding magnetic field operators to is completely encoded by that of the current operators~\cite{Shear},
\begin{align}
\hat{\vec{B}}^{\|}_{\vec{q}}(z, t)
&= \frac{\mu _0}{2} {\rm sgn} (z) e^{-q|z|} \hat{\vec{j}}_{\| ,\vec{q}}(t) \times \hat{z}, \label{Eq.Bpllop} \\
\hat{\vec{B}}^{\perp}_{\vec{q}}(z, t)
&= \frac{\mu _0}{2} e^{-q|z|}  \hat{\vec{j}}_{\perp ,\vec{q}}(t) \times  \left( {\rm sgn} (z) \hat{z} + i \hat{q}\right).\label{Eq.Bperpop}
\end{align}
In practice, it is more useful to express the components of the magnetic field operator in a reference frame in which $\hat{q}$ is defined by its angle $\theta _q$ from the $x$-axis.
\begin{eqnarray}
\hat{\vec{B}}_{\vec{q}}(z,t) &=& \hat{\vec{B}}^{\|}_{\vec{q}}(z,t) + \hat{\vec{B}}^{\perp}_{\vec{q}}(z,t) = \hat{B}_{\vec{q},x}(z,t)\hat{x} + \hat{B}_{\vec{q},y}(z,t)\hat{y} + \hat{B}_{\vec{q},z}(z,t)\hat{z}, \\
\hat{B}_{\vec{q},x}(z,t) &=& \frac{\mu _0}{2} e^{-q|z|}  {\rm sgn} (z) \left(\hat{j}_{\perp ,\vec{q}}(t) \cos \theta _q + \hat{j}_{\parallel ,\vec{q}}(t) \sin \theta _q \right),\\
\hat{B}_{\vec{q},y}(z,t) &=& \frac{\mu _0}{2} e^{-q|z|}  {\rm sgn} (z) \left(\hat{j}_{\perp ,\vec{q}}(t) \sin \theta _q - \hat{j}_{\parallel ,\vec{q}}(t) \cos \theta _q \right),\\
\hat{B}_{\vec{q},z}(z,t) &=& -i\frac{\mu _0}{2} e^{-q|z|} \hat{j}_{\perp ,\vec{q}}(t),
\end{eqnarray}

Without loss of generality, let us consider the magnetic response at a point $x = y = 0$ and some finite distance $z > 0$ above the system,
\begin{eqnarray}
\chi _{B_\mu B_\nu}  (z, t) &=& -i \Theta (t) \left\langle \left[ \hat{B}_\mu (z, t), \hat{B}_\nu(z, 0)\right] \right\rangle , \label{Eq.chiBzBz}\\
\hat{B}_{\mu} (z, t) &=& \sum _{\vec{q}} \hat{B}_{\vec{q},\mu} (z, t),\label{Eq.BOp}
\end{eqnarray}
where $\mu, \nu = x, y, z$ and $\Theta (t)$ denotes the Heaviside function. Using the operator relations Eqs.~\eqref{Eq.Bpllop} and~\eqref{Eq.Bperpop}, the magnetic noise can be written explicitly in terms of current correlators, or equivalently conductivities,
\begin{align}
\chi _{j_\alpha j_\beta}  (\vec{q}, t)&= -i \Theta (t) \left\langle \left[ \hat{j}_{\alpha, \vec{q}} (t), \hat{j}_{\beta, -\vec{q}}\right] \right\rangle, \\
\chi _{j_\alpha j_\beta}  (\vec{q}, \omega) &= - i \omega \sigma _{\alpha \beta} (\vec{q},\omega), 
\end{align}
where $\alpha, \beta = \|, \perp$. 
It follows from Eq.~\eqref{Eq.anicondtensor} of the main text that in the quasi-static limit, that the diagonal components of the magnetic noise tensor are
\begin{eqnarray}
\chi ''_{B_x B_x}  (z, \omega \rightarrow 0) 
&=& \frac{\mu _0 ^2 \omega}{4}\int \frac{d^2 \vec{q}}{4\pi^2} e^{-2qz} \sigma '_{\perp ,0}  (\vec{q})\cos ^2 (\theta _q), \label{Eq.chiBxBx} \\
\chi ''_{B_y B_y}  (z, \omega \rightarrow 0) 
&=& \frac{\mu _0 ^2 \omega}{4}\int \frac{d^2 \vec{q}}{4\pi^2} e^{-2qz} \sigma '_{\perp ,0}  (\vec{q})\sin ^2 (\theta _q), \label{Eq.chiByBy}\\
\chi ''_{B_z B_z}  (z, \omega \rightarrow 0) 
&=& \frac{\mu _0 ^2 \omega}{4}\int \frac{d^2 \vec{q}}{4\pi^2} e^{-2qz} \sigma '_{\perp ,0}  (\vec{q}),\label{Eq.chiBzBz}
\end{eqnarray}
while the off-diagonal components are
\begin{eqnarray}
\chi ''_{B_x B_y}  (z, \omega \rightarrow 0)
&=& \chi ''_{B_y B_x}  (z, \omega)
= \frac{\mu _0 ^2 \omega}{4}\int \frac{d^2 \vec{q}}{4\pi^2} e^{-2qz} \sigma '_{\perp ,0}  (\vec{q}) \cos (\theta _q) \sin (\theta _q),\label{Eq.chiBxBy}\\
\chi ''_{B_x B_z}  (z, \omega \rightarrow 0) 
&=& -\chi ''_{B_z B_x}  (z, \omega) 
= {\rm sgn} (z)\frac{\mu _0 ^2 \omega}{4}\int \frac{d^2 \vec{q}}{4\pi^2} e^{-2qz} {\rm Im}~\sigma _{\perp \perp} (\vec{q}, \omega \rightarrow 0)  \cos (\theta _q),\\
\chi ''_{B_y B_z}  (z, \omega \rightarrow 0) 
&=& -\chi ''_{B_z B_y}  (z, \omega)
= {\rm sgn} (z)\frac{\mu _0 ^2 \omega}{4}\int \frac{d^2 \vec{q}}{4\pi^2} e^{-2qz} {\rm Im}~\sigma _{\perp \perp} (\vec{q}, \omega \rightarrow 0)  \sin (\theta _q),
\end{eqnarray}
where $\mathcal{F}'$ and $\mathcal{F}''$ denotes respectively the real and imaginary parts of a complex function $\mathcal{F} = \mathcal{F}' + i \mathcal{F}''$.

In the presence of a symmetry that enforces the quasiparticle dispersion to satisfy $\epsilon_{\bf p}= \epsilon_{-{\bf p}}$ such as time reversal or space inversion,
The diagonal components of the magnetic noise tensor are
\begin{eqnarray}
\chi ''_{B_x B_x}  (z, \omega) 
&=& \frac{\mu _0 ^2 \omega}{4}\int \frac{d^2 \vec{q}}{4\pi^2} e^{-2qz} \left( {\rm Re}~\sigma _{\perp}  (\vec{q}, \omega)\cos ^2 (\theta _q) + {\rm Re}~\sigma _{\|}  (\vec{q}, \omega)\sin ^2 (\theta _q)  \right), \label{Eq.chiBxBxinv} \\
\chi ''_{B_y B_y}  (z, \omega) 
&=& \frac{\mu _0 ^2 \omega}{4}\int \frac{d^2 \vec{q}}{4\pi^2} e^{-2qz} \left( {\rm Re}~\sigma _{\perp}  (\vec{q}, \omega)\sin ^2 (\theta _q) + {\rm Re}~\sigma _{\|}  (\vec{q}, \omega)\cos ^2 (\theta _q)  \right), \label{Eq.chiByByinv}\\
\chi ''_{B_z B_z}  (z, \omega) 
&=& \frac{\mu _0 ^2 \omega}{4}\int \frac{d^2 \vec{q}}{4\pi^2} e^{-2qz} {\rm Re}~\sigma _{\perp}  (\vec{q}, \omega).\label{Eq.chiBzBzinv}
\end{eqnarray}
while the off-diagonal components of the magnetic noise tensor are
\begin{eqnarray}
\chi ''_{B_x B_y}  (z, \omega) 
&=& \chi ''_{B_y B_x}  (z, \omega)
= \frac{\mu _0 ^2 \omega}{4}\int \frac{d^2 \vec{q}}{4\pi^2} e^{-2qz} \left( {\rm Re}~\sigma _{\perp}  (\vec{q}, \omega) - {\rm Re}~\sigma _{\|}  (\vec{q}, \omega) \right) \cos (\theta _q) \sin (\theta _q),\label{Eq.chiBxBy}\\
\chi ''_{B_x B_z}  (z, \omega) 
&=& -\chi ''_{B_z B_x}  (z, \omega) 
= {\rm sgn} (z)\frac{\mu _0 ^2 \omega}{4}\int \frac{d^2 \vec{q}}{4\pi^2} e^{-2qz} {\rm Im}~\sigma _{\perp}  (\vec{q}, \omega) \cos (\theta _q),\\
\chi ''_{B_y B_z}  (z, \omega) 
&=& -\chi ''_{B_z B_y}  (z, \omega)
= {\rm sgn} (z)\frac{\mu _0 ^2 \omega}{4}\int \frac{d^2 \vec{q}}{4\pi^2} e^{-2qz} {\rm Im}~\sigma _{\perp}  (\vec{q}, \omega) \sin (\theta _q).
\end{eqnarray}
In the quasi-static limit, the off-diagonal components $\chi ''_{B_x B_z}$ and $\chi ''_{B_y B_z}$ vanish since Im~$\sigma _{\perp} (\vec{q}, \omega \rightarrow 0) = 0$ [Eq.~\eqref{Eq.invsymmcondtensor}], such that the only non-trivial components can be summarized by
Eq.~\eqref{Eq.chiBzBzw0} and Eq.~\eqref{Eq.chiBiBjw0} in the main text, 
\begin{align}
\chi ''_{B_z B_z}  (z, \omega \rightarrow 0)
&\simeq \frac{\mu _0 ^2 \omega}{4}\int \frac{d^2\vec{q}}{4\pi^2} e^{-2qz} \sigma _{\perp ,0}  (\vec{q}) + \mathcal{O}(\omega^3), \\
\chi ''_{B_i B_j}  (z, \omega \rightarrow 0)
&\simeq \frac{\mu _0 ^2 \omega}{4}\int \frac{d^2\vec{q}}{4\pi^2} e^{-2qz} \sigma  _{\perp ,0}  (\vec{q}) \hat{q}\cdot \hat{e}_i \hat{q}\cdot \hat{e}_j + \mathcal{O}(\omega^3),
\end{align}
with corrections $\mathcal{O}(\omega ^3)$. The leading term in the noise spectrum therefore captures geometric properties of the FS, a result that is common to both metals and spinon FS stats.

In the quantum transport regime, the out-of-plane component $\chi ''_{B_{z} B_{z}} (z, \omega \rightarrow 0)$ in particular can be evaluated even for an anisotropic FS as we show in the following. Substituting Eq.~\eqref{Eq.anitranscond} in the main text to Eq.~\eqref{Eq.chiBzBz}, we have
\begin{eqnarray}
\chi ''_{B_z B_z}  (z, \omega \rightarrow 0) 
&\simeq & (2S+1)\frac{e^2 \mu _0 ^2}{32 \pi h} \frac{\omega}{z}
\int_0 ^{2\pi} \frac{d \theta _q}{2\pi} \sum_i \mathcal{R}_{\rm F} |_{\vec{p}_i^* (\theta_q)} +\mathcal{O} (\omega ^3).
\end{eqnarray}
Notice that the set of points $\left\lbrace \vec{p}_i^* (\theta_q) \right\rbrace = \left\lbrace \vec{p}_i^* (\theta_q + \pi) \right\rbrace$; the integration with respect to $\theta_q$ corresponds to integrating $|\mathcal{R}_{\rm F}|$ 
around the FS twice. For a convex FS, one can perform this integration by first choosing an arbitrary interior point as the origin and then constructing a support function $r(\varphi)$ characterized by the distance $r$ and angle $\varphi$ of the shortest line drawn from the origin to the tangent curve for every point along the curve. The points $\left\lbrace \vec{p}_i^* (\theta_q) \right\rbrace$ are characterized by the angles $\varphi = \theta _q \pm \pi /2$, such that $d \varphi = d\theta _q$ and 

\begin{eqnarray}
\int_0 ^{\pi} d \theta _q \sum_i \mathcal{R} _{\rm F} |_{\vec{p}_i^* (\theta_q)} &=& \int_0 ^{2\pi} d \varphi \left( r(\varphi)  + \partial ^2 _\varphi r(\varphi) \right) = \mathcal{P}_{\rm FS}, \quad \text{convex FS}.
\end{eqnarray} 
which is a known mathematical result~\cite{alex2005,resnikoff2015}, where $\mathcal{P}_{\rm FS}$ denotes the perimeter of the (convex) FS.

\begin{figure}[h]
\includegraphics[scale=1.0]{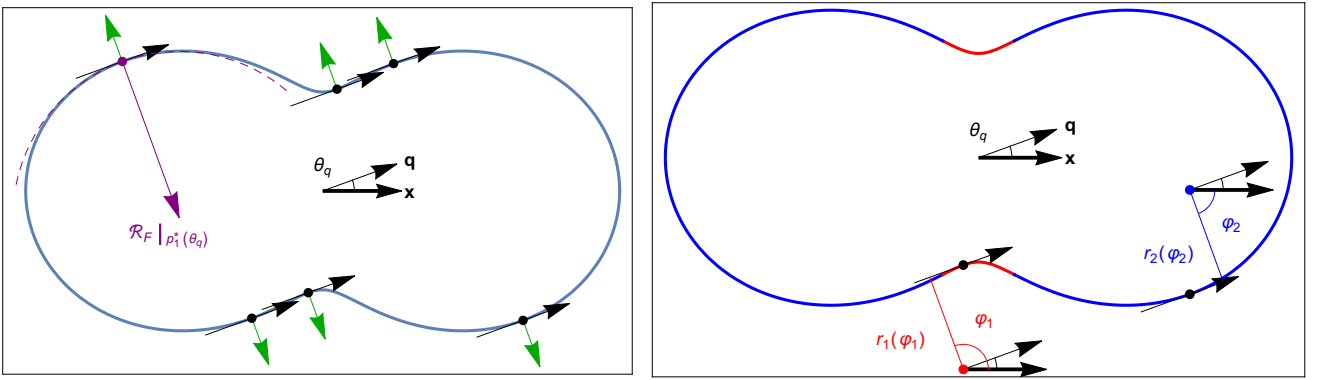}
\caption{\label{Fig.FSP} Left: Schematic showing the set of points $\lbrace \vec{p} ^*_{i}(\theta_q) \rbrace$ on the same anisotropic FS as that shown in Fig.~\ref{Fig.cartoon}(c) of the main text, with Fermi velocities (green arrows) orthogonal to $\vec{q}$ along a different direction. Right: Construction of the support angles $\varphi _m$ and functions $r_m (\varphi _m)$ for two of the six points in the left panel. The origin (red dot) for the concave region $m = 1$ (red) is a point outside of the FS while the origin (blue dot) for convex region $m = 2$ (blue) is a point inside the FS. All angles are defined relative to the $x$-axis (bold black line).}
\end{figure}

In fact, this result holds for a generic FS with concave regions. To see this, we split the $|\mathcal{R}_{\rm F}|$ integration into concave and convex regions of the FS. For each convex region, we choose an interior point as the origin as before. For each concave region, we choose an \textit{exterior} point as the origin relative to which the region is convex. Support functions for each region $r_m$ can be defined with respect to their support angles $\varphi_m$ about their respective origins. The crucial point in this construction is that the angles characterizing the points within each region $m$ is once again related to $\theta _q$ via $\varphi _m = \theta _q \pm \pi /2$ such that $d \theta _q = d \varphi _m$. The integration over each region gives its arc length $\ell_m$, the sum of which corresponds to the perimeter of the FS,
\begin{eqnarray}
\int_0 ^{\pi} d \theta _q \sum_i \left|\mathcal{R} _{\rm F} |_{\vec{p}_i^* (\theta_q)}\right| &=& \sum _m \int_{\phi _{m,0}} ^{\phi _{m,f}} d \varphi_m \left( r(\varphi_m)  + \partial ^2 _{\varphi_m} r_m(\varphi_m) \right) = \sum _m \ell _m = \mathcal{P}_{\rm FS}.
\end{eqnarray}
The construction procedure is illustrated in Fig.~\ref{Fig.FSP}.

Consequently,
\begin{eqnarray}
\chi ''_{B_z B_z}  (z, \omega \rightarrow 0) 
\simeq \frac{e^2 \mu _0 ^2}{16 \pi h} \frac{\omega}{z} \frac{(2S+1)}{2\pi}\mathcal{P}_{\rm FS}+ \mathcal{O}(\omega ^3),
\end{eqnarray}

For a circular FS, $ \mathcal{P}_{\rm FS} = 2\pi p_{\rm F,0}$ so that 
\begin{equation}\label{Eq.ImchiBzBzisosupp}
\chi ''_{B_z B_z}  (z, \omega \rightarrow 0) \simeq (2S+1)\frac{e^2 \mu _0 ^2}{16 \pi h} \frac{\omega p_{\rm F,0}}{z} + \mathcal{O}(\omega ^3), \quad \text{isotropic FS.}
\end{equation}


\section{Derivation of low-frequency noise contribution from spin correlations}\label{Sec.suppspinnoise}

In this appendix we derive the contribution of the spin fluctuations to the magnetic noise spectrum arising from spin-$\frac{1}{2}$ fermions and show that it is subdominant compared to contribution from current fluctuations, Eqs.~\eqref{Eq.spinnoise}--\eqref{Eq.spinnoiseratio} of the main text. The vector potential at points $\vec{x}$ due to a given magnetization $\vec{M} (\vec{x}_s)$ is
\begin{eqnarray}
\vec{A} (\vec{x}) &=& \frac{\mu _0}{4\pi} \int d^3 \vec{x}_s \frac{\vec{M}(\vec{x}_s) \times (\vec{x}-\vec{x}_s)}{|\vec{x}-\vec{x}_s|^3} 
=\frac{\mu _0}{4\pi} \left(\int d^3 \vec{x}_s \frac{\nabla _s \times \vec{M}(\vec{x}_s)}{|\vec{x}-\vec{x}_s|}  - \int d^3 \vec{x}_s \nabla _s \times  \frac{\vec{M}(\vec{x}_s)}{|\vec{x}-\vec{x}_s|} \right),
\end{eqnarray}
where the second term can be written as a surface integral which vanishes for physical distributions $\vec{M} (\vec{x}_s)$. This vector potential gives rise to a magnetic field,
\begin{eqnarray}
\vec{B} (\vec{x}) &=& \nabla \times \vec{A} (\vec{x}) = -\frac{\mu _0}{4\pi} \int d^3 \vec{x}_s (\vec{x}-\vec{x}_s) \times \frac{\nabla _s \times \vec{M}(\vec{x}_s)}{|\vec{x}-\vec{x}_s|^3} 
\end{eqnarray}

Consider the magnetization of a 2D system
\begin{eqnarray}
\vec{M} (\vec{x}_s) = \vec{M} _{\vec{k}} \delta (z_s) e^{-i \vec{k} \cdot \vec{r}_s}, \quad \vec{x}_s = (\vec{r}_s, z_s), \quad \vec{r}_s = (x_s, y_s),
\end{eqnarray}
where $\vec{k} = (k_x, k_y)$ is a momentum in the plane parallel to the system. It can be shown that this gives rise to a magnetic field
\begin{eqnarray}
\vec{B} (\vec{x}) 
&=& k \frac{\mu _0}{2} e^{-i \vec{k}\cdot \vec{r}}  e^{-k|z|} \left(  i{\rm sgn} (z)  M_k^z \hat{k}
+ \vec{M}_{\vec{k}} - \hat{k} \vec{M}_{\vec{k}}\cdot \hat{k} \right).
\end{eqnarray}
The intrinsic spin magnetic dipole moment of an electron is given by $\vec{\mu}_ s = -g_s \mu _B \vec{S}/\hbar$. The magnetization operator in momentum space is
\begin{equation}
\vec{\hat{M}}_{\vec{k}} = -\frac{1}{2} g_s \mu _B  \sum _{\vec{p}} \hat{c}^\dagger _{\vec{p + k}}  \hat{\vec{\sigma} } \hat{c} _{\vec{p}},
\end{equation}
where $\hat{\vec{\sigma}} = (\hat{\sigma} ^x, \hat{\sigma} ^y, \hat{\sigma} ^z)$ denote the pauli matrices. At $x = y =0$, the generic magnetic field operator is given by
\begin{eqnarray}
\vec{\hat{B}} (z) 
&=& -\frac{1}{4} \mu _0 g_s \mu_B \sum_{\vec{k, p}} \hat{\vec{b}} _{\vec{p, k}}(z), \quad \hat{\vec{b}} _{\vec{p, k}}(z) = 
 k  e^{-k|z|}  \hat{c}^\dagger _{\vec{p + k}}   \left(  i{\rm sgn} (z)  \hat{\sigma} ^z \hat{k}
+ \vec{\hat{\sigma}} - \hat{k} \vec{\hat{\sigma}} \cdot \hat{k} \right) \hat{c} _{\vec{p}}.
\end{eqnarray}
Note that only the first term changes sign under $\vec{k} \rightarrow - \vec{k}$. At time $t > 0$,
\begin{equation}
\vec{\hat{B}} (z , t) = e^{i \mathcal{H}  t} \vec{\hat{B}} (z) e^{-i \mathcal{H} t}.
\end{equation}
For simplicity, let us consider a paramagnetically ordered system with an isotropic FS described by the the free fermion Hamiltonian $\mathcal{H} = \sum _{\vec{k}, s} c^\dagger _{\vec{k}, s} c_{\vec{k}, s}$ so that the spin and momentum sectors are decoupled.
The time-dependent magnetic noise at the origin $x=y=0$ at zero temperature is then
\begin{eqnarray}
\chi _{{\rm spin},B_i B_j} (z,t) &=& - \frac{i}{\hbar} \Theta (t) \left\langle \left[ \hat{B}_i (z,t), \hat{B}_j (z) \right] \right\rangle _0,
\end{eqnarray}
where the expectation value is taken over the ground state. In frequency space, one finds
\begin{eqnarray}
\chi _{{\rm spin},B_i B_j} (z,\omega) 
&=&  \frac{1}{16} \mu ^2 _0 g^2_s \mu^2_B \alpha _{ij}
\int _0 ^{\infty} \frac{dk}{2\pi}
k^3 e^{-2k|z|}
\chi _{0 \sigma} (k, \omega), \\
\alpha _{ij} &=& \alpha _{ji}  = {\rm Tr} \left[ \int _0 ^{2\pi} \frac{d\theta}{2\pi} \left(  -i{\rm sgn} (z)  \hat{\sigma} ^z \hat{k}
+ \vec{\hat{\sigma}} - \hat{k} \vec{\hat{\sigma}} \cdot \hat{k} \right) _i
 \left(  i{\rm sgn} (z)  \hat{\sigma} ^z \hat{k}
+ \vec{\hat{\sigma}} - \hat{k} \vec{\hat{\sigma}} \cdot \hat{k} \right) _j \right] = 2 \delta _{ij},\\
\chi _{0 \sigma} (k, \omega) &=& \frac{m^*}{2\pi}\frac{p_{\rm F,0}}{k}\left[\Psi _2 \left(\frac{\omega + i \delta}{k v_{\rm F}} - \frac{k}{2 p_{\rm F,0}}\right) - 
\Psi _2 \left(\frac{\omega + i \delta}{k v_{\rm F}} + \frac{k}{2 p_{\rm F,0}}\right) \right], \quad \Psi _2 (\zeta) = \zeta - {\rm sgn} ({\rm Re}~\zeta)\sqrt{\zeta^2 - 1},\nonumber \\
\end{eqnarray}
where the trace in $\alpha _{ij}$ is performed over spin degrees of freedom, $f_{\vec{k}} =\Theta ( p_{\rm F,0} - k) $ is the zero temperature Fermi distribution, $\chi _{0 \sigma} (k, \omega)$ is the 2D one-spin Lindhard function, and $m^*$ the quasiparticle mass.
%
The imaginary part is
%
\begin{eqnarray}
\chi ''_{{\rm spin},B_i B_j} (z,\omega)
&=& -\delta _{ij} \frac{1}{16} \mu ^2 _0 g^2_s \mu^2_B \frac{m^*}{4\pi ^2} p_{\rm F,0} ^4 \cdot 2
\int _0 ^{\infty} d \tilde{k}
\tilde{k}^2 e^{-2\tilde{k} p_{\rm F,0}|z|} 
\left[ \Theta (1 - s^2_-) \sqrt{1 - s^2_-} 
- \Theta (1 - s^2_+) \sqrt{1 - s^2_+} \right], \nonumber \\ \\
s_\pm &=& s \pm \frac{k}{2p_{\rm F,0}} \equiv  \frac{\tilde{\omega}}{\tilde{k}} \pm \frac{\tilde{k}}{2}, \quad \tilde{\omega} = \frac{\omega}{2 E_{\rm F}} = \frac{\omega}{v_{\rm F}p_{\rm F,0}}, \quad \tilde{k} = \frac{k}{p_{\rm F,0}}.
\end{eqnarray}
Solving for the constraints provided by the Heaviside function, the integral becomes
\begin{eqnarray}
\chi ''_{{\rm spin},B_i B_j} (z,\omega)
&=&  -\delta _{ij} \frac{1}{16} \mu ^2 _0 g^2_s \mu^2_B \frac{m^*}{4\pi ^2} p_{\rm F,0} ^4 \cdot 2\sum _{s = \pm} (-s)
\int _{\tilde{k}_{s, <}} ^{\tilde{k}_{s, >}} d \tilde{k}
\tilde{k} e^{-2\tilde{k} p_{\rm F,0}|z|} 
\sqrt{\left(\tilde{k}^2 - \tilde{k}_{s, <}^2\right)\left(\tilde{k}_{s , >}^2 - \tilde{k}^2\right)},\\
\tilde{k}_{+, <} &=& 1 - \sqrt{1 - 2\tilde{\omega}} = \tilde{\omega} + \mathcal{O}(\tilde{\omega}^2),\quad \hspace{1.5ex}\tilde{k}_{+, >} = 1 + \sqrt{1 - 2\tilde{\omega}} = 2 - \tilde{\omega} + \mathcal{O}(\tilde{\omega}^2),\\
\tilde{k}_{- , <} &=& -1 + \sqrt{1 + 2\tilde{\omega}} = \tilde{\omega}+ \mathcal{O}(\tilde{\omega}^2), \quad \tilde{k}_{-, >} = 1 + \sqrt{1 + 2\tilde{\omega}} = 2 + \tilde{\omega}+ \mathcal{O}(\tilde{\omega}^2).
\end{eqnarray}

For $z > 0$ and in the large distance limit, $2\tilde{k}_{\pm, >} p_{\rm F,0}z \simeq 4 p_{\rm F,0}z \gg 1$, i.e. $z \gg 1/4p_{\rm F,0}$, the above can be approximated by
\begin{eqnarray}
\chi ''_{{\rm spin},B_i B_j} (z,\omega)
&\simeq &  \delta _{ij} \frac{1}{16} \mu ^2 _0 g^2_s \mu^2_B \frac{m^*}{4\pi ^2} p_{\rm F,0} ^4 \cdot 2\sum _{s = \pm} s
\tilde{k}_{s , >} \int _{\tilde{k}_{s, <}} ^{\infty} d \tilde{k}
\tilde{k} e^{-2\tilde{k} p_{\rm F,0}z} 
\sqrt{\tilde{k}^2 - \tilde{k}_{s, <}^2} \\
&\simeq &  \delta _{ij} \frac{1}{16} \mu ^2 _0 g^2_s \mu^2_B \frac{m^*}{4\pi ^2} p_{\rm F,0} ^4 \cdot 2\sum _{s = \pm} 2s
(1 - s \tilde{\omega})
\left[ \frac{\tilde{k}^3_{s, <}}{\alpha } K_2 \left(\alpha \right) \right], \quad \alpha = 2\frac{\omega z}{v_{\rm F}}.
\end{eqnarray}
where $K_2$ denotes the K-Bessel function of the second kind. 
Expanding $K_2$ in the $\alpha \ll 1$ limit, or equivalently, the low-frequency limit $\omega \ll v_{\rm F}/2z$,
\begin{eqnarray}
\chi ''_{{\rm spin},B_i B_j} (z,\omega)
&\simeq & -\delta _{ij} \frac{e^2 \mu ^2 _0}{16 \pi ^2} \left(\frac{g_s}{2}\right)^2 \left(\frac{m^*}{2m_0}\right)^2 \cdot
 \frac{\omega p_{\rm F,0}}{z}  \frac{2}{p_{\rm F,0}^2z^2},
\end{eqnarray}
where $m_0$ is the electron rest mass.
Comparing this to the current contribution [Eq.~\eqref{Eq.ImchiBzBzisosupp} with $S = 1/2$], we find
\begin{eqnarray}
\left|\frac{\chi ''_{{\rm spin}, B_{z} B_{z}} (z,\omega \rightarrow 0)}{\chi ''_{B_{z} B_{z}} (z,\omega \rightarrow 0)}\right|
&\simeq & \left(\frac{g_s}{2}\right)^2 \left(\frac{m^*}{2m_0}\right)^2 \cdot
\frac{2}{p_{\rm F,0}^2z^2},
\end{eqnarray}
i.e. a $1/p_{\rm F,0}^2z^2$ suppression of the spin fluctuation contribution relative to the current fluctuation contribution and can therefore be neglected at large distances $z \gg 1/p_{\rm F,0}$.

\clearpage

\subsection*{Comments on Ref.~\onlinecite{Chatterjee2019}}

Our results for the spin fluctuations above are identical for the $\mathbb{Z}_2$ and U(1) spin liquids with a spinon FS. This disagrees with some results obtained in Ref.~\onlinecite{Chatterjee2019}. Specifically, Table I of that reference claims the noise in the two types of spin liquids has a different dependence on temperature in the clean case when $T\gg \omega$. In contrast, we obtain a linear in $T$ behavior from Eq.~(\ref{Eq.noise}) in both cases. In addition, Table II of Ref.~\onlinecite{Chatterjee2019} predicts a different dependence on height $z$ (called $d$ therein) for the two spin liquids in clean systems in the limit $T\ll \omega$. We obtain a $z^{-3}$ dependence of the noise from spin fluctuations in both cases. In the following we point out specific mistakes in Ref.~\onlinecite{Chatterjee2019}, that led to the erroneous conclusions.

The error in the derivation of the $z$-dependence originates from the incorrect approximation of the following integral in Eq.~(B7) in the appendix of Ref.~\onlinecite{Chatterjee2019}, 
\begin{eqnarray}
\int_{(2\mu - \omega)/v_{\rm F} q} ^{(2\mu + \omega)/v_{\rm F} q} du~\sqrt{u^2 - 1}~\cancel{\simeq}~ \frac{\omega}{v_{\rm F} q}\left(\frac{\mu}{v_{\rm F}q}\right)^2.
\end{eqnarray}
Writing $u = u_0 + x$, $u_0 = 2\mu/v_{\rm F}q \simeq p_{\rm F}/q \gg 1$ and $s = \omega/v_{\rm F}q \ll u_0$, we find instead
\begin{eqnarray}
\int_{(2\mu - \omega)/v_{\rm F} q} ^{(2\mu + \omega)/v q} du~\sqrt{u^2 - 1} 
\simeq 2 u_0 s - \frac{s}{u_0}
\simeq \frac{4 \mu \omega}{(v_{\rm F} q)^2} + \frac{\omega}{2\mu},
\end{eqnarray}
so that the noise [Eq.~(B8) in the appendix of Ref.~\onlinecite{Chatterjee2019}] in the $\omega \rightarrow 0$ limit should read
\begin{eqnarray}
\chi ''_{\rm spin} \propto \int _{\omega /v_{\rm F}} ^{\infty} dq~q^3 e^{-2qz} \frac{q^2}{\sqrt{v_{\rm F}^2q^2 - \omega^2}}\frac{\omega \mu }{(v_{\rm F}q)^2} 
\propto \omega \int _0 ^{\infty} dq~q^2 e^{-2qz}
\approx \frac{\omega}{z^3},
\end{eqnarray}
identical to the result they obtained for the U(1) quantum spin liquid with spinon FS and our result above.

The error in the $T$ dependence originates from an incorrect approximation of the integral in Eq.~(B9) in the appendix of Ref.~\onlinecite{Chatterjee2019}, 
\begin{align}
 \int_1^\infty du\frac{\sqrt{u^2-1}}{\cosh^2[(v_Fqu-2\mu)/2T]}\neq \Bigl(\frac{\mu}{v_Fq}\Bigr)^2.
\end{align}
Because the $\cosh^2$ is exponentially large in its argument, the main contribution to this integral originates from values $u$ in an interval of width $\sim T/v_F q$ around $u\simeq 2\mu/v_Fq$. For $\mu\gg T,v_Fq$ this restricts the integral to values $u\gg 1$ and we can approximate
\begin{align}
 \int_1^\infty du\frac{\sqrt{u^2-1}}{\cosh^2[(v_Fqu-2\mu)/2T]}
 \simeq&  \int_1^\infty du\frac{u}{\cosh^2[(v_Fqu-2\mu)/2T]}\\
 \sim &\frac{\mu T}{(v_Fq)^2},
\end{align}
where we have used the indefinite integral
\begin{align}
 \int du \frac{u}{\cosh^2( bu-a)}=-\frac{\log[\cosh(a - b u)] + b u \tanh(a - b u)}{b^2}
\end{align}
and expanded to leading order in $v_Fq/\mu$ and $T/\mu$.
This means that there should be an extra factor of $T/\mu$ in the noise in Eq.~(29) and Eq.~(B10) of Ref.~\onlinecite{Chatterjee2019} and the noise for the the $\mathbb{Z}_2$ FS should also be linear in $T$ in Table I in agreement with our results.

\clearpage

\section{Effects of collision on low-frequency noise}\label{Sec.suppnoisewcoll}

In this appendix, we analyze in detail the low-frequency noise obtained in the presence of collisions leading to the discussion in Sec.~\ref{Sec.noisecoll} of the main text. For simplicity, we consider isotropic systems with circular FSs. In this case, the non-trivial components of the magnetic noise from current fluctuations following Eqs.~\eqref{Eq.chiBxBxinv}--\eqref{Eq.chiBzBzinv} are 
\begin{align}
\chi ''_{B_i B_i}  (z, \omega) 
&= \frac{\mu _0 ^2 \omega}{16 \pi}\int dq q e^{-2qz} \left( {\rm Re}~\sigma  _{\|} (q, \omega) + {\rm Re}~\sigma _{\perp} (q, \omega)  \right), \label{Eq.chiBBiso} \\
\chi ''_{B_z B_z}  (z, \omega)
&= \frac{\mu _0 ^2 \omega}{8 \pi}\int dq q e^{-2qz} {\rm Re}~\sigma _{\perp}  (q, \omega). \label{Eq.BBresp}
\end{align}
%
Of particular interest is the out-of-plane component, which can be approximated by approximating the respective transverse conductivities as
\begin{eqnarray}
{\rm Re}~\sigma _{\perp}  (q, \omega) &\simeq& \Theta \Big(q - \tilde{q}_{(s)}(\omega)\Big)\sigma  _\perp (q, 0^+) + \Theta \Big(\tilde{q}_{(s)}(\omega) - q\Big) {\rm Re}~\sigma _\perp (q, \omega).
\end{eqnarray}
As per the discussion in Ref.~\onlinecite{Chatterjee2019}, the distance $z$ sets a cutoff scale $q_z = 1/2z$ at which the system's response is probed. For cases when the various momentum scales $\lbrace q_z, q_*, q_{**}, \tilde{q}_{(s)}(\omega) \rbrace$ (also $q_0$ for the SFSS) are well separated, the noise can be approximated by integrating the corresponding expression for the transverse conductivity, found in Tables.~\ref{Table.transcondwmetal}--\ref{Table.momcutoff} and Table.~\ref{Table.transcond} in the main text, over the region of $q$ in which $q_z$ lies. 
This approximation gives rise to a $\omega ^{l+1} z^{-(n+2)}$ dependence of the noise when the corresponding transverse conductivity scales as Re~$\sigma _\perp (q, \omega) \propto \omega ^{l} q^n$ for $n > -2$. For the special case of $n = -2$, i.e. the quasi-static transverse conductivity in the hydrodynamic transport regime. In this case, instead of the naive $z^{0}$-dependence, the $z$-dependence is well approximated by 
\begin{equation}\label{Eq.hydrozscale}
f(z) = {\rm Ei}(-2q_{*}z) - {\rm Ei}(-2q_{**}z),
\end{equation}
where Ei$(x)$ denotes the exponential integral function. Consequently, the $z$-dependence of the noise is given by the distance scale at which it is probed at with respect to the associated distance scales set by the above momentum scales, $z_i = 1/2q_i$. This accounts for the different $z$-dependences in the various regimes shown in Fig.~\ref{Fig.Bperpnoise} and Tables.~\ref{Table.chiBzBzscalemetal}--\ref{Table.chiBzBzscalespinon} below.
As an illustration, we consider the case of the SFSS with $q_0 \ll q_{**} \ll \tilde{q}_{(s)}(\omega) \ll q_*$, for which case the noise at distances $\tilde{z}_s (\omega) \ll z \ll z_{**}$ can be approximated as
\begin{eqnarray}
\chi ''_{B_z B_z}  (z, \omega)
&\simeq & \frac{\mu _0 ^2 \omega}{8 \pi}\int _{q_{**}} ^{\tilde{q}_{(s)}(\omega)}dq~q e^{-2qz} {\rm Re}~\sigma _{\perp}  (q, \omega) \nonumber \\
&\simeq & \frac{\mu _0 ^2 \omega}{8 \pi} \int _{0} ^{\infty}dq~q^{7} e^{-2qz} g_S \frac{e^2}{h}\frac{(1+F_1)^2}{8} \frac{v_{\rm F}^2 p_{\rm F,0}}{q_C \omega ^2}\frac{c^4}{\omega _p^4} \nonumber \\
&\simeq & g_S (1+F_1)^2\frac{e^2\mu _0 ^2}{64 \pi h} \frac{v_{\rm F}^2 p_{\rm F,0}}{q_C \omega }\frac{c^4}{\omega _p^4} \frac{\Gamma (9)}{(2z)^8}, 
\end{eqnarray}
where $\Gamma$ denotes the gamma function. 

Therefore, in order to access the system's response in the quasi-static quantum regime, the noise should be probe at distances $z \ll z_*$, where $z_* \sim l_{\rm mfp}$ the system mean free path in the $T \rightarrow 0$ limit.


\begin{table}[h]
\begin{tabular}[b]{cccc}
\hline\\[-1.em]
\hline\\[-1.em]
Metals & Diffusive \quad & Hydrodynamic \quad & Quantum \\ [.1em]
& $z \gg z_{**}$ & $z_{**} \gg z \gg z_{*}$ & $z_{*} \gg z$\\ [.5em]
\hline\\[-1.em]
$\chi ''_{B_z B_z} (z \gg \tilde{z} (\omega)) $ & {\large $\frac{\omega}{z^2}$} & $\omega f(z)$ & {\large $\frac{\omega}{z}$} \\ [1.em]
\hline\\[-1.em]
$\chi ''_{B_z B_z} (z \ll \tilde{z}(\omega)) $ & {\large $\frac{1}{\omega z^2}$} & {\large $\frac{1}{\omega z^4}$} & {\large $\frac{1}{\omega ^3 z^4}$} \\ [1.em]
\hline\\[-1.em]
\hline\\[-1.em]
\end{tabular}
\caption{Frequency ($\omega $) and distance ($z$) dependence of magnetic noise from current fluctuations in various transport regimes in metals, where $f(z)$ is given in Eq.~\eqref{Eq.hydrozscale}.
}\label{Table.chiBzBzscalemetal}
\end{table}

\begin{table}[t]
\begin{tabular}[b]{cccc}
\hline\\[-1.em]
\hline\\[-1.em]
SFSSs & Diffusive & Hydrodynamic & Quantum \\ [.1em]
& $z \gg z_{**}$ & $z_{**} \gg z \gg z_{*}$ & $z_{*} \gg z$\\ [.5em]
\hline\\[-1.em]
$\chi ''_{B_z B_z} (z \gg \tilde{z}_s (\omega)) $ & {\large $\frac{\omega}{z^2}$} & $\omega f(z)$ & {\large $\frac{\omega}{z}$} \\ [1.em]
\hline\\[-1.em]
$\chi ''_{B_z B_z} \big(z_0 \ll z \ll \tilde{z}_s(\omega) \big) $ & {\large $\frac{1}{\omega z^6}$} & {\large $\frac{1}{\omega z^8}$} & {\large $\frac{1}{\omega z^7}$} \\ [1.em]
\hline\\[-1.em]
$\chi ''_{B_z B_z} (z \ll z_0) $ & {\large $\frac{\omega ^3}{z^2}$} & {\large $\frac{\omega ^3}{z^4}$} & {\large $\frac{\omega^{13/3}}{z^2}$} \\ [1.em]
\hline\\[-1.em]
\hline\\[-1.em]
\end{tabular}
\caption{
Analogue of Table.~\ref{Table.chiBzBzscalemetal} for SFSSs.
}\label{Table.chiBzBzscalespinon}
\end{table}


\begin{figure}[t]
\includegraphics[scale=1.0]{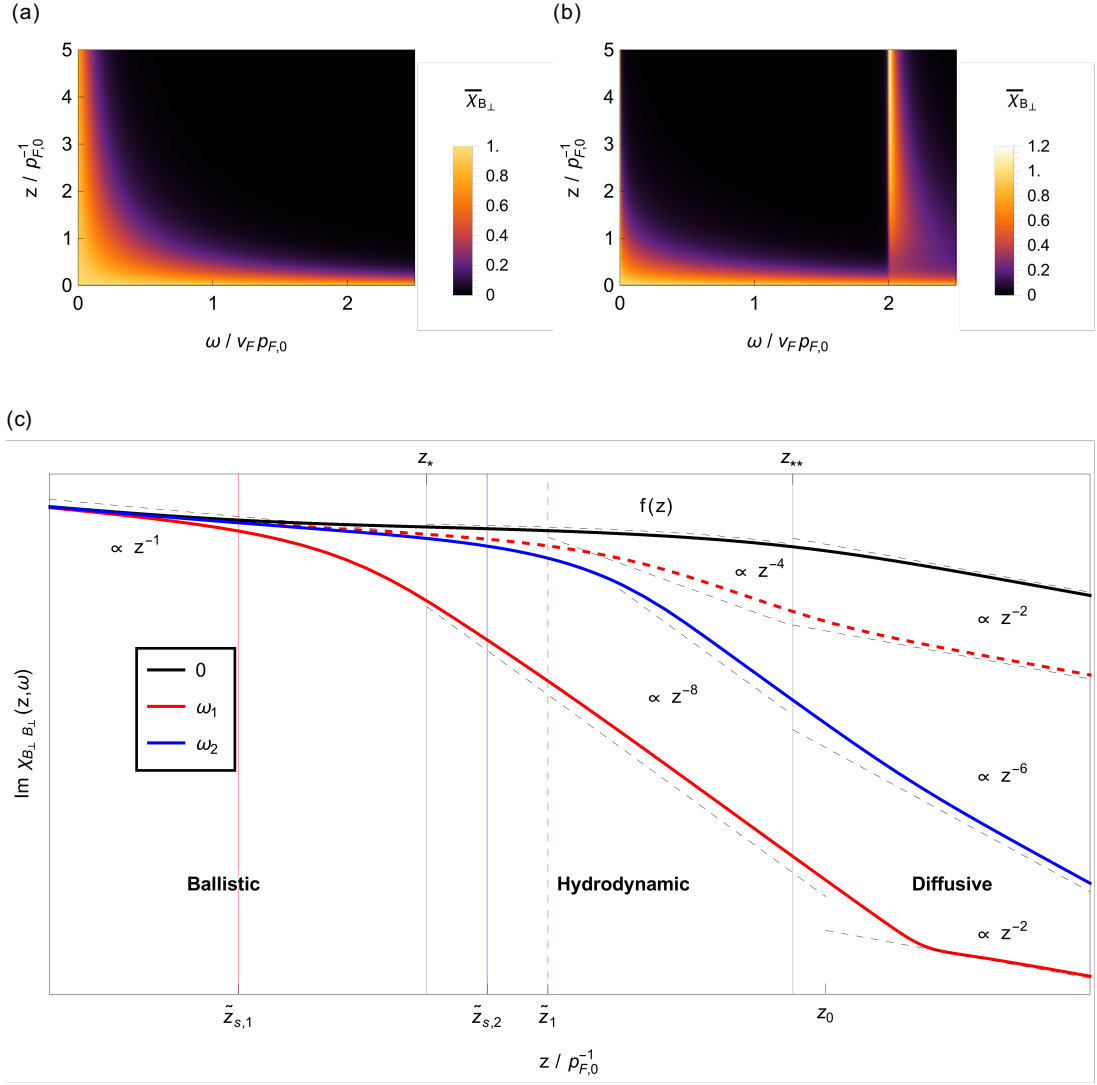}\\
\caption{
Plots of the rescaled magnetic noise $\bar{\chi} _{B_z} \propto \chi ''_{B_z B_z} z/\omega p_{\rm F,0}$, as a function of the out-of-plane distance $z$ from the 2D sample, as well as frequency $\omega$, for (a) an isotropic metal and (b) an isotropic SFSS with $\omega _p = 2 v_{\rm F} p_{\rm F,0}$, for which there is an additional peak.
(c) Magnetic noise due to current fluctuations obtained from transverse conductivities at the same frequencies $\omega_1$ and $\omega_2$ shown in Fig.~\ref{Fig.condspinonsupp} for SFSS (solid plots) and metals (dashed plot).
}
\label{Fig.Bperpnoise}
\end{figure}

\end{document}